\documentclass[review]{elsarticle}

\pdfoutput=1

\usepackage{lineno,hyperref,amsmath,amssymb,graphicx}
\usepackage{subfigure,url}

\journal{Physics Reports}

\biboptions{compress}

\def\be{\begin{equation}}
\def\ee{\end{equation}}
\def\bea{\begin{align}}
\def\eea{\end{align}}
\def\nn{\nonumber}

\DeclareMathOperator{\real}{Re}
\DeclareMathOperator{\imag}{Im}
\def\ket#1{|{#1}\rangle}
\def\bra#1{\langle{#1}|}

\newcommand{\eqn}[1]{Eq.~(\ref{#1})}

\newcommand{\secdec}{{\textsc{SecDec}}}

\newcommand{\pysecdec}{py{\textsc{SecDec}}}

\newcommand{\eps}{\epsilon}
\newcommand{\rd}{{\mathrm{d}}}

\newcommand{\mhh}{m_{\mathrm{hh}}}

\begin{document}

\begin{frontmatter}

\title{Collider Physics at the Precision Frontier}

\author{Gudrun Heinrich}
\address{Karlsruhe Institute of Technology, Institute for Theoretical
  Physics, Wolfgang-Gaede-Str.~1, 76131 Karlsruhe, Germany}


\ead{gudrun.heinrich@kit.edu}


\begin{abstract}
The precision frontier in collider physics is being pushed at
impressive speed, from both the experimental and the theoretical
side. The aim of this review is to give an overview of recent
developments in precision calculations within the Standard Model of
particle physics, in particular in the Higgs sector.
While the first part focuses on phenomenological results, the second part reviews some of the techniques which allowed the
rapid progress in the field of precision calculations.
The focus is on analytic and semi-numerical techniques for multi-loop
amplitudes, however fully numerical methods as well as subtraction schemes for infrared divergent real radiation
beyond NLO are also briefly described.
 \end{abstract}



\end{frontmatter}

\newpage

\tableofcontents


\section{Introduction}

\subsection{Motivation}

The Higgs boson, discovered in
2012~\cite{Aad:2012tfa,Chatrchyan:2012ufa}, is the ``youngest'' in the
discovery list of 
elementary particles, and also the most peculiar. So far it is the only
scalar particle which seems not to be composite, and if it was
the only one of its kind to exist this would be intriguing from a
theory point of view. However, it seems more natural to think of it as
the first milestone towards a new journey, and therefore exploring the
Higgs sector is one of the main pillars in the planning process of the
future collider physics program~\cite{deBlas:2019rxi,Strategy:2019vxc}.

The ``precision frontier'' plays a vital role in this program, for at least two reasons.
On one hand to match the upcoming increased experimental precision, in particular 
at the High-Luminosity LHC and even more so at future colliders,
on the other hand also because current measurements seem to indicate that
a possible New Physics sector is associated with energy scales that are high enough to leave
rather subtle imprints on observables near the electroweak scale.
Therefore the theoretical predictions need to be precise enough to tell apart New Physics effects from
discrepancies to the data due to insufficient modeling.
The latter can have various origins, like missing higher orders in the
description of the hard scattering, large logarithms spoiling the
convergence of the perturbative series, 
uncertainties related to parton showers and parton densities, parametric uncertainties in the couplings and masses,
or non-perturbative effects.
The last few years have seen enormous progress to push the precision in the different aspects.
The purpose of this article is to review parts of this endeavour, focusing in particular on the Higgs sector and
techniques for higher order perturbative calculations.

\subsection{Brief history}
The progress in the calculation of higher order corrections to
processes relevant for LHC physics can be illustrated by considering
the so-called ``Les Houches Wishlist''. This wishlist actually had its origin in
a table put together in 2004 by experimentalists for a workshop at Fermilab
to identify processes where NLO predictions would be ``nice to have''
for the Tevatron Run II. It contained multi-particle final state
processes such as 
$WWW+b\bar{b}\,+\leq 3$\,jets. At the Les Houches 2005 Workshop at TeV
Colliders, a {\it realistic} wish list was created, that contained
$2\to 4$ processes at NLO as the cutting edge~\cite{Buttar:2006zd}. In the years to follow,
a ``NLO revolution'' took place, which led to the development of
automated tools to calculate NLO corrections to
high-multiplicity-processes.
In 2011, the original NLO wishlist was not only ticked, but processes
that have not even been considered feasible in 2005 had been
calculated at NLO, for example 4--jet production in hadronic collisions~\cite{Bern:2011ep},
and in the years to follow even processes such as $W$+5--jet
production~\cite{Bern:2013gka} or  di-photon plus 3--jet
production~\cite{Badger:2013ava} became available at NLO.
Therefore, in the years after, the list has been continued as {\it
  precision wish list}, where NNLO accuracy,  the need for NLO
electroweak (EW) corrections and the description of decays beyond the
narrow width approximation were the main subjects in the field of fixed-order
precision calculations.
Impressive NLO calculations in this context are for example the
calculation of NLO EW and QCD corrections to Higgs production in
association with off-shell top-antitop pairs~\cite{Denner:2016wet}, or
the full NLO QCD corrections to off-shell $t\bar{t}b\bar{b}$
production~\cite{Denner:2020orv}, which is a $2\to 8$ process.
For a recent review on electroweak radiative corrections we refer to Ref.~\cite{Denner:2019vbn}.

It is often said that an ``NNLO revolution'' took place in the
years after 2015. This was mainly due to the development of efficient
subtraction schemes for IR divergent real radiation at NNLO.
Despite the enormous progress, this subject is currently still under intense development, also in view of extensions to N$^3$LO.

On the loop side, the main challenges in fixed order precision calculations are
to increase the number of final state particles in NNLO calculations,
which mostly means going from $2\to 2$ to $2\to 3$ multiplicities at two loops, as
well as increasing the number of mass scales in processes at
two loops and beyond.
In Section~\ref{sec:amplitudes} we will report on progress in the
technical aspects of such calculations.

\subsection{Purpose and structure of this review}

The purpose of this review is to give an overview of the current state
of the art in precision calculations for  high
energy collider processes, in particular processes that can be
measured at the LHC.
Its main focus is of a rather technical nature, presenting various methods and techniques that led
to the achievements in precision phenomenology which are described in
the first part of the review. The phenomenological part, Section~\ref{sec:Higgspheno},  is limited to
the Higgs sector, first because it is the least explored part of the
Standard Model so far, second because the intention is to rather give a more detailed account of Higgs physics
than a superficial account of precision calculations in all sectors of
the Standard Model. The description of precision calculations beyond
the Standard Model (BSM) would exceed the scope of this review and
therefore BSM physics is not treated here.
The aim of Section~\ref{sec:Higgspheno} is not to review Higgs
phenomenology in full detail, but rather to highlight the
calculations available in the literature that led to the
state-of-the-art predictions for the considered processes
and to point to open issues in
the quest for higher precision.
Precision phenomenology is a very
rapidly developing field. 
Recent phenomenological studies and results in the Higgs sector are
reviewed in
Refs.~\cite{Dawson:2018dcd,DiMicco:2019ngk,Spira:2016ztx}, for an overview
not limited to the Higgs sector we point to
Ref.~\cite{Amoroso:2020lgh}, containing also the ``Les Houches 2019 wishlist''.

Section~\ref{sec:amplitudes} contains the description of techniques to
calculate integrals and amplitudes beyond one loop. It focuses on
recent developments and describes  both analytic and numerical
approaches. Among the numerical approaches, sector decomposition is
described in some more detail.

In Section~\ref{sec:IRsubtraction}, methods to treat infrared
divergent real radiation at NNLO and beyond are listed briefly.
A summary and an outlook are given in Section~\ref{sec:summary}.

\section{Current status of precision Higgs phenomenology}
\label{sec:Higgspheno}

\begin{figure}[htb]
  \centering
    \includegraphics[width=0.65\textwidth]{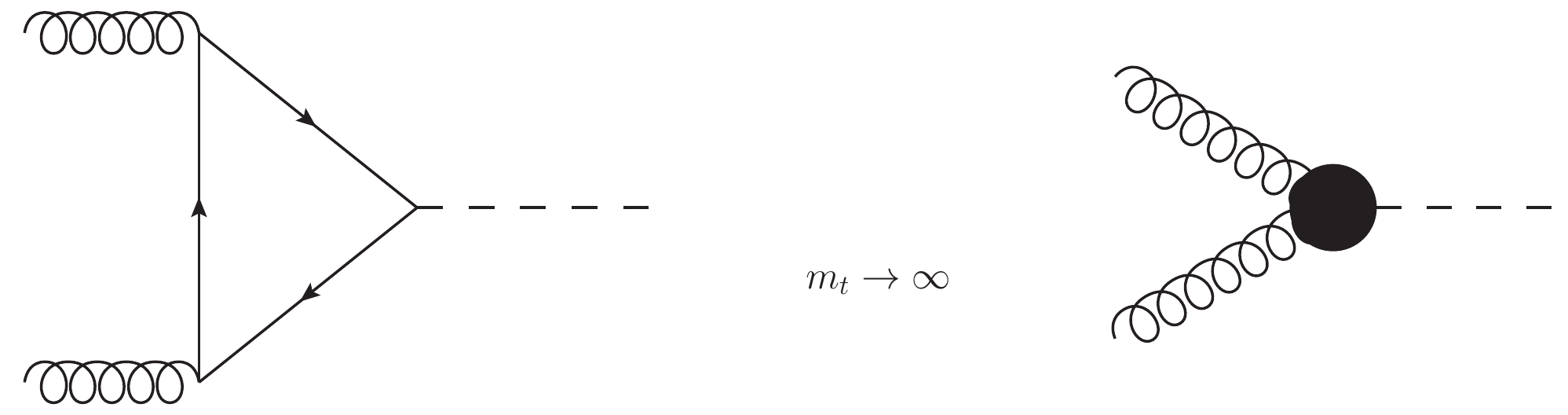}
\caption{Illustration of the heavy top limit (HTL).\label{fig:HTL}}
\end{figure}

QCD corrections in the Higgs sector are special as the Higgs boson does not couple directly to gluons.
Therefore, the precision calculations for  loop-induced processes, such as Higgs (plus jets) production in gluon fusion, can be roughly divided into
two categories: (a) calculations based on an effective Lagrangian
containing effective Higgs-gluon couplings which arise in the $m_t\to
\infty$ approximation, also called ``heavy top limit (HTL)'', and (b) calculations in the full Standard
Model. As the HTL shrinks the top quark loop mediating Higgs-gluon
interactions to a point (see Fig.~\ref{fig:HTL}), calculations in the HTL start at tree level
and involve only massless partons in the five-flavour scheme,
in contrast to the full SM where the leading order Higgs-gluon interaction is loop-induced.
Therefore, calculations in the HTL can be pushed to higher orders more
easily than calculations in the full SM.
In fact, most calculations of highest available order for
observables at hadron colliders are in the Higgs sector.

\subsection{N$^3$LO corrections and beyond}


The predictions for hadron collider cross sections beyond NNLO available to date are
mostly at the inclusive cross section level.
However, very recently, fully differential predictions for Higgs boson production via gluon fusion up to N$^{3}$LO  in QCD have been achieved~\cite{Chen:2021isd}, finding that the N$^3$LO corrections for
fiducial distributions of two-photon final states can be non-uniform and partly larger than the corrections in the inclusive case.
N$^3$LO  results for the Higgs boson rapidity distribution have been calculated prior to Ref.~\cite{Chen:2021isd} in Refs.~\cite{Cieri:2018oms,Dulat:2018bfe}.
The resummed $q_T$-spectrum at N$^3$LL$^\prime$+N$^3$LO, both inclusively and with fiducial cuts, has been presented in Ref.~\cite{Billis:2021ecs}, featuring the highest precision achieved so far for a transverse momentum distribution at a hadron collider.

For the inclusive case in the threshold approximation, the N$^{3}$LO corrections to Higgs production in gluon fusion in the HTL,  have been calculated in Refs.~\cite{Anastasiou:2015ema,Anastasiou:2016cez}.
The threshold approximation has been overcome in Ref.~\cite{Mistlberger:2018etf}, presenting the exact  N$^{3}$LO inclusive cross section in the HTL.
Furthermore, N$^4$LO soft and virtual corrections to inclusive Higgs production have been calculated recently~\cite{Das:2020adl,Ahmed:2020nci}.

Higgs production in bottom quark fusion is also known to  N$^{3}$LO~\cite{Mondini:2019gid,Duhr:2019kwi}, including matching the 4- and 5-flavour schemes to third order in the strong coupling~\cite{Duhr:2020kzd}.

Inclusive Higgs boson production in vector boson fusion (VBF) at N$^{3}$LO has been calculated in Ref.~\cite{Dreyer:2016oyx},
Higgs boson pair production in VBF in Ref.~\cite{Dreyer:2018qbw}, both based on the projection-to-Born method~\cite{Han:1992hr,Cacciari:2015jma} described in Section \ref{sec:IRsubtraction}.
N$^{3}$LO corrections to Higgs boson pair production in gluon fusion in the HTL have been calculated in Refs.~\cite{Chen:2019lzz,Banerjee:2018lfq,Chen:2019fhs}.

The Drell-Yan cross section to third order in the strong coupling has been presented in Ref.~\cite{Duhr:2020seh}, charged current Drell-Yan production at N$^{3}$LO in Ref.~\cite{Duhr:2020sdp}.
Very recently,  predictions for fiducial cross section of the Drell-Yan process at N$^3$LL+N$^3$LO have been calculated in Ref.~\cite{Camarda:2021ict}.

N$^{3}$LO corrections to jet production in deep inelastic scattering~\cite{Currie:2018fgr} as well as charged current DIS~\cite{Gehrmann:2018odt} also have been calculated,  using the projection-to-Born method.

It should also be mentioned that, apart from hadron collider cross sections, more inclusive quantities calculated  at N$^3$LO  or beyond  are known
since some time, such as deep inelastic structure functions~\cite{Vermaseren:2005qc} or the Higgs decay to hadrons,
which is available at N$^{4}$LO~\cite{Baikov:2005rw,Baikov:2008jh,Baikov:2010je,Baikov:2012er,Davies:2017xsp,Herzog:2017dtz}.
For more details about multi-loop results we refer to Section~\ref{sec:multi-loop}.

\subsection{Higgs production in gluon fusion}

While in the heavy top limit N$^{3}$LO corrections are available as listed above, 
Higgs production with full $m_t$-dependence has been calculated up to NLO (two loops) in Refs.~\cite{Spira:1995rr,Harlander:2005rq,Anastasiou:2006hc,Aglietti:2006tp,Anastasiou:2009kn}.
The NNLO corrections in the HTL have  been calculated in Refs.~\cite{Harlander:2002wh,Catani:2001cr,Anastasiou:2002yz,Ravindran:2003um}.
Finite top quark mass effects at NNLO have been addressed in Refs.~\cite{Harlander:2009bw,Pak:2009bx,Harlander:2009mq,Pak:2009dg,Harlander:2009my,Pak:2011hs} based on a $1/m_t$-expansion, finding corrections that do not exceed 1\% after rescaling with  exact NLO results.
The NNLO$_{HTL}$  corrections have been implemented in the programs {\sc HiGlu}~\cite{Spira:1995mt}, {\sc iHixs}~\cite{Anastasiou:2011pi,Anastasiou:2012hx}, {\tt SusHi}~\cite{Harlander:2012pb} and  {\sc ggHiggs}~\cite{Ball:2013bra,Bonvini:2014jma}. The latter can be combined with {\sc Troll}~\cite{Bonvini:2014joa}, which in its latest version provides threshold resummation up to N$^3$LL~\cite{Bonvini:2016frm} and joint large-$x$ and small-$x$ resummation~\cite{Bonvini:2018xvt}. 
In Ref.~\cite{Marzani:2008az}, the leading high energy behaviour with finite top mass has been calculated at NNLO, resulting in an estimate of the effect of the high energy $m_t$-dependence on the  NNLO K-factor to be of the order of a few per cent.
Three-loop corrections to the Higgs-gluon vertex with a massive quark loop have been obtained by combining information from the
large-$m_t$ and threshold expansions with the help of a conformal mapping and a Pad\'e approximation~\cite{Davies:2019nhm}.
For the subset of three-loop diagrams which contain a closed massless quark loop, analytic results  in terms of harmonic polylogarithms have been obtained~\cite{Harlander:2019ioe}.
The first N$^3$LO result which incorporates finite top quark mass terms in a large-mass expansion up to four-loop order has been presented in Ref.~\cite{Davies:2019wmk}.

The light quark loops and in particular effects of the interference of diagrams containing top quark loops with light quark loops have moved into the focus recently.
The leading (double) logarithmic corrections in $\ln(m_H/m_q)$
 have been evaluated to all orders in $\alpha_s$ in Refs.~\cite{Liu:2017vkm,Liu:2018czl}.
The effect of bottom quarks on the Higgs transverse momentum spectrum for  $m_b \lesssim p_T \lesssim m_t$ has been studied  in Ref.~\cite{Caola:2018zye} by matching the NLO calculation with NNLL transverse momentum resummation, using two different schemes.
 This led to an uncertainty of the order of 15--20\% on the top-bottom interference contribution to the $p_T$-spectrum.
 Since the interference amounts to about 5\% of the full $p_T$-spectrum, it was concluded that unknown higher order $b$-quark mass effects can modify the Higgs transverse momentum distribution by a few percent.
 QCD corrections to the interference contribution of $ggH$ and $c\bar{c}H$ Higgs production modes have been studied in Ref.~\cite{Bizon:2021nvf}.

 Exact results for the 3-loop form factor with a single massive quark and otherwise light quarks have been presented in Ref.~\cite{Czakon:2020vql},
 based on a numerical solution of a system of differential equations for the occurring master integrals.
 The results confirm that an approach based on Pad\'e approximants as used in Ref.~\cite{Davies:2019nhm} is sufficient to obtain sub-percent precision for physical observables as long as a vanishing $b$-quark mass is assumed.
In Ref.~\cite{Prausa:2020psw}, an analytic three-loop result for the
leading colour contribution to the Higgs-gluon form factor in QCD is presented. In contrast to the case with massless quarks~\cite{Harlander:2019ioe},  this calculation requires the introduction of a new class of iterated integrals with integration kernels involving elliptic integrals which are not iterated integrals of modular forms, see Section~\ref{sec:elliptic} for more details.

\begin{figure}[htb]
  \centering
    \includegraphics[width=0.5\textwidth]{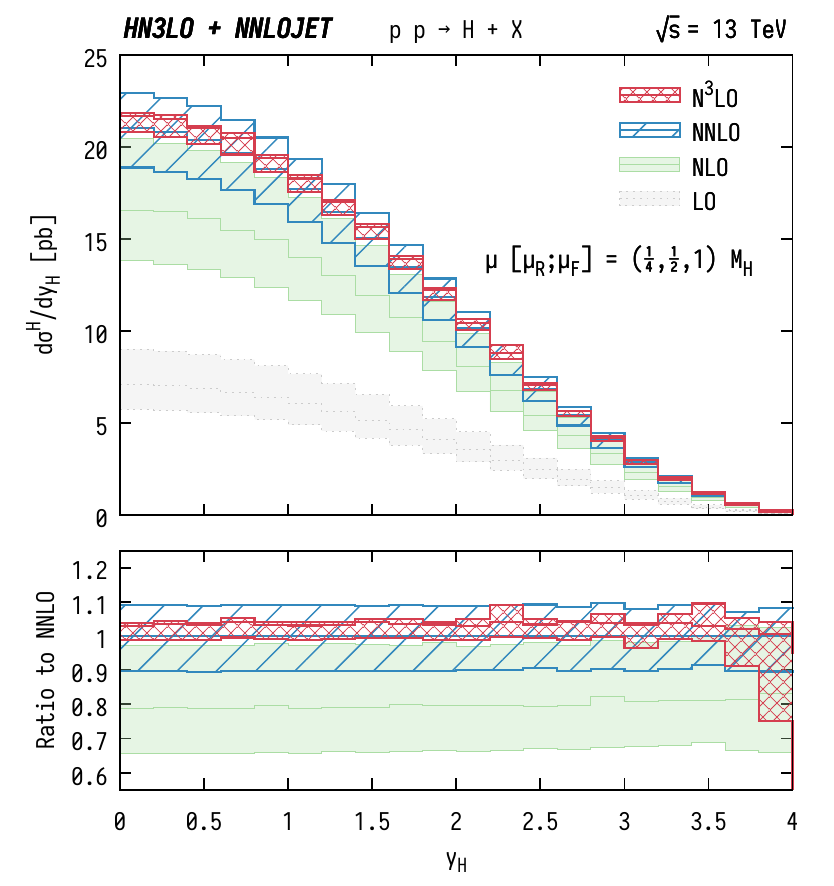}
    \includegraphics[width=0.45\textwidth]{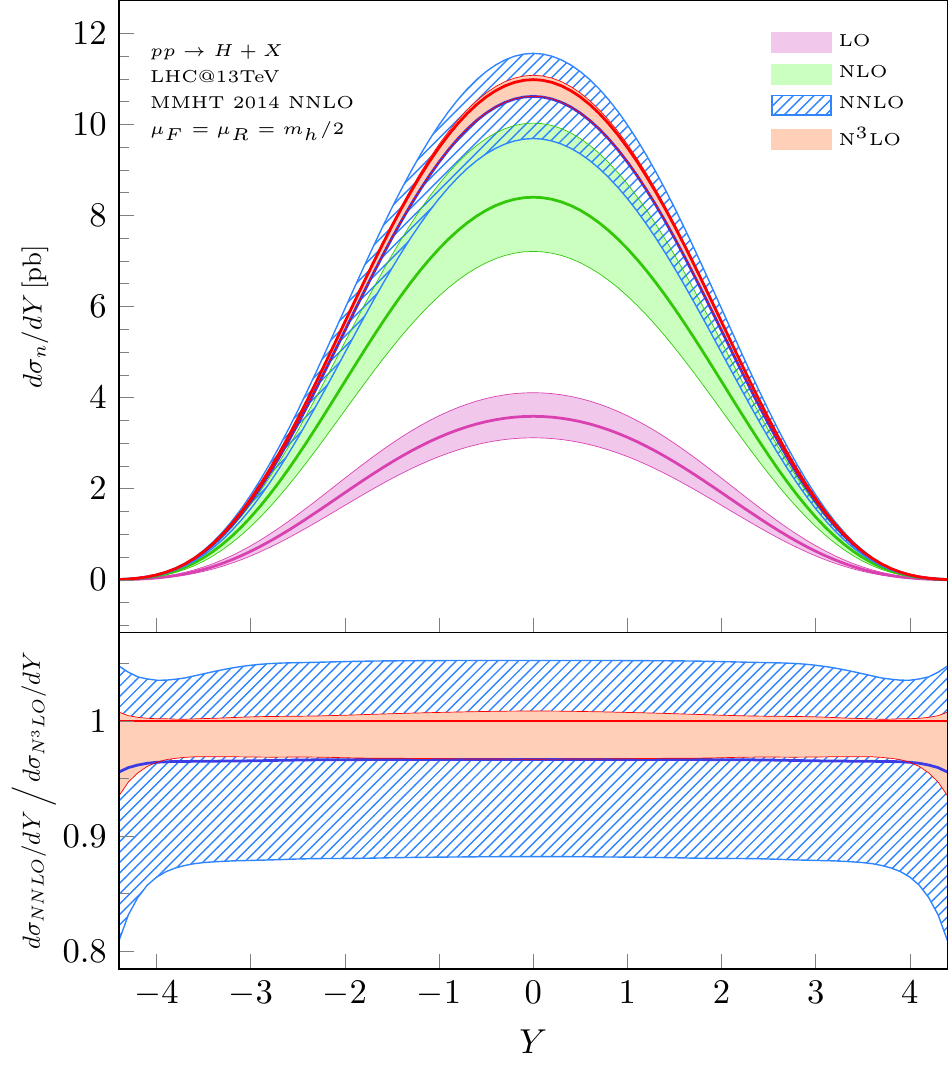}
\caption{Higgs boson rapidity distribution at different orders in perturbation theory, including scale uncertainties. Figures from Refs.~\cite{Cieri:2018oms} (left) and \cite{Dulat:2018bfe} (right).\label{fig:yHn3lo}}
\end{figure}

The N$^{3}$LO$_{HTL}$ inclusive cross section in the threshold
approximation~\cite{Anastasiou:2015ema,Anastasiou:2016cez} has been implemented in the codes {\sc iHixs}~\cite{Anastasiou:2011pi,Anastasiou:2012hx} and {\tt SusHi}~\cite{Harlander:2016hcx}.
The code, {\sc iHixs\,2}~\cite{Dulat:2018rbf} contains the exact N$^{3}$LO$_{HTL}$ corrections as well as the  the leading EW corrections from Ref.~\cite{Actis:2008ug} in a factorising approach and the option to perform threshold resummation to N$^3$LL.
Differential results at  N$^{3}$LO$_{HTL}$ have been presented in Refs.~\cite{Cieri:2018oms,Dulat:2018bfe}.
Results beyond the threshold approximation have been calculated in
Ref.~\cite{Mistlberger:2018etf}, see also ~\cite{Dulat:2017prg}.
Threshold resummation in two-dimensional Mellin space up to NNLO+NNLL for the rapidity distribution has been performed in Ref.~\cite{Banerjee:2017cfc}. The rapidity distribution calculated at N$^{3}$LO$_{HTL}$ in
Refs.~\cite{Cieri:2018oms,Dulat:2018bfe} shows a relatively flat K-factor of about
1.03 compared to the previous order, calculated in Refs.~\cite{Harlander:2002wh,Anastasiou:2002yz,Ravindran:2003um,Catani:2007vq,Grazzini:2008tf,Dulat:2017brz}.
In the region $|y_H\leq 3.6|$,  the scale uncertainties are reduced by more than 50\%, see Fig.~\ref{fig:yHn3lo}.

\begin{figure}
\hfill
\includegraphics[width=0.49\textwidth]{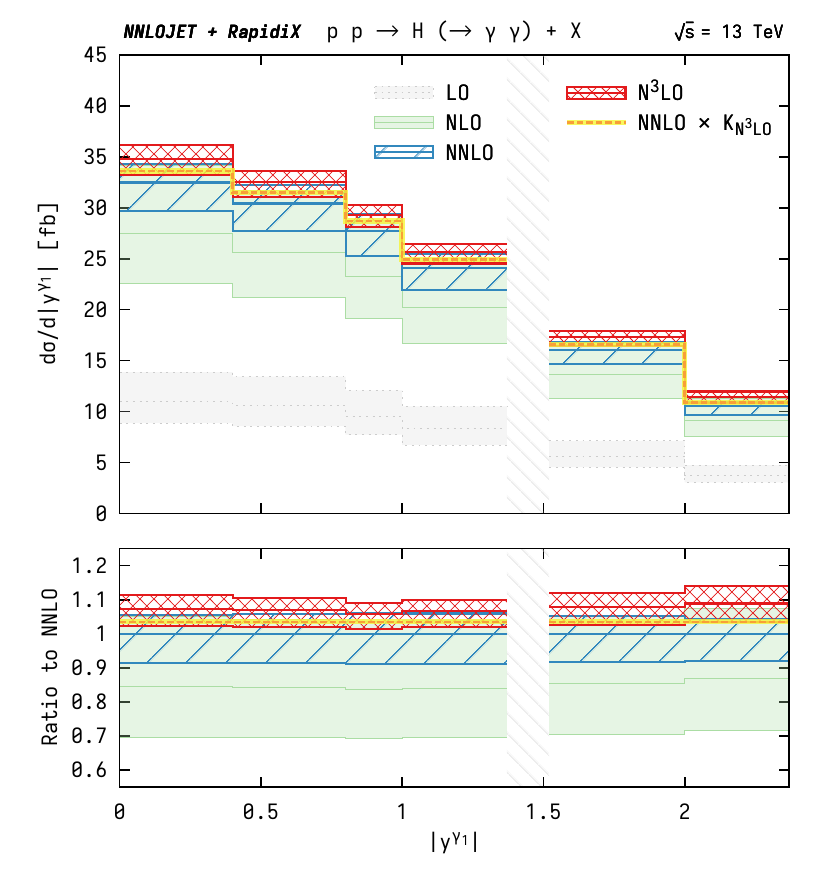}
\hfill
\includegraphics[width=0.49\textwidth]{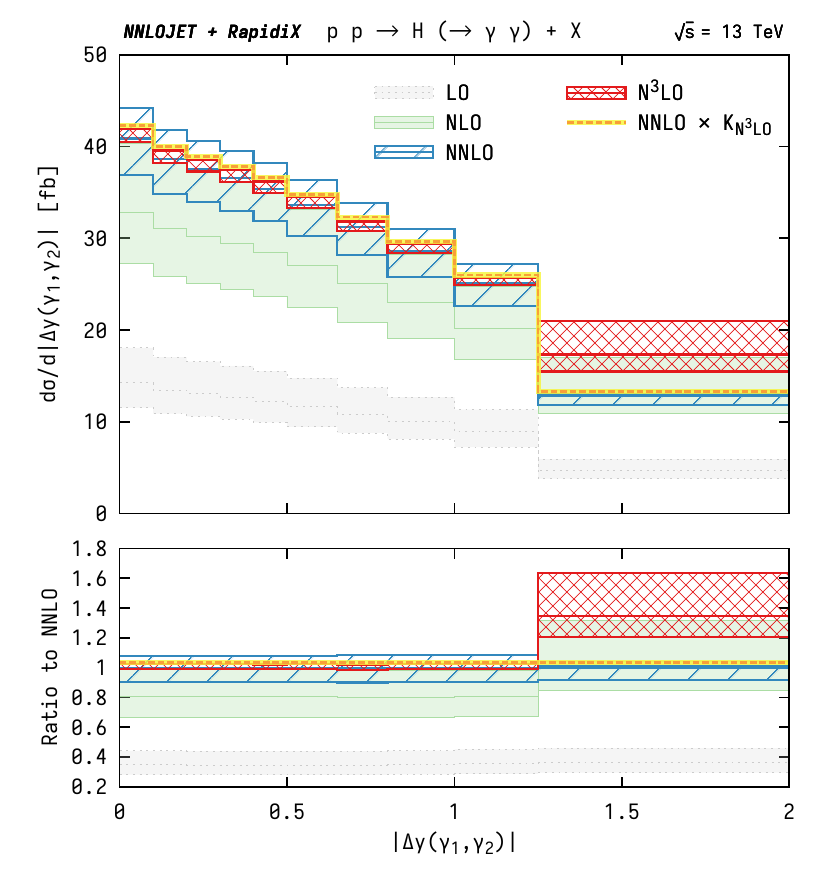}
\hspace*{\fill}
\caption{
  Differential predictions for the rapidity of the leading photon (left) and the absolute value of the rapidity difference of the two photons (right).
  Predictions are shown at LO~(grey), NLO~(green), NNLO~(blue), N$^3$LO~(red), and for the NNLO prediction rescaled by the inclusive $K_{N3LO}$-factor~(orange). The shaded vertical band in the left plot corresponds to the region excluded by the fiducial cuts.
  Figures from Ref.~\cite{Chen:2021isd}.
\label{fig:Hfiducial}}
\end{figure}

It is interesting to compare the rapidity distribution shown in Fig.~\ref{fig:yHn3lo} to the fiducial case, i.e. the distributions of the photon final states from the Higgs boson decay which became available recently~\cite{Chen:2021isd}.
From Fig.~\ref{fig:Hfiducial} one can see that  that the differential N$^3$LO predictions are larger than expected from the inclusive \(K\)-factor, in particular the kinematic enhancement at large rapidity differences, arising from the fiducial cuts, cannot be captured by reweighting with the inclusive K-factor.

The calculation of the N$^4$LO soft and virtual corrections to inclusive Higgs production~\cite{Das:2020adl}
revealed that at 14\,TeV, these corrections enhance the cross section by 2.7\% for the scale choice $\mu_R=m_H$, while the enhancement is 0.2\% for $\mu_R=m_H/2$, both with $\alpha_s$ and PDFs evaluated at $\mu=m_H$.

With these corrections available, the uncertainties due to missing
higher orders, which were considered as the dominant theoretical
uncertainties for a long time, are now  at a level where other uncertainties, which  are estimated to be in the
ballpark of 1\%~\cite{deFlorian:2016spz},  play an important role, see Table~\ref{tab:YR4}.
The PDF and $\alpha_s$ uncertainties are now the dominant uncertainties.
\begin{table}
\hspace*{-1.3cm}
\begin{tabular}{cccccccc}
\hline
\begin{tabular}{c} $\delta$(scale) \end{tabular} &
\begin{tabular}{c} $\delta$({\sc PDF-TH}) \end{tabular} &
\begin{tabular}{c} $\delta$(EW) \end{tabular} & 
\begin{tabular}{c} $\delta(t,b,c)$  \end{tabular} & 
\begin{tabular}{c} $\delta(1/m_t)$ \end{tabular}&
 \begin{tabular}{c} $\delta$(PDF) \end{tabular}&
\begin{tabular}{c} $\delta(\alpha_s)$ \end{tabular}&
  \\
  \hline
${}^{+0.10\textrm{ pb}}_{-1.15\textrm{ pb}} $ &  $\pm$0.56 pb & $\pm$0.49 pb& $\pm$0.40 pb& $\pm$0.49 pb&$\pm$ 0.89 pb&${}^{+1.25\textrm{ pb}}_{-1.26\textrm{ pb}} $
\\ \hline
${}^{+0.21\%}_{-2.37\%}$  & $\pm 1.16\%$ & $\pm 1\%$ & $\pm 0.83\%$ & $\pm 1\%$ &$\pm 1.85\%$&${}^{+2.59\%}_{-2.62\%}$\\
 \hline
\end{tabular}
\caption{Status of the theory uncertainties on Higgs boson production in gluon fusion at $\sqrt{s}=13\,$TeV.
  The table is taken from Ref.~\cite{deFlorian:2016spz} and the LHC Higgs WG1 TWiki, with $\delta$(trunc) removed after the work of Ref.~\cite{Mistlberger:2018etf}. The value for $\delta$(EW) was a rough estimate when Ref.~\cite{deFlorian:2016spz} was published. Meanwhile the order of magnitude has been confirmed by the calculations of  Refs.~\cite{Bonetti:2016brm,Bonetti:2017ovy,Bonetti:2018ukf,Anastasiou:2018adr,Bonetti:2020hqh}.\label{tab:YR4}
}
\end{table}

Two-loop electroweak corrections to Higgs production in gluon
fusion were calculated in Refs.~\cite{Aglietti:2004nj,Degrassi:2004mx,Actis:2008ug}.
The mixed QCD-EW corrections which appear at two loops for the first
time were calculated directly 
 in Ref.~\cite{Anastasiou:2008tj}, where however the unphysical
 limit $m_Z, m_W\gg m_H$ was employed.
 In Refs.~\cite{Bonetti:2016brm,Bonetti:2017ovy,Bonetti:2018ukf}, this restriction was
 lifted and the mixed QCD-EW corrections at order $\alpha^2\alpha_s^2$
 were calculated, where the real radiation contributions were included in the soft gluon approximation. It was found that 
the increase in the total cross section between pure NLO QCD and NLO
QCD+EW is about 5.3\%.
The calculation of Ref.~\cite{Bonetti:2018ukf} has been confirmed by
Ref.~\cite{Anastasiou:2018adr}, where also the hard real radiation was calculated, in the limit of small vector
boson masses, corroborating the validity of the soft gluon
approximation. Finally, the two-loop mixed QCD-EW corrections with full dependence on the Higgs and on the vector boson masses have been presented in Ref.~\cite{Bonetti:2020hqh}, as well as the exact NLO QCD corrections to the light-quark part of the mixed QCD-EW contributions~\cite{Becchetti:2020wof}, overcoming the approximations of the previous calculations.

Another ingredient needed for the full mixed QCD-EW corrections at order $\alpha^2\alpha_s^2$ are the non-factorising mixed QCD-EW corrections from the one-loop partonic subprocess $gg\to Hq\bar{q}$, calculated in Ref.~\cite{Hirschi:2019fkz} and indicating that these corrections can be neglected.

 An analytic result for the two-loop amplitude for the process $gg\to H$ to order $\epsilon^2$, which is a building block for the infrared subtraction terms
 which would enter an  N$^3$LO calculation with full top quark mass dependence, has been presented in Ref.~\cite{Anastasiou:2020qzk}.
 Combined with the results of Refs.~\cite{Davies:2019nhm,Czakon:2020vql}  for the 3-loop $m_t$-dependent form factor as well as the results of Ref.~\cite{Frellesvig:2019byn} entering the two-loop real-virtual contributions, a full N$^3$LO result seems to emerge on the horizon.

 The Abelian part of the double-logarithmic corrections for the $gg \to Hg$ amplitude has been obtained in Ref.~\cite{Melnikov:2016emg}.
 In Ref.~\cite{Anastasiou:2020vkr} an all-order next-to-leading logarithmic approximation for the light-quark-loop-mediated amplitude of Higgs boson production in gluon fusion has been calculated, in order to assess the higher order bottom quark contribution to the Higgs boson production cross section in the threshold approximation.
 However, the conclusion is that an actual accuracy of the logarithmic and threshold approximations is difficult to estimate and an exact computation of quark mass effects is therefore expected to be important in consolidating the theoretical precision of the top-bottom interference contribution to the inclusive Higgs production cross section.

The transverse momentum spectrum of the Higgs boson has been studied extensively, for a recent method in momentum space,  used  to obtain NNLL+NNLO predictions for the $p_T^H$ spectrum see Ref.~\cite{Monni:2016ktx}.
Resummed predictions further have been produced at NNLO+N$^3$LL both
at the inclusive level~\cite{Chen:2018pzu} and in the $H\rightarrow\gamma\gamma$ channel
with fiducial cuts~\cite{Bizon:2018foh}.
The resummation considerably reduces the scale uncertainties and makes the result
for $p_T \lesssim 40\ \mathrm{GeV}$ more reliable, while the size of the N$^3$LL corrections is small, 
amounting to 5\% only at very small $p_T$.
Recently, the Higgs boson transverse momentum spectrum in the presence of a jet veto has been studied at doubly differential level
at NLO+NNLL~\cite{Monni:2019yyr}, i.e. a joint resummation of large logarithms  involving either $p_t^H/m_H$ or $p_t^{j,\rm{veto}}/m_H$ has been  performed.

Important developments are also combinations of NNLO predictions with
a parton shower~\cite{Hamilton:2013fea,Hoche:2014dla}, including finite top and bottom mass effects~\cite{Hamilton:2015nsa,Bizon:2019tfo,Hu:2021rkt}.

The experimental uncertainty on the total Higgs boson cross section is expected to be 3\% or less with a data sample
of 3000~fb$^{-1}$~\cite{Amoroso:2020lgh}. 
To achieve the desired theoretical precision, more complete calculations of 
finite-mass effects, including light quark masses,  as well as  PDFs at N$^3$LO accuracy seem to be the most important issues concerning the inclusive cross section.

\subsection{Higgs production in bottom quark fusion}

Higgs production in bottom quark fusion has been calculated up to N$^3$LO in the five-flavour scheme~\cite{Duhr:2019kwi} and matched to the four-flavour scheme at order $\alpha_s^3$~\cite{Duhr:2020kzd}.
In the four-flavour scheme the bottom quark is treated as massive and is produced in the hard process.
This leads to higher final-state multiplicities and more complicated matrix elements.
Therefore Higgs production in bottom quark fusion is only known up to NLO in the four-flavour scheme~\cite{Dittmaier:2003ej,Dawson:2003kb,Wiesemann:2014ioa}, see also Refs.~\cite{Dawson:2005vi,Liu:2012qu,Dittmaier:2014sva}.
The four-flavour scheme is plagued by large logarithms involving the $b$-quark  mass,  which can  spoil  the  convergence  of  the  perturbative  series.   It  is  therefore desirable to combine the two schemes into a single prediction.  Several methods to perform such a combination have been proposed in the literature~\cite{Forte:2016sja,Bonvini:2016fgf,Bonvini:2015pxa,Forte:2015hba,Harlander:2011aa}.
These prescriptions however have suffered from the fact that the equivalent of the NNLO result in the five-flavour scheme is only the leading order in the four-flavour scheme.  Ref.~\cite{Duhr:2020kzd} offers the first consistent matching of the four- and five-flavour schemes up to third order in the strong coupling.  This is achieved by combining the N$^3$LO result for the cross section~\cite{Duhr:2019kwi} with the matching procedure of Refs.~\cite{Forte:2016sja,Forte:2015hba}.
The combination of the N$^3$LO cross section with the resummation of threshold logarithms at N$^3$LL has been performed in Ref.~\cite{H:2019dcl},
the NNLO QCD$\oplus$QED corrections have been calculated in Ref.~\cite{H:2019nsw}.

We also comment briefly on the decay of a Higgs boson to a bottom quark pair here, as the involved matrix elements are related.
$H\to b\overline{b}$ is the dominant Higgs decay mode, however due to the 
large QCD backgrounds, experimental analyses in the past mostly have focused on associated ($VH$) production modes.
Due to jet-substructure techniques~\cite{Asquith:2018igt}, it recently became possible to measure $gg\to H\to b\bar{b}$ in the boosted regime~\cite{Sirunyan:2017dgc,ATLAS:2018hzj}.

The QCD corrections to the decay rate $H\to b\overline{b}$ are known up to N$^4$LO~\cite{Baikov:2005rw,Davies:2017xsp,Herzog:2017dtz}. The NLO electroweak  corrections are known for some time~\cite{Dabelstein:1991ky,Kataev:1997cq}, as well as the mixed QCD$\times$EW corrections at $\mathcal{O}(\alpha \alpha_s)$~\cite{Kataev:1997cq,Mihaila:2015lwa}.
At the level of differential results, NNLO QCD was the state of the art~\cite{Anastasiou:2011qx,DelDuca:2015zqa,Bernreuther:2018ynm,Primo:2018zby} until recently, before in Ref.~\cite{Mondini:2019gid} differential results at  N$^3$LO were presented, complemented  by the corrections proportional to $y_ty_q\,(q=b,c)$ at order $\alpha_s^3$~\cite{Mondini:2020uyy}.
Furthermore, resummed NNLL$^\prime$+NNLO resummed predictions matched to the {\sc Geneva}{} parton shower framework have been obtained in Ref.~\cite{Alioli:2020fzf}.
Phenomenological studies~\cite{Ferrera:2017zex,Caola:2017xuq,Behring:2019oci,Gauld:2019yng,Alioli:2019qzz,Bizon:2019tfo,Behring:2020uzq} have focused on interfacing the decay at this order to $VH$ production, which is also known at NNLO in QCD~\cite{Ferrera:2011bk,Ferrera:2014lca,Campbell:2016jau}, see  Section~\ref{sec:VH}.


Partial NLO EW corrections to $H b\overline{b}$ production have been calculated in Ref.~\cite{Zhang:2017mdz}.
Very recently, the complete NLO QCD and EW predictions for $H b\overline{b}$ production in the four-flavour scheme have been calculated~\cite{Pagani:2020rsg}, i.e. all contributions of order $\alpha_s^m\alpha^{n+1}$ with $m+n\leq 3$.
Terms with $n\geq 2$ include $ZH$ production, where the $Z$-boson decays into $b\overline{b}$ pair. In addition, at order $\alpha_s\alpha^3$, VBF configurations with $Z$-bosons arise. It is shown that the impact of these additional contribution is sizeable. The conclusion of Ref.~\cite{Pagani:2020rsg} is therefore  that experimental cuts to suppress these contributions would reduce the cross section to an extent which renders the idea of directly extracting the Yukawa coupling $y_b$ from the measurement of  $H b\overline{b}$ at the LHC hopeless.


\subsection{Higgs plus jet production}

Higgs boson production  in association with a jet has been calculated in the heavy top limit up to NNLO  in Refs.~\cite{Chen:2014gva,Boughezal:2015dra,Boughezal:2015aha,Caola:2015wna,Chen:2016zka,Campbell:2019gmd} using different methods for the subtraction of IR divergent double real radiation (see section~\ref{sec:IRsubtraction}), thereby also shedding light on the role of power corrections within the N-jettiness subtraction method~\cite{Campbell:2019gmd}.

\begin{figure}[htb]
  \centering
 \vspace*{-7mm}
  \includegraphics[width=0.5\textwidth]{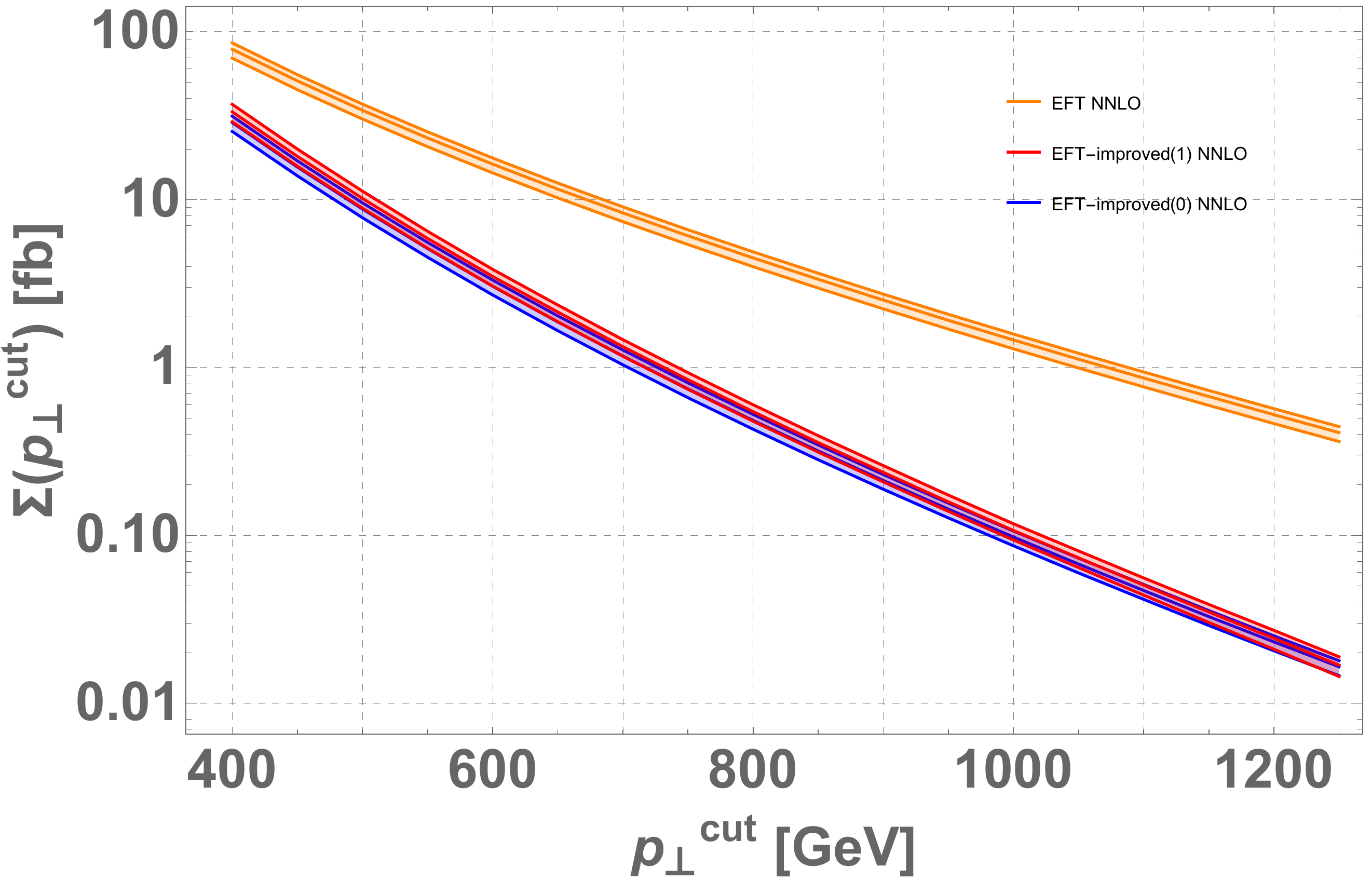}
  \includegraphics[width=0.47\textwidth]{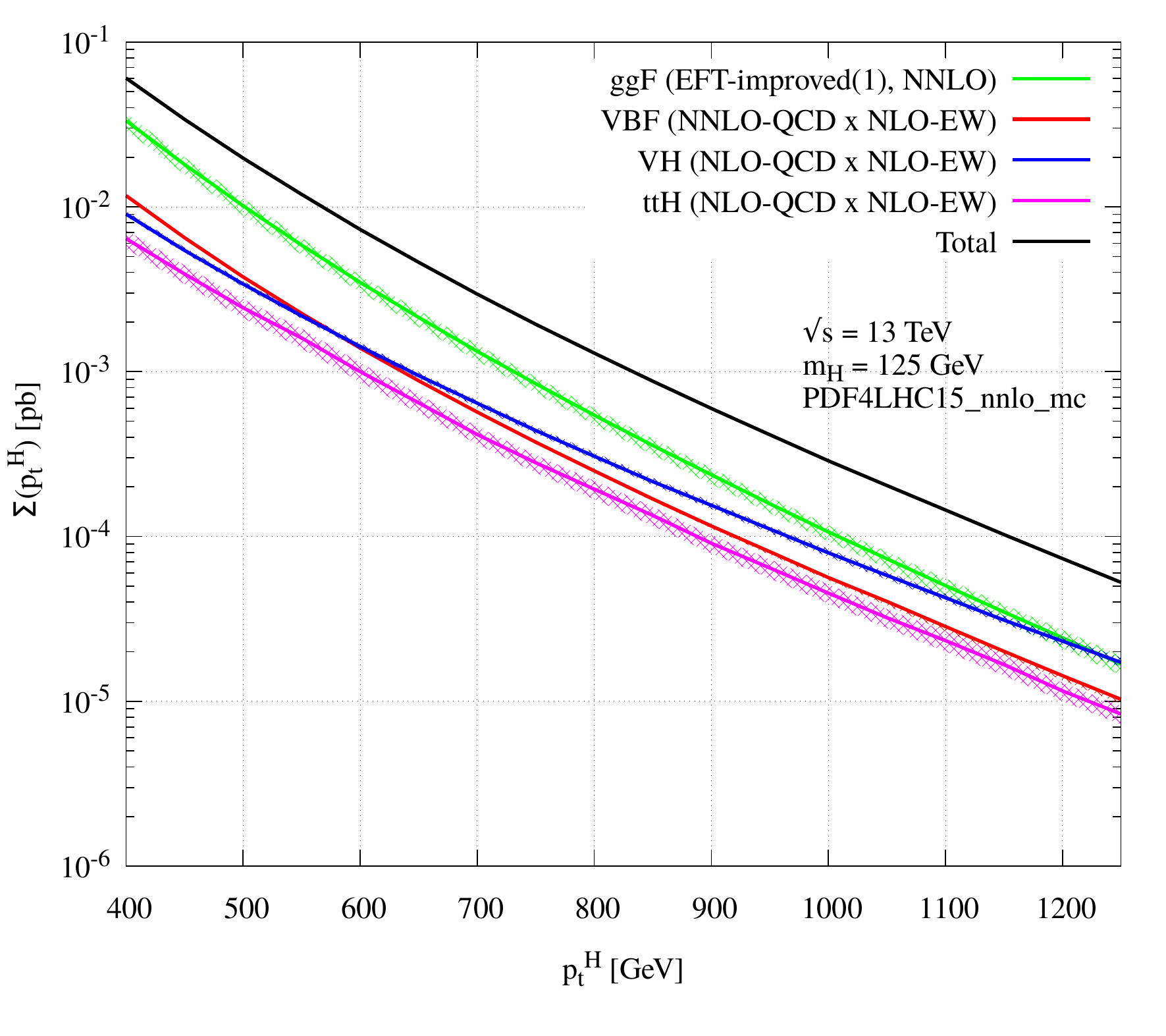}
  \caption{Left: Cumulative cross section
as a function of the $p_\perp$-cut at NNLO in the HTL, as well as rescaled by the LO (NLO) full-SM
spectrum labelled by EFT-improved$(0)$ (EFT-improved$(1)$). 
    Right: Comparison of cross sections from different production channels:  gluon-fusion (green), VBF (red), vector
  boson associated (blue) and top-quark pair associated (magenta).
Figures from Ref.~\cite{Becker:2020rjp}.\label{fig:Hpt}}
\end{figure}

Top quark mass effects have been included at NLO in various approximations~\cite{Buschmann:2014sia,Hamilton:2015nsa,Frederix:2016cnl,Neumann:2016dny} before the NLO calculation with full top quark mass dependence was presented~\cite{Jones:2018hbb}, providing more precise predictions at large $p_T$, a region which is sensitive to New Physics effects due to additional heavy fermions in the loop or other sources of modifications of the Higgs couplings, see Fig.~\ref{fig:Hpt}.

At low $p_T$ values, bottom quark effects play a role and have been  calculated in Refs.~\cite{Melnikov:2016qoc,Melnikov:2017pgf,Lindert:2017pky}.
Top mass effects in the large transverse momentum
expansion $p_T\gg 2m_t$ also have been studied~\cite{Lindert:2018iug,Neumann:2018bsx}, using two-loop integrals calculated in Ref.~\cite{Kudashkin:2017skd}.
Bottom quark-induced contributions to Higgs plus jet production at NNLO have been calculated in Ref.~\cite{Mondini:2021nck}
in the five-flavour scheme, retaining the bottom quark mass only in the coupling to the Higgs boson.

Higgs boson decays in combination with NNLO HTL results have been included in Ref.~\cite{Caola:2015wna} ($H\to\gamma\gamma$, $H\to W^+W^-\to e^+\mu^-\nu\bar{\nu}$), Ref.~\cite{Chen:2016zka} ($H\to\gamma\gamma$) and Ref.~\cite{Chen:2019wxf} ($H\to 4$\,leptons), where the detailed comparison to ATLAS and CMS measurements in the 4 lepton channel  revealed the important role of different lepton isolation prescriptions~\cite{Chen:2019wxf}.

An extensive study of Higgs production at large transverse momentum, $p_T^H\geq 400$\,GeV, has been performed in Ref.~\cite{Becker:2020rjp}. It contains the currently best prediction for Higgs+jet production in the boosted regime, obtained by a combination of the NNLO prediction computed in the HTL (denoted by ``EFT'' in Fig.~\ref{fig:Hpt}) with the exact NLO prediction.
The result is shown in Fig.~\ref{fig:Hpt} (left), where the cumulative cross section as a function of the lower cut on $p_T^H$ is shown, comparing NNLO HTL, the combinations with the exact NLO ($\Sigma^{\text{EFT-improved (1), NNLO}}$) and the exact LO ($\Sigma^{\text{EFT-improved (0), NNLO}}$) calculations, where the combination is defined as
\begin{equation}
\label{eq:approx_NNLO}
\Sigma^{\text{EFT-improved (1), NNLO}}(p_{\perp}^{\rm cut})\equiv \frac{\Sigma^{\text{SM,
      NLO}}(p_{\perp}^{\rm cut})}{\Sigma^{\text{EFT, NLO}}(p_{\perp}^{\rm cut})} \,\Sigma^{\text{EFT, NNLO}}(p_{\perp}^{\rm cut})\,, 
\end{equation}
analogous for LO.
Ref.~\cite{Becker:2020rjp} also contains a comparison of results from Monte Carlo tools~\cite{Alioli:2008tz,Campbell:2012am,Hamilton:2012rf,Hamilton:2013fea,Hamilton:2015nsa,Frederix:2016cnl}  with the best fixed order prediction. It was found that 
the predictions obtained with the 
generators  {\tt HJ-MiNLO}~\cite{Hamilton:2012rf} and {\tt
  MG5$_{-}$MC@NLO}~\cite{Frederix:2016cnl} are in very good agreement with one
another and that they both reproduce the best fixed order
prediction within uncertainties, however the latter are of the order of 20-30\%.
Another interesting outcome of this work is the conclusion that other Higgs production channels, in particular $VH$ production, also play a non-negligible role in the boosted regime, see Fig.~\ref{fig:Hpt} (right).

The assessment of the uncertainties will remain incomplete until the NLO electroweak corrections in the gluon fusion channel  as well as a study of the top quark mass scheme dependence are available.

A first step towards the NLO electroweak corrections to Higgs+jet production in gluon fusion has been made
in Ref.~\cite{Becchetti:2018xsk}, where the planar master integrals for the  light fermion contributions have been calculated analytically.
A big step is also provided by the master integrals for the two-loop mixed QCD-EW corrections to $gg \to Hg$, calculated in Ref.~\cite{Bonetti:2020hqh}.
From the results for inclusive Higgs production discussed above it is expected that the
light fermion contributions represent the bulk of the electroweak corrections to Higgs plus jet production in the gluon fusion channel.

The Higgs boson transverse momentum is certainly a very important observable at the LHC, as the Higgs $p_T$-spectrum is sensitive to New Physics effects. For example, it allows to break the degeneracy along $c_t+c_g=$\,const. which is present when measuring the inclusive Higgs boson production cross section~\cite{Schlaffer:2014osa,Dawson:2014ora}.


 \subsection{Higgs plus two or more jets}

 The production of a Higgs boson in association with two jets has two main production modes, vector boson fusion (VBF) and genuine QCD production. The VBF channel is the second largest Higgs production channel, and as it probes gauge boson scattering, it directly probes our understanding of perturbative unitarity  as guaranteed by the SM Higgs mechanism.

 The NLO QCD corrections to Higgs production in VBF were first calculated differentially in Refs.~\cite{Figy:2003nv,Berger:2004pca} and have been implemented into the programs {\tt VBFNLO}~\cite{Arnold:2008rz,Arnold:2011wj,Baglio:2014uba} and {\tt MCFM}~\cite{Campbell:2011bn,Campbell:2015qma,Campbell:2019dru}.

 The total cross section in the VBF production mode is known to N$^3$LO in the structure function approach~\cite{Dreyer:2016oyx},
 differing only at the per-mil level from the inclusive NNLO result~\cite{Bolzoni:2010xr}.
The NNLO corrections are also known differentially~\cite{Cacciari:2015jma,Cruz-Martinez:2018rod} and in general do not exceed 5\% of the NLO corrections, while the latter can be as large as 30\% and lead to substantial modifications of the shape of the jet distributions.
 Nonetheless, there is a kinematic dependence of the  NNLO  corrections in  the  $p_T$- and rapidity  distributions  of  the  two  tagging  jets, and since it is precisely through cuts on these observables that the VBF cross section is selected, the NNLO effects may have an impact on the efficiency of the VBF cuts and thus on the precision of the measurements.

 The calculations have been performed in the so-called ``structure function approach''~\cite{Han:1992hr,Cacciari:2015jma}, also called ``projection-to-Born method'', which neglects non-factorising contributions due to interactions between the two quark lines radiating the vector bosons.
 Non-factorisable effects have been studied in the eikonal approximation in Ref.~\cite{Liu:2019tuy} and estimated to amount to 1\% at most.
 In Ref.~\cite{Dreyer:2020urf}, the validity of the approximation employed in Ref.~\cite{Liu:2019tuy} outside tight VBF cuts is studied for both single and double Higgs production in VBF. For single Higgs production, it was found that for typical selection cuts the non-factorisable NNLO corrections are small and mostly contained within the scale uncertainty bands of the factorisable part. However,  for the fully inclusive VBF phase space, the non-factorisable NNLO corrections were estimated to be of the same order as the NNLO factorisable ones, and moderately larger than the factorisable N$^3$LO corrections.
Nonetheless, as also emphasised in Ref.~\cite{Dreyer:2020urf}, this estimate is based on an extrapolation of the eikonal approximation into a regime where it ceases to be valid, and should therefore only be taken as an indication of the true size of non-factorisable NNLO corrections to fully inclusive VBF Higgs production.
 Interference effects between the VBF production channel and QCD production are negligible~\cite{Andersen:2007mp,Ciccolini:2007jr}. 
 The dependence of the differential NNLO cross section on the definition of the tagging jets has been studied in Ref.~\cite{Rauch:2017cfu}. 

The NLO EW corrections~\cite{Ciccolini:2007jr,Ciccolini:2007ec,Figy:2010ct,Denner:2011id,Denner:2014cla} are at the 5\% level for the total cross section, but increase in the high-energy tails of distributions such as the transverse momentum of the leading jet.
The EW corrections have negative sign, such that they largely cancel the QCD corrections at the inclusive level.
The program {\sc Hawk}~\cite{Denner:2011id,Denner:2014cla} contains the NLO EW and QCD corrections to both Higgs production in VBF as well as Higgs production in association with a vector boson, which also leads to a $H+2$\,jets final state if the weak boson is decaying hadronically.

VBF Higgs production at NLO with up to 3 jets has been calculated in Refs.~\cite{Campanario:2013fsa,Campanario:2018ppz}.
A phenomenological study of the impact of the gluon fusion channel for $H + \le 3$\,jets at NLO in the HTL as a background to VBF measurements has been performed in Ref.~\cite{Greiner:2015jha}, 
top- and bottom quark mass effects in $H + \le 3$\,jets have been studied in 
Ref.~\cite{Greiner:2016awe}.
Mass effects in $H+2j$ at high energies have also been studied~\cite{Andersen:2018kjg} within the ``High Energy Jets''
framework~\cite{Andersen:2009he,Andersen:2011hs,Andersen:2017kfc,Andersen:2018tnm}.
The one-loop amplitude for $H+4$\,jets with full mass dependence has been calculated in Ref.~\cite{Budge:2020oyl}.

Parton shower effects in VBF Higgs production have been studied in detail in Ref.~\cite{Jager:2020hkz}, comparing various showers and generators. It was found that with typical VBF cuts, the uncertainties resulting from predictions by different tools are at the 10\% level for observables that are accurate to NLO. For observables sensitive to extra radiation effects,  uncertainties up to about 20\% have been found. It was also emphasised that in processes of VBF type, 
employing a global recoil scheme for the generation of initial-state radiation, as done by default in most parton shower Monte Carlo programs, is not appropriate in combination with a calculation in the structure function approach, where the radiation is always along one of the quark lines. Therefore the recoil should take place along the quark line where the gluon was emitted, also known as dipole recoil scheme~\cite{Cabouat:2017rzi}.
Considering the fact that, within a single generator at NLO+PS accuracy, the theoretical uncertainties estimated by the usual renormalisation and factorisation scale variations (and  possibly variations of a scale  that  controls  the  shower  hardness) were found to be at the few percent level,  there is a clear need for a better understanding of the differences between the various tools providing NLO+PS accuracy.

\subsection{Higgs production in association with a vector boson}
\label{sec:VH}

The production of a Higgs boson in association with a vector boson, also known as {\em Higgs-Strahlung}, is an important process
at the LHC, as well as at lepton colliders.
Even though the cross sections  at the LHC  are smaller than for Higgs production in gluon fusion and VBF, this process has many appealing features.
For example, for $ZH$ production  combined with a leptonic $Z$-decay, triggering is straightforward, independent of the Higgs decay.
This makes this channel especially attractive in combination with challenging Higgs decays, like invisible or hadronic Higgs decays, in particular $H\to b\bar{b}$~\cite{Dawson:2018dcd}. Furthermore, $VH$ production provides the opportunity to probe the Higgs couplings to gauge bosons.

Inclusive NNLO  QCD corrections are available since quite some time~\cite{Harlander:2002wh} and are implemented 
in the program {\sc VH@NNLO}~\cite{Brein:2003wg,Brein:2011vx,Brein:2012ne,Harlander:2018yio}.
NLO electroweak corrections have been first calculated in Ref.~\cite{Ciccolini:2003jy}.
Combined NLO QCD+EW corrections are also available~\cite{Denner:2014cla,Granata:2017iod,Obul:2018psx}, where
the program {\sc Hawk}~\cite{Denner:2011id,Denner:2014cla}  also contains the VBF mode to account for all possible channels in the case of a hadronically decaying vector boson, $pp\to VH\to H+2$\,jets.
Ref.~\cite{Granata:2017iod} contains NLO QCD+EW predictions for $HV$ and $HV$+jet production including parton shower effects, within the {\tt Powheg-Box-Res}+{\tt OpenLoops} framework.

Differential QCD corrections have  been calculated up to NNLO, including $H\to b\bar{b}$ decays at different orders.
Refs.~\cite{Ferrera:2011bk,Ferrera:2014lca,Campbell:2016jau} include Higgs decays at NLO, while they are included up to NNLO in Refs.~\cite{Ferrera:2017zex,Caola:2017xuq}. 

The $VH$ process is also an ideal testing ground for infrared subtraction schemes at NNLO, being less basic than Drell-Yan but still containing only colourless particles in the final state. This subject will be treated in Section \ref{sec:IRsubtraction}, however it should be mentioned here that the differential NNLO results calculated in Refs.~\cite{Ferrera:2011bk,Ferrera:2014lca,Ferrera:2017zex} are based on $q_T$-subtraction~\cite{Catani:2007vq}, where Ref.~\cite{Ferrera:2017zex} in addition uses the {\it ColourfulNNLO} method for the $H \to b\bar{b}$ decay at NNLO~\cite{DelDuca:2015zqa}.
The results of Ref.~\cite{Campbell:2016jau} are based on N-jettiness, implemented in MCFM~\cite{Boughezal:2016wmq}, the ones of Ref.~\cite{Caola:2017xuq,Behring:2019oci,Behring:2020uzq} are based on nested soft-collinear subtraction~\cite{Caola:2017dug}.

The combination of fixed-order QCD computations with parton showers has been studied at NLO+PS in association with up to one jet~\cite{Luisoni:2013cuh} and at NNLOPS~\cite{Astill:2016hpa,Astill:2018ivh,Bizon:2019tfo} using the  MiNLO procedure in {\sc Powheg}~\cite{Hamilton:2012np}.
Merged results of the NLO Drell-Yan contribution and the loop-induced gluon initiated contribution for 0+1 jet multiplicities are also available in {\sc Sherpa}~\cite{Goncalves:2015mfa}, see also~\cite{Hespel:2015zea} for {\sc MG5\_MC@NLO}.
The NNLO corrections have been combined with NNLL resummation in the 0-jettiness variable and matched
to a parton shower within the {\sc Geneva} Monte Carlo framework~\cite{Alioli:2019qzz}.

The differential NNLO predictions for $VH$ observables in combination with $H\to b\bar{b}$ have been calculated in Ref.~\cite{Ferrera:2017zex} (for $V=Z, W$) and in Ref.~\cite{Caola:2017xuq} (for $V=W$), considering massless $b$-quarks except in the Higgs Yukawa coupling and using the same-flavour-$k_T$ algorithm~\cite{Banfi:2006hf} to define $b$-jets.
The Higgs boson decay is treated in the narrow width approximation.

Ref.~\cite{Gauld:2019yng}  presents a computation of $VH$ observables for the processes $V=Z,W^\pm$, including NNLO corrections to both production and decay subprocesses.
It provides a fully differential description of the final states, i.e. including the decays of the vector boson into leptons and the Higgs boson into $b$-quarks with off-shell propagators of the vector- and Higgs bosons.
It was found that an independent variation of the production and decay scales results in percent-level uncertainties.
In Ref.~\cite{Behring:2020uzq}, 
associated $WH$ production with $H \to b\bar{b}$ decays has been calculated at NNLO QCD with massive $b$-quarks. Comparing the massive and massless descriptions, differences of ${\cal O}(5\%)$ in fiducial cross sections were found, and even larger differences for differential results in kinematic regions where the $b$-jets have large transverse momenta.
Ref.~\cite{Gauld:2020ced} presents predictions for $WH$+jet production up to order ${\cal O}(\alpha_s^3)$ and ${\cal O}(\alpha_s^2y_t)$.

Threshold corrections up to N$^3$LO for the Drell-Yan-type part of the  inclusive cross section have been calculated in Ref.~\cite{Kumar:2014uwa}
and included in the program  {\sc VH@NNLO}~\cite{Brein:2012ne}.
Soft-gluon resummation of both threshold logarithms and logarithms which are important at low transverse momentum of the $V H$ pair have been considered up to N$^3$LL in Ref.~\cite{Dawson:2012gs} and have been found to be very close to the fixed order NNLO result.
Threshold resummation for the $gg\to ZH$ process has been calculated in Ref.~\cite{Harlander:2014wda}.


The loop-induced gluon-initiated contributions are finite and enter at order $\alpha_s^2$, i.e. formally at NNLO considering the $pp\to ZH$ process.
Due to the dominance of the gluon PDFs at the LHC they are however sizeable,  contributing about 6\% to the total NNLO cross section, and the contribution can be twice as large in the boosted Higgs boson regime $p_T^H\gtrsim 150$\,GeV~\cite{Harlander:2013mla,Englert:2013vua}.
The gluon-initiated subprocess is very sensitive to modified Yukawa couplings and/or non--SM particles running in the loop, 
therefore the NLO corrections to this process, calculated at LO in Ref.~\cite{Kniehl:1990iva},  are important.
However, these NLO corrections  contain two-loop integrals involving $m_t, m_H$ and $m_Z$, and such integrals are currently unknown analytically.
Therefore the NLO QCD corrections  have been calculated in various approximations.
In Ref.~\cite{Altenkamp:2012sx} they have been calculated in the $m_t\to \infty$ limit, leading to a K-factor to reweight the full one-loop result.
In addition, top quark mass effects in $gg\to ZH$ at NLO QCD have been considered in the framework of a $1/m_t$-expansion in Ref.~\cite{Hasselhuhn:2016rqt}.
However, the $1/m_t$-expansion becomes unreliable for invariant masses of the $HZ$ system larger than the top quark pair production threshold $2m_t$ and therefore cannot be used for precise predictions of the high energy tails of $m_{ZH}$ or $p_T^H$ distributions, which are most sensitive to new physics effects.
In Ref.~\cite{Harlander:2018yns}, a data-driven strategy to extract the gluon-initiated component
(or, more precisely, the non-Drell-Yan component) for $ZH$ production has been suggested, 
based on the comparison of the $ZH$ to the $WH$ cross section and the corresponding invariant mass distributions of the $VH$ system.

Very recently, the two-loop corrections to $gg\to ZH$ with full top quark mass dependence became available: based on a high-energy expansion supplemented by Pad\'e-approximants~\cite{Davies:2020drs}, based on a transverse momentum expansion~\cite{Alasfar:2021ppe} complementing the kinematic region covered by~\cite{Davies:2020drs}, and numerically with exact top quark mass dependence~\cite{Chen:2020gae}.

The process $b\bar{b}\to ZH$ in the five-flavour scheme, but with a non-vanishing
bottom-quark Yukawa coupling, has been calculated in the soft-virtual approximation at NNLO QCD in Ref.~\cite{Ahmed:2019udm}, the polarised $q\bar{q}\to ZH$ two-loop amplitudes have been calculated in Ref.~\cite{Ahmed:2020kme}.

Finally, Higgs production in association with a photon has gained some attention recently. As the LO amplitude for $gg\to H\gamma$ vanishes due to Furry's theorem, the relative contributions from light quarks is enhanced, such that this process may offer sensitivity to the light quark Yukawa couplings at the high-luminosity LHC~\cite{Aguilar-Saavedra:2020rgo,Gabrielli:2016mdd}.
The NLO QCD corrections $H+\gamma$ production in VBF have been calculated in Ref.~\cite{Arnold:2010dx}, the process 
$gg\to ZH\gamma$ has been studied in Ref.~\cite{Agrawal:2014tqa}.

\subsection{Top quark pair associated Higgs production, H+single top}

The process $t\bar{t}H$ is particularly interesting due to its direct sensitivity to the top-Yukawa coupling $y_t$.
However, this process suffers from large systematic uncertainties due to the very complicated final states.
Currently the combination with $H\to \gamma\gamma$ is the most promising channel~\cite{Dawson:2018dcd,Aad:2020ivc,Sirunyan:2020sum}, however the $H\to b\bar{b}$ channel  is of increasing importance as deep learning methods gain momentum as a way to improve the signal-to-background ratio~\cite{Erdmann:2017hra,Kasieczka:2019dbj,Ren:2019xhp,Abdughani:2019wuv}.

NLO QCD corrections for on-shell $t\bar{t}H$ production are known for many
years~\cite{Beenakker:2001rj,Reina:2001sf,Beenakker:2002nc,Dawson:2002tg,Dawson:2003zu}.
They have been matched to parton showers in Refs.~\cite{Frederix:2011zi,Garzelli:2011vp,Hartanto:2015uka}.
NLO EW corrections have first been calculated in Ref.~\cite{Frixione:2014qaa}, they have been 
combined with NLO QCD  corrections within the narrow-width-approximation (NWA) for top-quark decays in Refs.~\cite{Yu:2014cka,Frixione:2015zaa}.
NLO QCD corrections to off-shell top quarks in $t\bar{t}H$ production with leptonic decays have been calculated in Ref.~\cite{Denner:2015yca}.
A combination of these NLO QCD corrections with NLO EW corrections has been presented in Ref.~\cite{Denner:2016wet}, the $t\bar{t}b\bar{b}$
background with full off-shell effects has been calculated in Ref.~\cite{Denner:2020orv}.
These calculations involve  one-loop amplitudes with up to 10 external legs, and thus are showpieces of the one-loop matrix-element generator {\sc Recola}~\cite{Actis:2016mpe,Denner:2017wsf}.

Soft gluon resummation at NLO+NNLL has been performed in
Refs.~\cite{Kulesza:2015vda,Broggio:2015lya,Broggio:2016lfj,Kulesza:2017ukk}, soft and Coulomb corrections have been resummed in Ref.~\cite{Ju:2019lwp}.
The NLO+NNLL  resummed results have been further improved  including also the processes $t\bar{t}W^\pm, t\bar{t}Z$~\cite{Kulesza:2018tqz,Broggio:2019ewu}, where Ref.~\cite{Broggio:2019ewu} also includes EW corrections. 

\begin{figure}[tb]
		\begin{center}
			\includegraphics[width=.48\textwidth]{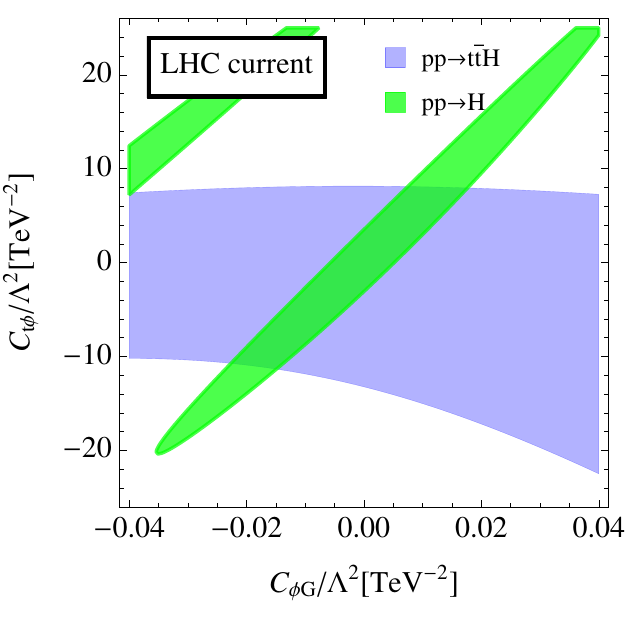}
			\includegraphics[width=.48\textwidth]{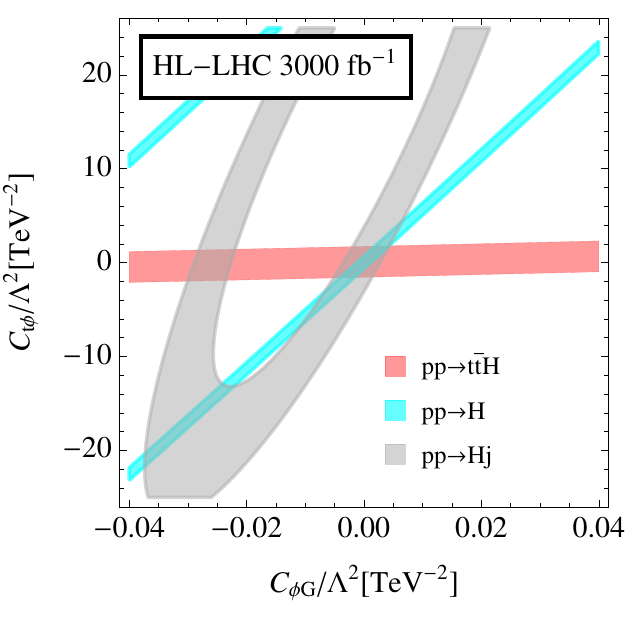}
		\end{center}
		\caption{Allowed region in the $C_{t\phi}$-$C_{\phi G}$ plane at 95\% confidence level,
                  where $C_{t\phi}$ and $C_{\phi G}$ denote the Wilson coefficients of the top-Higgs and Higgs-gluon operators, respectively.
		Left: current constraints.  Right: Projection for the HL-LHC. The theoretical uncertainties are not included. Figures from Ref.~\cite{Maltoni:2016yxb}.}
		\label{fig:ttH}
\end{figure}

NLO QCD results in the Standard Model Effective Field Theory (SMEFT) have been calculated in Ref.~\cite{Maltoni:2016yxb},
including also a study of the processes $pp\to H$, $pp\to H\,j$ and $pp\to HH$, which also involve the Higgs-top and Higgs-gluon operators.
Fig.~\ref{fig:ttH} shows how a combined fit based on these processes can break the degeneracies in the coupling parameter space present in the individual processes.

The $H\to \gamma\gamma$ decay channel of the $t\bar{t}H$ process is particularly well suited to measure a possible CP-violating phase of the top Yukawa coupling~\cite{Gunion:1996xu,Artoisenet:2013puc,Ellis:2013yxa,Demartin:2014fia,Boudjema:2015nda,Buckley:2015vsa,Gritsan:2016hjl,AmorDosSantos:2017ayi,Goncalves:2018agy,Ferroglia:2019qjy}
and experimental constraints are already available~\cite{Sirunyan:2020sum,Aad:2020ivc}.

Using single top plus Higgs production to probe the CP-properties of the top Yukawa coupling has been studied e.g. in  Refs.~\cite{Campbell:2013yla,Ellis:2013yxa,Demartin:2015uha,AmorDosSantos:2017ayi,Barger:2019ccj}.

In Ref.~\cite{Deutschmann:2018avk} NLO results for the process $pp\to b\bar{b}H$ have been calculated and analysed in view of constraining both $y_t$ and $y_b$.

Given  the projection that the statistical uncertainty will shrink to the order of 2-3\% after 3000 fb$^{-1}$~\cite{Azzi:2019yne} (while currently statistical and systematic uncertainties are of about the same size, depending on the Higgs decay channel), the measurement of $t\bar{t}H$ will be dominated by systematics.
As the dominant systematic uncertainties currently come from modelling uncertainties of signal and backgrounds~\cite{Aaboud:2018urx,Sirunyan:2018hoz}, there is a clear need to reduce the 
theory uncertainties.
At NLO QCD the scale uncertainties are of the order of 10-15\%, therefore NNLO QCD corrections are necessary to match the experimental precision at the HL-LHC. First steps towards the corresponding real radiation corrections have been made in Ref.~\cite{Catani:2021cbl}.

\subsection{Higgs boson pairs}

\begin{figure}[htb]
  \centering
 \includegraphics[width=0.7\textwidth] {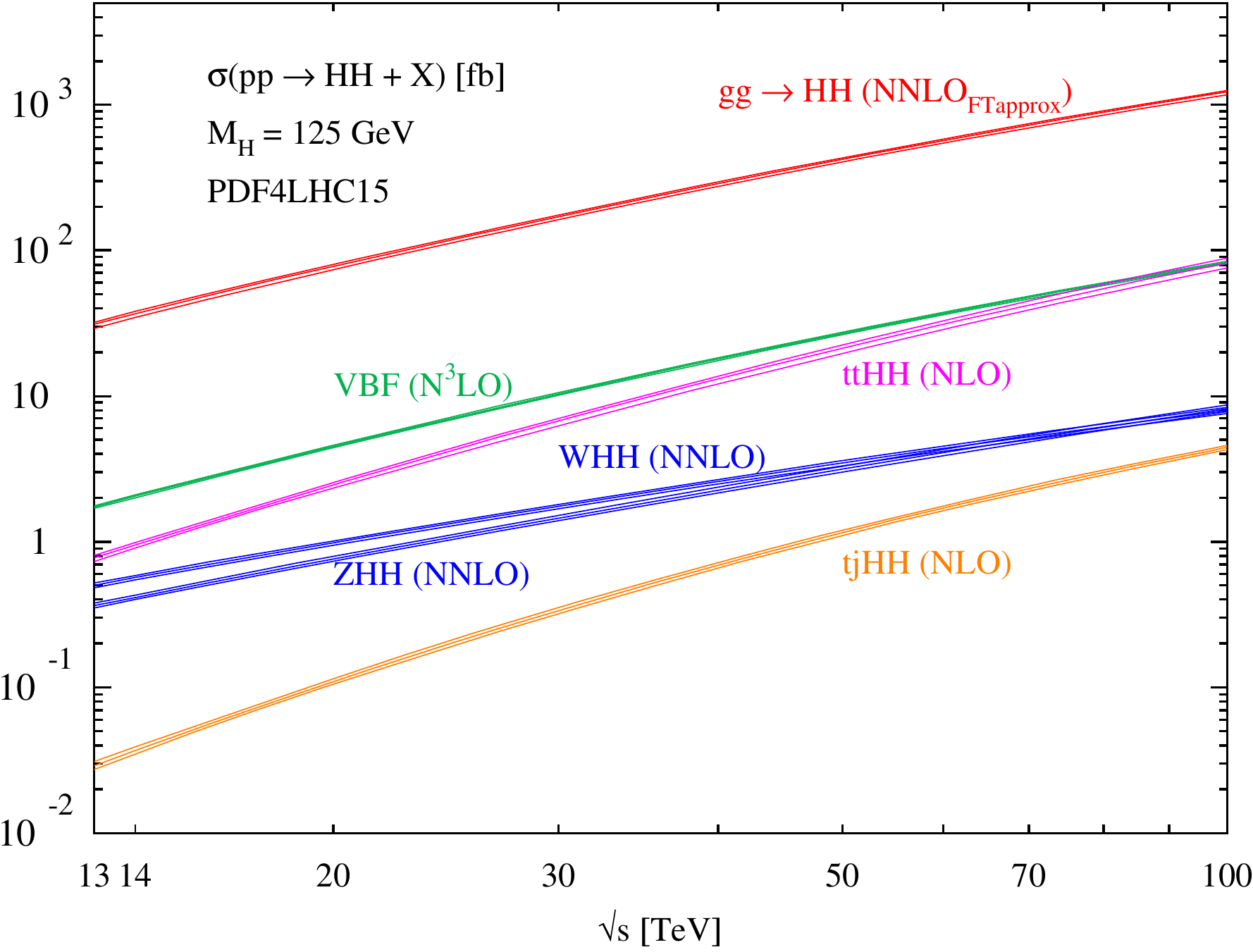}
  \caption{The main production channels for Higgs boson pair production. Figure from Ref.~\cite{deFlorian:2016spz}.}
  \label{fig:HHprodchannels}
\end{figure}

Similar to single Higgs production, the main Higgs boson pair production channels are gluon fusion, VBF, associated production with a vector boson and associated production with a top quark pair. The corresponding cross sections as functions of energy are shown in Fig.~\ref{fig:HHprodchannels}.

Higgs boson pair production is an interesting process due to its potential to access the trilinear Higgs boson self-coupling, which is one of the few SM parameters being still largely unconstrained. The projection for the HL-LHC is to constrain it to about 50\% uncertainty after 3000\,fb$^{-1}$, combining different decay channels and the two experiments ATLAS and CMS~\cite{Cepeda:2019klc,Adhikary:2017jtu}. At a future FCC-hh operating at $\sqrt{s}=100$\,TeV, an accuracy of about 5\% could be achieved with 30\,ab$^{-1}$~\cite{Mangano:2020sao,Goncalves:2018qas}.
As in the SM the Higgs boson self-couplings are completely determined by the Higgs VEV and its mass, any statistically significant deviation from the predicted value would be a clear sign of New Physics.
Currently  the most stringent 95\% CL limit
on the total $gg\to HH$ cross section at $\sqrt{s}=13$\,TeV is $\sigma^{HH}_{\rm{max}}= 6.9\times \sigma_{SM}$, constraining trilinear coupling modifications to the range $-5.0\leq \lambda/\lambda_{\rm{SM}}\leq 12.0$~\cite{Aad:2019uzh} with the assumption that all other couplings are SM-like.

The trilinear Higgs couplings can also be constrained in an indirect way, 
through measurements of processes which are sensitive to these couplings via electroweak corrections~\cite{McCullough:2013rea,Gorbahn:2016uoy,Degrassi:2016wml,Bizon:2016wgr,Maltoni:2017ims,Kribs:2017znd,Degrassi:2017ucl,Nakamura:2018bli,Kilian:2018bhs,Maltoni:2018ttu,Vryonidou:2018eyv,Gorbahn:2019lwq}.
Such processes offer important complementary information. However the number of operators entering the loop corrections 
at the same level in an EFT expansion is in general larger than the leading set of operators entering Higgs boson pair production.
Therefore the limits on $c_{hhh}=\lambda/\lambda_{\rm{SM}}$ extracted this way risk to be weakened by the larger number of parameters to fit. 
An experimental analysis based on single Higgs boson production processes has been performed to derive combined constraints from single and double Higgs boson production~\cite{ATLAS:2019pbo}.
Under the  assumption that all
deviations from the SM expectation are stemming from a modification of the trilinear coupling, 
the derived bounds on $c_{hhh}$ at 95\% CL from the combined analysis are 
$-2.3 \leq c_{hhh}\leq 10.3$~\cite{ATLAS:2019pbo}. However once the couplings to vector bosons and/or fermions are allowed to vary as well, these bounds deteriorate significantly.

The idea of indirect constraints through loop corrections also has been employed trying to constrain the quartic Higgs boson self-coupling from (partial) EW corrections to Higgs boson pair production~\cite{Maltoni:2018ttu,Liu:2018peg,Bizon:2018syu,Borowka:2018pxx}.


Higgs boson pair production in gluon fusion in the SM has been calculated at leading order in Refs.~\cite{Eboli:1987dy,Glover:1987nx,Plehn:1996wb}, and at NLO in the $m_t\to\infty$ limit (HTL), rescaled with the full Born matrix element,  in Ref.~\cite{Dawson:1998py}. 
An approximation called ``FT$_{\rm{approx}}$", was introduced in Ref.~\cite{Maltoni:2014eza},  which contains the full top quark mass dependence in the real radiation, 
while the virtual part is calculated in the Born-improved $m_t\to\infty$ limit.
The NLO QCD corrections with full top quark mass dependence became available more recently~\cite{Borowka:2016ehy,Borowka:2016ypz,Baglio:2018lrj,Baglio:2020ini},
both groups calculated the corresponding two-loop integrals numerically. In Refs.~\cite{Borowka:2016ehy,Borowka:2016ypz} the program {\sc SecDec}~\cite{Borowka:2015mxa,Borowka:2017idc} was used, which will be discussed in more detail in Section~\ref{sec:secdec}.
Implementations of the full NLO QCD corrections in parton shower Monte Carlo programs are also available~\cite{Heinrich:2017kxx,Jones:2017giv,Heinrich:2019bkc,Heinrich:2020ckp}.

NNLO QCD corrections have been computed in the HTL in Refs.~\cite{deFlorian:2013uza,deFlorian:2013jea,Grigo:2014jma,Grigo:2015dia,deFlorian:2016uhr}, where Ref.~\cite{deFlorian:2016uhr} contains fully differential results.
The results of Ref.~\cite{Grigo:2015dia} also contain an expansion in $1/m_t^2$, soft gluon resummation has been performed at NNLO+NNLL level in Ref.~\cite{deFlorian:2015moa}. 
The calculation of Ref.~\cite{deFlorian:2016uhr} has been combined with results including the top quark mass dependence as far as available in Ref.~\cite{Grazzini:2018bsd}, and the latter has been supplemented by soft gluon resummation in Ref.~\cite{deFlorian:2018tah}.
The scale uncertainties at NLO are still at the 10\% level, while they are decreased to about 5\% when including the NNLO corrections.
The uncertainties due to the chosen top mass scheme have been assessed in Refs.~\cite{Baglio:2020ini,Baglio:2020wgt} and found to be larger than the NNLO scale uncertainties.
The N$^3$LO corrections to Higgs boson pair production in gluon fusion in the heavy top limit  recently became available~\cite{Chen:2019lzz}, and
top mass effects at NLO obtained from the {\tt ggHH} code~\cite{Heinrich:2017kxx,Heinrich:2019bkc,Heinrich:2020ckp}  have been included by a reweighting procedure in Ref.~\cite{Chen:2019fhs}.
The new results reduce the scale uncertainties to the level of a few percent,
such that the remaining uncertainties are mainly related to the missing mass effects at this order, as well as to the  top-quark mass scheme dependence.
The PDF uncertainties are estimated to be at the 2\% level~\cite{Cepeda:2019klc}.

Analytic approximations for the top quark mass dependence of the two-loop amplitudes in the NLO calculation have been studied in Refs.~\cite{Grober:2017uho,Bonciani:2018omm,Xu:2018eos,Mishima:2018olh,Davies:2018ood}.
In Ref.~\cite{Grober:2017uho} the large-$m_t$ expansion is combined with a threshold expansion with a Pad\'e ansatz, 
in Ref.~\cite{Bonciani:2018omm},  an expansion in $p_T^2+m_h^2$ has been worked out.
Ref.~\cite{Xu:2018eos} considers an expansion around the limit $m_h^2\to 0$, constructed to comprise both large-$m_t$ and high energy expansions.
In Refs.~\cite{Mishima:2018olh,Davies:2018ood} the high energy limit for Higgs boson pair production is considered.
Complete analytic results in the high energy limit have been presented in Ref.~\cite{Davies:2018qvx}.

At very high energies, such as for $p_T^h$ values $\gtrsim$ 1\,TeV, the results of the {\tt ggHH} program of Refs.~\cite{Heinrich:2017kxx,Heinrich:2019bkc,Heinrich:2020ckp} become unreliable due to lack of support from the grid encoding the virtual two-loop amplitude.
To remedy this, the results of Ref.~\cite{Davies:2018qvx} have been combined with the full NLO results to obtain reliable results over the full kinematic range~\cite{Davies:2019dfy}.
In Ref.~\cite{Davies:2019djw},  $1/m_t^2$ corrections to the HTL three-loop $gg \to HH$ amplitude have been calculated.
Results for the real-virtual corrections which would enter a full NNLO calculation have been presented in Ref.~\cite{Davies:2019xzc}.
Four-loop corrections to the effective coupling of two Higgs bosons and two, three or four gluons have been provided in Refs.~\cite{Gerlach:2018hen,Spira:2016zna}.

\begin{figure}[htb]
  \centering
 \includegraphics[width=0.49\textwidth] {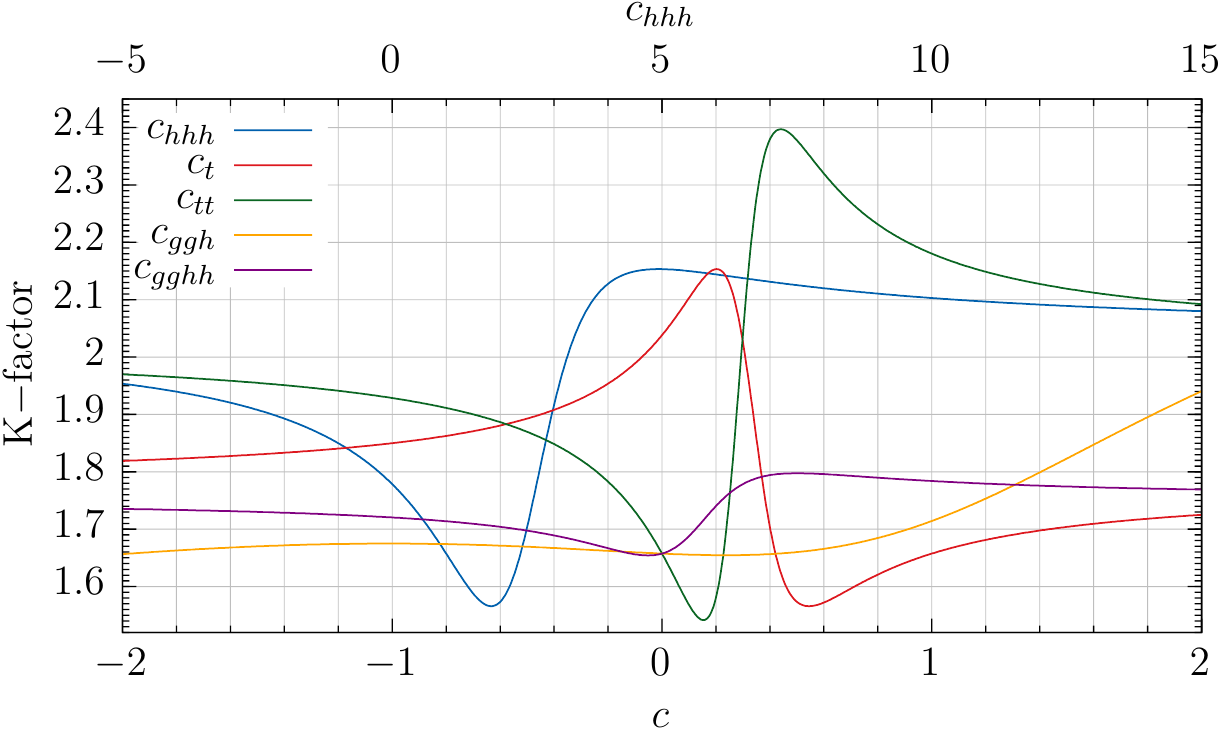}
 \includegraphics[width=0.48\textwidth] {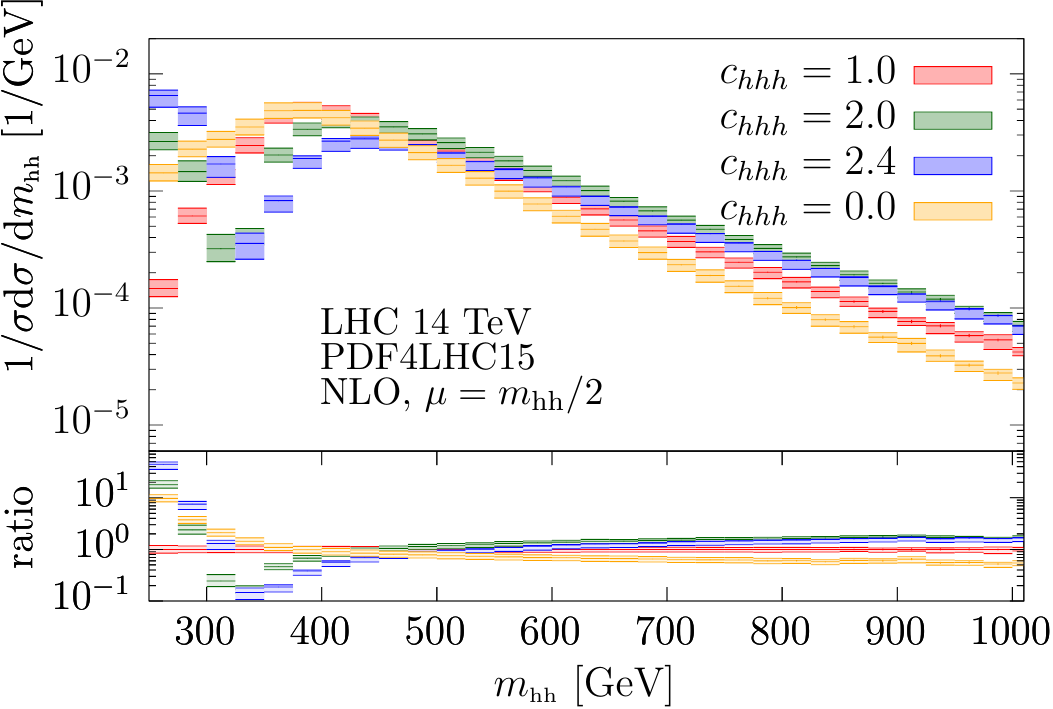}
  \caption{Left: NLO K-factors as functions of the anomalous couplings. Right: Normalised Higgs boson pair invariant mass distributions for various
  values of $c_{hhh}$  at $\sqrt{s}=14$\,TeV. The uncertainty bands are from
  3-point scale variations around the central scale $\mu_0 =\mhh/2$. Figures from Refs.~\cite{Buchalla:2018yce,Heinrich:2019bkc}.}
  \label{fig:EWChL}
\end{figure}

The effects of operators within an Effective Field Theory (EFT) description of Higgs boson pair production and the prospects to constrain the Higgs boson self-couplings 
have been studied extensively in the literature, see e.g. Refs.~\cite{Contino:2012xk,Goertz:2014qta,Chen:2014xra,Azatov:2015oxa,Dawson:2015oha,Carvalho:2015ttv,Cao:2015oaa,Cao:2016zob,DiVita:2017eyz,deBlas:2018tjm,Kim:2019wns,Barducci:2019xkq,Cheung:2020xij,Abdughani:2020xfo,Adhikary:2020fqf}.
For dedicated NLO (HTL) studies see Refs.~\cite{Grober:2015cwa,Grober:2016wmf,Maltoni:2016yxb}, including also CP-violating operators~\cite{Grober:2017gut}.
EFT studies at NNLO in the $m_t\to\infty$ limit are also available~\cite{deFlorian:2017qfk,Amoroso:2020lgh}.
In Ref.~\cite{Buchalla:2018yce} for the first time the full NLO QCD corrections have been combined with an EFT approach to study BSM effects.
These predictions have been used to investigate the influence of anomalous couplings on the $m_{hh}$ shape at NLO~\cite{Capozi:2019xsi}, and they have been implemented in the publicly available program {\tt ggHH}~\cite{Heinrich:2020ckp}. The effect of anomalous couplings on the inclusive NLO K-factors and on the $\mhh$ distribution is illustrated in Fig.~\ref{fig:EWChL}.

Higgs pair production from bottom quark annihilation has been calculated to NNLO in QCD in the soft-virtual approximation in the five-flavour scheme~\cite{H:2018hqz}.

Electroweak corrections should also be considered at this level of precision, in particular as they can distort  differential cross sections.
The number of mass scales in the corresponding two-loop amplitudes currently does not allow an analytic calculation of the latter. A subclass of integrals occurring in these corrections has been calculated numerically in Ref.~\cite{Borowka:2018pxx}.

\subsection*{HH in vector boson fusion}

For Higgs boson pair production in vector boson fusion, results in the structure function approach are available for the total cross section at N$^3$LO~\cite{Dreyer:2018qbw} and
fully differentially at NNLO~\cite{Dreyer:2018rfu}.
The NNLO corrections were found to be at the level of $3-4\%$ after typical VBF cuts. The  N$^3$LO
corrections turned out to be negligible at the central scale choice, however
they reduce the NNLO scale uncertainty by a factor of four.
It was found that the changes to the cross section due to variations of $c_{hhh}$ can be substantial, however the K-factor as a function of $c_{hhh}$ does not show large variations.

Strategies to disentangle the VBF and direct production modes have been discussed in Ref.~\cite{Dolan:2015zja}.
In Ref.~\cite{Dreyer:2020urf} the non-factorisable diagrams, i.e. the ones not captured by the structure function approach, have been estimated, based on the eikonal approximation.
The calculation reveals that that the non-factorisable corrections are sizeable. This fact can be attributed to a delicate cancellation of the various Born diagrams, which is spoiled by the radiative corrections. Because the factorisable corrections to the di-Higgs process are smaller than in the single Higgs process, the non-factorisable corrections are of the same order of magnitude or even dominant compared to the NNLO factorisable ones. For the fiducial volume studied in Ref.~\cite{Dreyer:2020urf} the two corrections have opposite sign and partially cancel each other.

In Ref.~\cite{Dreyer:2020xaj}, the NLO EW corrections combined with NNLO QCD corrections are presented, including also non-factorisable contributions. 
It was found that  the NLO EW corrections within the fiducial volume amount to about $-6$\% and can be
up to $-20$\%  in differential cross sections. The EW corrections combined with the NNLO QCD corrections lead to a total correction of $-14.8$\% compared to the LO prediction. It was also confirmed that the structure function approximation becomes unreliable in rather inclusive setups or in extreme regions of  phase space.

\subsection*{HHV}

Associated production of a Higgs boson pair with a vector boson has been calculated at NNLO for the inclusive case in Ref.~\cite{Baglio:2012np}. The gluon fusion channel, entering $ZHH$ production at order $\alpha_s^2$, was found to contribute ${\cal O}(20\%)$ to the total NNLO cross section.
Differential NNLO results were calculated in Ref.~\cite{Li:2017lbf} for $ZHH$ and in  Ref.~\cite{Li:2016nrr} for $WHH$, based on $q_T$-subtraction.

\subsection*{HH\,$t\bar{t}$}

In Refs.~\cite{Englert:2014uqa,Liu:2014rva} it has been pointed out that the $HHt\bar{t}$  channel at the HL-LHC may provide additional information for a determination of the trilinear Higgs coupling. In fact, as can be seen from Fig.~\ref{fig:HHprodchannels}, the cross section in this channel also rises faster with centre-of-mass energy than for the other channels. The potential of this channel at FCC-hh energies recently has been analysed in Ref.~\cite{Banerjee:2019jys}.
An important feature of this process is that in models where the Higgs sector is non-linearly realised, i.e. described by the Electroweak Chiral Lagrangian (also called ``Higgs Effective Field Theory'' (HEFT))~\cite{Feruglio:1992wf,Alonso:2012px,Buchalla:2013rka}, the $t\bar{t}H$ and $t\bar{t}HH$ couplings are not linearly dependent as they are in SMEFT, as has been discussed e.g. in Refs.~\cite{Buchalla:2018yce,DiMicco:2019ngk,Banerjee:2019jys,Capozi:2019xsi}.
This is due to the fact that in HEFT, the leading terms in the Lagrangian contributing to fermion-Higgs interactions are
\begin{align}
{\cal L}\supset 
-m_t\left(c_t\frac{h}{v}+c_{tt}\frac{h^2}{v^2}\right)\,\bar{t}\,t \;,
\label{eq:ewchl}
\end{align}
where $c_t$ and $c_{tt}$ are unrelated.
In SMEFT, if $M$ is a high mass scale, and
 $c_t=1+a v^2/M^2+{\cal O}(\frac{1}{M^4})$, where $a$ is a parameter of
 order one, it follows that $c_{tt}=\frac{3}{4}\,a v^2/M^2+{\cal O}(\frac{1}{M^4})$ up to $c_H$-terms, such that $c_{tt}$
 is naturally small if the corrections to $c_t$ are small.
 This can also be seen from the translation between the two schemes in the conventions of Refs.~\cite{Grober:2015cwa,Buchalla:2018yce}:
\begin{align}
 c_t&=1-\bar c_u-\frac{\bar c_H}{2}\; ,\;
c_{tt}=-\frac{ 3\bar c_u}{4}-\frac{\bar c_H }{4}\;. 
\end{align}
Therefore an independent measurement of the effective $t\bar{t}HH$ coupling is of great importance to shed light on the EW symmetry breaking sector.
In Ref.~\cite{Cohen:2020xca}, which elucidates in detail  the relations between SMEFT and HEFT and their domains of validity, the conclusions state: ``It would be valuable to identify measurements one could make that would test if our low energy EFT could be SMEFT or must be HEFT''.
A measurement of  the effective $t\bar{t}HH$ coupling would certainly fall into this category.

\subsection{Multi-Higgs}

The SM cross section for triple Higgs production in gluon fusion at approximate NNLO is about 5$fb$ at 100\,TeV~\cite{deFlorian:2019app}, so it is smaller than the one for double Higgs production at 100\,TeV by about a factor of about 250.
Therefore a measurement of the quartic Higgs boson coupling from this process is not possible at the LHC and challenging even at a 100\,TeV collider.
However, the smallness of the cross section is related to delicate cancellations between contributions involving $\lambda_3$ and $\lambda_4$ for SM values of all couplings. Therefore, New Physics effects could enhance the cross section significantly~\cite{Agrawal:2017cbs,Agrawal:2019bpm,Fuks:2017zkg,Kilian:2017nio,Liu:2018peg,Papaefstathiou:2015iba,Papaefstathiou:2019ofh}, such that it is important to have the theory predictions under control.

As this process already at Born level contains one-loop pentagon diagrams with massive loops, the calculation of higher order corrections is challenging. At leading order it has been calculated in Refs.~\cite{Plehn:2005nk,Binoth:2006ym}.
NLO corrections were calculated in Ref.~\cite{Maltoni:2014eza}, using the approximation FT$_{\rm{approx}}$, where the virtual amplitude is calculated in the HTL, while keeping the exact dependence on the top quark mass is used for the real emission contributions.
In Ref.~\cite{deFlorian:2016sit} the two-loop virtual amplitudes where computed in the HTL to obtain the soft-virtual approximation of the NNLO corrections.
These results were extended in Ref.~\cite{deFlorian:2019app}
beyond the soft-virtual approximation, computing the complete set of NNLO QCD corrections for triple Higgs production within the HTL.  Partial finite top mass  effects are included by taking into account the NLO FT$_{\rm{approx}}$ results of Ref.~\cite{Maltoni:2014eza}.

Triple Higgs production in association with a vector boson also has been studied~\cite{Dicus:2016rpf}, as well as multi-Higgs production in Vector Boson Fusion~\cite{Kilian:2018bhs}. 
The possibility to obtain combined constraints on both the triple and the quartic Higgs coupling at future $e^+e^-$ colliders has been discussed in Ref.~\cite{Maltoni:2018ttu} considering  $e^+e^-\to ZH^n$ and $e^+e^-\to \nu\bar{\nu}H^n$ with $n=1,2,3$, see also~\cite{DiVita:2017vrr,Barklow:2017awn}.
In Ref.~\cite{Chiesa:2020awd} the prospects of measuring the quartic Higgs self-coupling at a multi-TeV muon collider are investigated, Ref.~\cite{Han:2020pif} discusses the prospects to measure the couplings of  Higgs bosons to electroweak gauge bosons at a muon collider.

\subsection{Tools for  differential NNLO predictions}

While dedicated NNLO programs to provide differential results for specific processes became publicly available some time ago already, such as
{\sc FehiPro}~\cite{Anastasiou:2005qj,Anastasiou:2009kn}, {\sc Hnnlo}~\cite{Catani:2007vq,Grazzini:2008tf,Grazzini:2013mca}
for Higgs boson production in gluon fusion,  {\sc 2$\gamma$nnlo}~\cite{Catani:2011qz} for diphoton production,
{\sc eerad3}~\cite{Ridder:2014wza} for 3-jet production in $e^+e^-$ annihilation, 
 {\sc Fewz}~\cite{Melnikov:2006kv,Gavin:2010az}, {\sc DYnnlo}~\cite{Catani:2009sm} and {\sc DYturbo}~\cite{Camarda:2019zyx,Camarda:2021ict} for vector-boson production or {\sc ProVbfH}~\cite{Cacciari:2015jma} and {\sc ProVbfH}\,v1.2~\cite{Dreyer:2016oyx,Dreyer:2020urf} for Higgs production in vector boson fusion or {\sc ProVbfHH}~\cite{Dreyer:2018qbw,Dreyer:2018rfu,Dreyer:2020urf} for Higgs pair production in vector boson fusion,
 multi-purpose programs are rare for obvious reasons.
 A summary of publicly available programs that can provide differential NNLO results for hadron colliders is given in Table~\ref{tab:nnlotools}.
 \begin{table}[htb]
 \hspace*{-1cm}
   \begin{tabular}{|c|c|c|l|}
     \hline
     process& name & Refs. & remarks\\
     \hline
     $pp\to H$& {\sc FehiPro}&\cite{Anastasiou:2005qj,Anastasiou:2009kn}&$t$- and $b$-quark masses up to ${\cal O}(\alpha_s^3)$\\
     $pp\to H$& {\sc Hnnlo}&\cite{Catani:2007vq,Grazzini:2008tf,Grazzini:2013mca}&$t$- and $b$-quark masses up to ${\cal O}(\alpha_s^3)$\\
     $pp\to H$&{\sc Hqt} &\cite{Bozzi:2005wk,deFlorian:2011xf}& $p_T$-dist., NNLL resummation\\
 $pp\to\gamma\gamma$& {\sc 2$\gamma$nnlo}&\cite{Catani:2011qz,Catani:2018krb} & also available in MCFM~\cite{Campbell:2016yrh}\\
   $t\to W(l\nu)\,b$&{\sc nnTopDec}&  \cite{Gao:2012ja}& {\tt https://nntopdec.hepforge.org/}\\
     Drell-Yan& {\sc Fewz}&\cite{Melnikov:2006kv,Gavin:2010az,Gavin:2012sy}&\\
     Drell-Yan&{\sc DYnnlo}&\cite{Catani:2009sm}&\\
     Drell-Yan&{\sc DYturbo}&\cite{Camarda:2019zyx,Camarda:2021ict,Bozzi:2010xn} &merge of {\sc DYnnlo}, {\sc DYres}, {\sc DYqt}\\
     &&& {\tt https://dyturbo.hepforge.org/}\\
      $pp\to VH$&{\sc MCFM}& \cite{Campbell:2016jau}&\\
     VBF H prod.& {\sc ProVbfH}&\cite{Cacciari:2015jma,Dreyer:2016oyx,Dreyer:2020urf} &\\
     VBF HH prod.&  {\sc ProVbfHH}&\cite{Dreyer:2018qbw,Dreyer:2018rfu,Dreyer:2020urf}   &\\            
  \hline
     colour singlet $2\to 1, 2$& {\sc Matrix}+{\sc OpenLoops}&\cite{Grazzini:2017mhc,Cascioli:2011va,Buccioni:2019sur,Kallweit:2019zez} & {\tt https://matrix.hepforge.org}\\
            &&&incl. NLO EW corrections for VV$^{(\prime)}$ \\
        colour singlet $2\to 1,2$& {\sc Matrix}+{\sc Radish}&\cite{Kallweit:2020gva} &incl. $q_T$ resummations  \\
     colour singlet $2\to 1, 2$&{\sc MCFM}&\cite{Boughezal:2016wmq,Campbell:2019dru}&{\tt https://mcfm.fnal.gov/}\\
       \hline          
     \end{tabular}
\caption{{\bf Public} programs which can provide differential results at NNLO relevant for hadron colliders.\label{tab:nnlotools}}
\end{table}   
 
 Currently the publicly available tools providing differential NNLO predictions for a number of different processes are {\sc Matrix}~\cite{Grazzini:2017mhc,Kallweit:2019zez}+{\sc OpenLoops}~\cite{Cascioli:2011va,Buccioni:2019sur} and {\sc MCFM}~\cite{Boughezal:2016wmq,Campbell:2019dru}.
 The NNLO processes available in {\sc MCFM} and {\sc Matrix} were initially focused on colour-singlet final states, however this restriction has been lifted e.g. in Ref.~\cite{Campbell:2019gmd} for NNLO H+jet production in  {\sc MCFM},  or in Ref.~\cite{Catani:2020tko} providing NNLO results for top quark pair production within {\sc Matrix}.
 The {\sc NNLOJet} collaboration started out with NNLO predictions for single jet inclusive~\cite{Currie:2016bfm} and dijet production~\cite{Currie:2017eqf}, however in the meantime has tackled a considerable number of processes involving  vector bosons, Higgs bosons and jets.
A very recent highlight is the fully differential calculation of $pp\to H$ at N$^3$LO~\cite{Chen:2021isd} by the {\sc NNLOJet} collaboration.
 Some of the results for NNLO jet cross sections are available in the form of grids, see e.g.~\cite{Britzger:2019kkb}. Similarly, results for $t\bar{t}$ production at NNLO are available in the form of {\tt fastNLO}~\cite{Kluge:2006xs,Britzger:2012bs} tables~\cite{Czakon:2017dip,Cooper-Sarkar:2020twv}.

 Making calculations at high perturbative orders available in a fast and flexible way, for example in the form of grids which allow the convolution with different PFDs, is certainly a subject that will be of increasing importance. Besides the {\tt fastNLO}~\cite{Kluge:2006xs,Britzger:2012bs} and {\tt APPLgrid}~\cite{Carli:2010rw} projects, the {\tt pineAPPL}~\cite{Carrazza:2020gss} framework has been developed recently, which can 
compute fast interpolation grids accurate to NLO QCD as well as NLO QCD+EW, particularly important for LHC processes where EW corrections have a sizeable effect. Approaches based on N-tuples or matrix elements expanded in a basis of orthogonal phase-space functions are also promising~\cite{Maitre:2020blv}.

 Last but not least, it should be mentioned that one of the main challenges to move forward in precision phenomenology is to combine fixed-order calculations beyond NLO consistently with parton showers. Several approaches are being worked on and have been applied already to colour singlet final states, for example the NNLO+PS matching implemented in the parton shower program {\sc Geneva}~\cite{Alioli:2012fc,Alioli:2013hqa,Alioli:2015toa,Alioli:2019qzz,Alioli:2021qbf}, {\sc UNnlops}~\cite{Hoeche:2014aia,Hoche:2014dla,Hoche:2018gti} or {\sc MiNnlo}~\cite{Monni:2019whf,Monni:2020nks,Lombardi:2020wju}, which is a new development based on {\sc Nnlops}+{\sc MiNlo}~\cite{Hamilton:2012rf,Hamilton:2015nsa,Re:2018vac,Astill:2018ivh}.

\section{Overview of modern techniques for loop amplitudes}
\label{sec:amplitudes}

\subsection{Multi-loop Amplitudes}

Scattering amplitudes can be considered as the core of any
perturbative calculation of a physical quantity relevant to particle
interactions, in collider experiments as well as in a wider context
including also gravity.
The calculation of scattering amplitudes beyond the leading order in
perturbation theory has seen immense progress in the last decade,
which led to a deeper mathematical understanding of the structure of both tree- and
loop amplitudes, and opened the door to many important
phenomenological applications.

We do not cover one-loop amplitudes here because the techniques for
one-loop calculations and NLO automation are in a rather mature state. 
For reviews see e.g. Refs.~\cite{Campbell:2006wx,Bern:2007dw,Ellis:2011cr,Dixon:2013uaa,Denner:2019vbn}.
More specialised reviews about modern methods for one-loop
QCD amplitudes can be found in Refs.~\cite{Britto:2010xq,Bern:2011qt,Ita:2011hi,Elvang:2013cua}.
One of the main reasons why the one-loop case is substantially
different from the multi-loop case is that for one-loop amplitudes,
all scalar products between loop momenta and external momenta
appearing in the numerator can be expressed in terms of propagators
and external kinematic invariants, which means there are no ``irreducible scalar products (ISPs)''.
Further, for any number of external legs, the integral basis can be expressed in
terms of $N$-point functions with $N\leq 4$ in $D=4$ dimensions, and the so-called ``rational parts'',
related to the $(D-4)$-dimensional components of the loop momentum, are easily calculable.
As a consequence the class of
analytic functions the results must belong to is predictible and
simple: the most complicated analytic functions that can occur up to order $\eps^0$ are
dilogarithms. This will be different once we go beyond one loop.


\subsubsection{Basic concepts}

Before we discuss recent developments and highlights in amplitude
calculations, let us introduce some notation and basic concepts which will
be essential in order to describe various techniques for the
calculation of multi-loop amplitudes.

\subsubsection*{Integral families}

We consider  an integral with $L$ loops in $D$ dimensions
with propagator powers $\nu_j$,  
\begin{eqnarray}\label{eq:intfamily}
F(\nu_1\ldots\nu_n)  &=&
\int \prod\limits_{l=1}^{L} \frac{\rd^Dk_l}{i\pi^{\frac{D}{2}}}\;
\prod\limits_{j=1}^{n} 
\frac{1}{P_{j}^{\nu_j}(\{k\},\{p\},m_j^2)}\;,
\end{eqnarray}
where the propagators $P_{j}(\{k\},\{p\},m_j^2)$ 
depend on the  loop momenta $k_{l}$, the external momenta
$\{p_1,\dots p_E\}$ and the (not necessarily nonzero) masses $m_j$.
We use dimensional regularisation~\cite{tHooft:1972tcz,Bollini:1972ui} to regulate both ultraviolet and infrared divergences,
such that, with $D=4-2\eps$, the singularities in four dimensions show up as $1/\eps\,$-terms to some power.
The set of propagators can also contain propagators which depend linearly on
the loop momentum rather than quadratic, as they occur for example in
heavy quark effective field theory or in non-covariant gauges.
A scalar integral with no loop-momentum dependence in the numerator would correspond to  $\nu_j\ge 0$ for all $j$.
The set of integrals $F(\nu_1\ldots\nu_n)$ is called an {\em integral family} if
it contains all integrals with a propagator configuration such that any scalar
product of a loop momentum with another loop momentum or with an external momentum 
can be  expressed as a linear combination of inverse propagators contained in the same family.
The integrals in a family are in general linearly dependent. 
Finding a convenient basis in the vector space formed by the
integrals, in terms of which all integrals of a given family can be
expressed, corresponds to a reduction to {\it master
  integrals}. Finding a convenient master integral basis is of crucial
importance for both a compact representation of the amplitude as well
as for the evaluation of the master integrals. Depending on the method
used for the evaluation (e.g. differential equations or a numerical
evaluation), what is understood as ``convenient'' can differ.

Note that for a process with $N$ external particles in $D$
dimensions, only $E = \mbox{min}(N-1,D)$ of these external momenta will be independent ($N$ is
reduced by one due to momentum conservation). In a $D$-dimensional space
any vector can be expressed as a linear combination of $D$ basis vectors. 
For an $N$-point process with $L$ loops, the number $n$ of genuinely different scalar
products of the type $k_i \cdot k_j$ or $k_i \cdot p_j$ is given by
\be
n = L(L + 1)/2 + LE \;,\label{eq:scalprods}
\ee
where the first term comes from contracting the loop momenta with themselves, and the second
from contracting the loop momenta with the external momenta.
%
%
A set of $t$ propagators of an integral family defines a {\em sector} of
this family~\cite{vonManteuffel:2012np}.
The number of different $t$-propagator sectors is $\binom{n}{t}$ and therefore
in principle, $\sum_{t=0}^{n} \binom{n}{t} = 2^n$ sectors are contained in an integral
family, however many of them will be zero or related by symmetries.

One purpose of an integral family is to label and classify loop
integrals.
To every $t$-propagator sector with propagator denominators $P_{j_1},\ldots, P_{j_t}$
we can define a set of integrals with different propagator powers by
%
\begin{equation}\label{eq:dim_reg_int}
I(t,r,s) = \int \prod\limits_{l=1}^{L} \frac{\rd^Dk_l}{i\pi^{\frac{D}{2}}}\;
\frac{P_{j_{t+1}}^{s_1} \ldots P_{j_n}^{s_{n-t}}}{P_{j_1}^{r_1} \ldots P_{j_t}^{r_t}}\;,
\end{equation}
with integer exponents $r_i \geq 1$ and $s_i \geq 0$.
An integral of a given family $F$ can therefore be characterised by
$t,r,s$ and the indices $\{\nu_1,\ldots,\nu_{n}\}=\{r_1, \ldots,
r_t,-s_1, \ldots -s_{n-t}\}$, 
where 
$r=\sum_{i=1}^t r_i$  and $s=\sum_{i=1}^{n-t} s_i$.
Positive $\nu_i$ denote powers of ``regular'' propagators, i.e. propagators in the denominator,
negative $\nu_i$ denote powers of inverse propagators, i.e. they form non-trivial numerators, 
and zero means the absence of a propagator.
The numbers $t$, $r$, $s$  can be calculated from the vector
$\vec{\nu}$, so they are redundant once $\vec{\nu}$ is given, but they facilitate a categorisation of the integrals.
For  a $t$-propagator sector of an $n$-propagator integral family,
the number of integrals that can be built for certain values of $r$ and $s$ is given by
$
N(n,t,r,s) = \binom{r-1}{t-1} \binom{s+n-t-1}{n-t-1} \, .
$
The two binomial factors count all possible ways to arrange the exponents of
the propagators in the denominator and numerator, respectively.
The integral with $r=t$ and $s=0$ of some sector is called 
\emph{corner integral} of this sector~\cite{vonManteuffel:2012np}.

Note that from \eqn{eq:scalprods} it becomes clear that all topologies
with a number of propagators $n_P>n$ are reducible, i.e. scalar products in the
numerator  can be expressed through linear
combinations of propagators of the same integral.
For $L=1$, one can show that the only 
irreducible numerators are of the type $k_i\cdot n_j$, where $n_j$ denotes
directions transverse to the hyperplane spanned by the physical
external momenta. These terms vanish after integration over the loop momenta
(in integer dimensions), a fact which is built into unitarity-based
reduction at integrand level. Starting from two loops, genuine
irreducible numerators can occur.

\subsubsection{Integration by parts identities}
It can be shown~\cite{tHooft:1978jhc,Tkachov:1981wb,Chetyrkin:1981qh,Smirnov:1991jn} that for Feynman integrals
regulated by dimensional regularisation, the integral over a total derivative
is zero. If ${\cal I}$ is the integrand of an integral of the form
(\ref{eq:intfamily}), taking derivatives as follows
\begin{equation}
 \int \prod\limits_{l=1}^{L}\rd^D k_{l} \, \frac{\partial}{\partial k_{i}^{\mu}} \, \big[ v^{\mu} \,
  {\cal I}(\vec{\nu}) \big] = 0
\label{eq:IBP}
\end{equation}
leads to identities between different integrals, so-called {\it integration by parts (IBP) identities}~\cite{Tkachov:1981wb,Chetyrkin:1981qh}.
The term $v^\mu$ can be a loop- or external momentum. More
generally, it can also be a linear function of loop- and external momenta which can
be chosen conveniently, for example such that propagators with powers
$\nu_i>1$ (also called ``propagators with dots'') do not occur~\cite{Gluza:2010ws,Schabinger:2011dz,Zhang:2014xwa,Ita:2015tya}.
If there are $L$ loop momenta and $E$ independent external
momenta one can therefore build $L\, (L+E)$ equations from one
integral. The integrals which are used as starting points to build a
system of equations  are called {\em seed integrals}.

The system of IBP identities is in general over-constrained, such that
 most of the integrals can be expressed as linear combinations of a small subset of integrals,
the master integrals (MIs). 
Lorentz-invariance identities~\cite{Gehrmann:1999as} of the form
\begin{equation}
  \sum\limits_{i=1}^{E}\left(p^{\nu}_{i}\frac{\partial}{\partial
      p_{i\mu}}-p_{i}^{\mu}\frac{\partial}{\partial
      p_{i\nu}}\right)\,F(\vec{\nu})=0.
\label{eq:lorentz}
\end{equation}
can be used to obtain additional relations (which are redundant, as
shown in~\cite{Lee:2008tj}, but can help convergence in solving the linear system).

The choice of the MIs is not unique, and as mentioned already, a convenient choice of the basis can make an
enormous difference in the calculation of amplitudes of a certain
complexity. Further, it is important to take symmetries into account,
as well as shifts of the loop momenta that do not change the kinematic invariants.
For more details we refer to Refs.~\cite{vonManteuffel:2012np,Maierhoefer:2017hyi}.

\subsubsection*{Example}

As an example for IBP reduction, we consider a massive vacuum bubble with a propagator power $\nu$,
\be
F(\nu)=\int \rd\kappa \frac{1}{(k^2-m^2+i\delta)^\nu}\;,\;\nu>0\;,
\ee
where we have used the abbreviation $\rd\kappa=\rd^Dk/(i\pi^{\frac{D}{2}})$. 
Using the IBP identity
$$
\int \rd\kappa\,
\frac{\partial}{\partial k_{\mu}} \left\{  \frac{k_{\mu}}{(k^2-m^2+i\delta)^\nu}  \right\}= 0
$$
leads to
\begin{align}
0&=\int \rd\kappa\left\{\frac{1}{(k^2-m^2+i\delta)^\nu}\,\frac{\partial}{\partial k_{\mu}}\left(k_{\mu}\right)-\nu\,k_{\mu}\,\frac{2k^{\mu}}{(k^2-m^2+i\delta)^{\nu+1}}\right\}\nn\\
&=D\,F(\nu)-2\nu\,\left[ F(\nu)+m^2\,F(\nu+1) \right]\nn\\
&\Rightarrow F(\nu+1)=\frac{D-2\,\nu}{2\,\nu\,m^2}\,F(\nu)\;.
\end{align}
We see that in this simple example there is only one master integral,
which we choose to be
\be
F(1)=-\Gamma(1-D/2)\,(m^2)^{\frac{D}{2}-1}\;.
\ee



In less trivial cases, an {\it order relation} among the 
integrals has to be introduced to be able to solve the system for a set of master integrals.
For example, an integral $T_1$ is considered to be smaller than an integral $T_2$ 
if $T_1$ can be obtained from $T_2$ by omitting some of the
propagators.
Within the same topology, the
integrals can be ordered according to the powers of their propagators,
using for example lexicographical ordering for the set of indices $\{-s_1, \ldots -s_{n-t},r_1, \ldots,
r_t\}$.

The first systematic approach of IBP reduction has been formulated 
by Laporta\,\cite{Laporta:2001dd}, therefore it is sometimes also
called {\em Laporta-algorithm}. The public tools where this algorithm
has been implemented will be listed in Section~\ref{subsec:reduction}, 
see also Refs.~\cite{Smirnov:2006ry,Grozin:2011mt} for brief reviews of the method.
Formulations of IBP reduction in terms of algebraic geometry have
proven very useful, for more details we refer to
Refs.~\cite{Mastrolia:2012wf,Zhang:2012ce,Larsen:2015ped,Zhang:2016kfo}
and Section~\ref{sec:status_amp}.


\subsubsection*{Amplitude calculations: workflow}

 The typical workflow to calculate an amplitude beyond one loop is the following:
\begin{enumerate}
\item amplitude generation, for example in terms of Feynman diagrams, 
\item bringing the amplitude to a convenient form,
\item reduction to master integrals,
\item calculation of the master integrals,
\item evaluation of the amplitude.
\end{enumerate}

We will discuss steps 1 and 2 in Section \ref{subsec:amp_rep}, step
3 in Section \ref{subsec:reduction}  and step 4 in Section
\ref{sec:masters}.
If step 5 involves integrals whose evaluation is time consuming (for
example non-trivial numerical integrations), it is conveniently performed in a way that takes into account the
weight of the integral coefficients in the amplitude, such that
integrals  whose numerical contribution to the amplitude is small
need not be calculated with the same precision as the ones with larger
numerical importance.


\subsubsection{Amplitude representations}
\label{subsec:amp_rep}

Step 1 above is usually relying at least partly on automated tools like
{\sc Qgraf}~\cite{Nogueira:1991ex}, {\sc
  FeynArts}~\cite{Hahn:1998yk,Hahn:2000kx} or {\sc
  FeynCalc}~\cite{Shtabovenko:2016sxi,Shtabovenko:2020gxv}, see also {\sc feyngen}~\cite{Borinsky:2014xwa} for high loop orders.
However, fully numerical approaches which avoid the combinatorial complexity
associated with the Feynman-diagrammatic approach are promising in
view of the growing loop/leg order of amplitudes to be tackled.
Numerical methods avoiding the reduction to master integrals will
be discussed in Section \ref{subsec:4dim}.

Step 2 is of course closely linked to step 1 and ideally both
steps are performed simultaneously, for example by generating the
amplitude using on-shell methods, thus avoiding in the first place the generation of
structures which will lead to cumbersome cancellations later.
A very promising approach is to use $D$-dimensional unitarity at two
loops in combination with  rational reconstruction to generate the
integrands, see e.g. Refs.~\cite{Badger:2016ozq,Peraro:2016wsq,Abreu:2017xsl,Abreu:2018rcw,Hartanto:2019uvl,Abreu:2020jxa} and Section~\ref{sec:status_amp}.

If the amplitude is generated in terms of Feynman diagrams, step
2 involves the saturation of open Lorentz- and spinor indices and the
mapping of the amplitude to smaller building blocks, and can have different forms.
For simple amplitudes which are not loop-induced, one solution is of course to directly interfere the virtual amplitude with the Born amplitude, however for more complicated processes this becomes quickly intractable.
A very general method is to identify all possible Lorentz and spinor
structures and to write the amplitude as a linear combination of  these structures, where the coefficients are called {\em form factors}. The latter can be extracted from the full amplitude by suitable projection operators, see e.g.~\cite{Binoth:2002xg,Glover:2003cm,Actis:2004bp} for early work  on the construction of projectors.
One of the advantages of this method is that all objects can be defined naturally in $D$ dimensions, while working with helicity amplitudes requires a very careful treatment of 4-dimensional versus $D$-dimensional objects. However, this method scales very badly with the number of external legs,
even though projectors for the case of five-gluon scattering have been derived~\cite{Boels:2018nrr}.

Recently, promising new approaches have been proposed~\cite{Chen:2019wyb,Peraro:2019cjj} which define {\em physical} projectors, thereby limiting the proliferation of terms.
When expressed in terms of the original tensors, only a subset of them will contribute and their number will correspond exactly to the number of independent helicity amplitudes in the process under consideration, thus getting rid of all the additional, unphysical degrees of freedom related to the $(D-4)$-dimensional space.
The power of the method described in Ref.~\cite{Peraro:2019cjj} is demonstrated by the construction of a complete set of physical projectors for the scattering of five gluons in QCD. The method described in Ref.~\cite{Chen:2019wyb} also exploits the simplifications stemming from the use of 4-dimensional external momenta to construct the projectors.
However, it requires to perform an explicit decomposition of external polarisation states in terms of 4-dimensional momenta, while the method of Ref.~\cite{Peraro:2019cjj} yields helicity projectors that are uniquely written as linear combinations of standard, $D$-dimensional projection operators.

\subsubsection{Amplitude reduction}
\label{subsec:reduction}

The reduction of integrals occurring in an amplitude (beyond one loop)
to a set of
linearly independent master integrals is most commonly performed using
integration-by-parts (IBP)
identities~\cite{Tkachov:1981wb,Chetyrkin:1981qh}  as
well as Lorentz invariance (LI) identities~\cite{Gehrmann:1999as} as described above.

Public implementations of the Laporta algorithm are available in the programs
{\sc Air}~\cite{Anastasiou:2004vj},
{\sc Reduze}~\cite{Studerus:2009ye,vonManteuffel:2012np},
{\sc Fire}~\cite{Smirnov:2008iw,Smirnov:2014hma,Smirnov:2019qkx},
{\sc LiteRed}~\cite{Lee:2012cn,Lee:2013mka}
and {\sc Kira}~\cite{Maierhoefer:2017hyi,Maierhofer:2018gpa,Klappert:2020nbg}.
These automated tools have been vital for the progress in the calculation of 3-point and
4-point amplitudes at two or more loops in the last few years.
However, as the number of scales involved in the reduction increases,
for example for two-loop 5-point amplitudes or $2\to2$ scattering with several
massive particles involved, the algebraic linear systems start to
become intractable. This led to important new developments, suggesting to perform the IBP reduction
numerically over finite fields~\cite{vonManteuffel:2014ixa}.
Tools which are useful to perform such a reduction are {\sc
  FiniteFlow}~\cite{Peraro:2016wsq,Peraro:2019svx} and {\sc
  FireFly}~\cite{Klappert:2019emp,Klappert:2020aqs}, {\sc MultivariateApart}~\cite{Heller:2021qkz} as well as
version 2.0 of {\sc Kira}~\cite{Klappert:2020nbg}.

It should also be mentioned that there are very powerful specialised reduction
programs, which can tackle massive tadpoles, like {\sc Matad}~\cite{Steinhauser:2000ry},  or two-point functions at three and four loops,
respectively, called {\sc Mincer}~\cite{Gorishnii:1989gt} and {\sc Forcer}~\cite{Ruijl:2017cxj}.

\subsubsection*{Basis choice for the master integrals}

The choice of the master integral basis is not unique, and for more
complicated amplitudes it can become instrumental to choose a
convenient basis.
Several criteria to define a ``good'' basis can be considered.
One of them is the requirement that the dimension $D$ factorises in the denominators at the end of the reduction process,
first used in Ref.~\cite{Melnikov:2016qoc}.
The coefficients of the master integrals are rational functions of the
kinematic invariants and the space-time dimension $D$. These
functions can be extremely cumbersome, and in some cases even
singular. However, vanishing denominators in IBP reductions are either spurious or 
related to singularities corresponding to solutions of the Landau equations.
Physical poles appearing as $1/(D-4)$-terms in the integral
coefficients cannot depend on the actual values of the kinematic invariants.
This implies that the
denominators can be decomposed into a product of functions depending
only on kinematic invariants and masses, and functions depending only
on the dimension $D$, which means that a representation must exist
where the dependence on $D$ is factorising.
In Ref.~\cite{Usovitsch:2020jrk}, a public Mathematica package is
presented to achieve such a $D$-factorising basis in combination with
{\sc Kira}~\cite{Klappert:2020nbg}, in Ref.~\cite{Smirnov:2020quc} an algorithm is presented which is
implemented in {\sc Fire}~\cite{Smirnov:2019qkx}.

For numerical evaluations of the master integrals, it can be
particularly convenient to use a so-called
quasi-finite basis~\cite{vonManteuffel:2014qoa,vonManteuffel:2015gxa}.
Such a basis is characterised by the fact that the integrals themselves
are finite, apart from poles stemming from $\Gamma$-functions appearing
as prefactors. These basis integrals  are related to the original
topologies or subtopologies  by  shifts to higher space-time
dimensions and allowing for higher powers of the propagators. Dimensional recurrence relations can
be implemented based on Refs.~\cite{Tarasov:1996br,Lee:2009dh} to achieve convenient dimension shifts.
In Ref.~\cite{Lee:2017ftw}, a program is presented which allows to
calculate master integrals based on dimensional recurrence relations and
analyticity in the dimension $D$.

For the analytic calculation of master integrals with the method
of differential equations, discussed in Section \ref{sec:DEanalytic},
it is very convenient to have integrals which are of uniform weight,
also called {\em ``UT''} for {\em uniform transcendentality}, where the
{\em weight} or {\em degree of transcendentality} $DT$ is defined
by~\cite{Heinrich:2007at}
\begin{align}
&DT(r)=0\,\mbox{ for rational } r,\,
DT(\pi^k)=DT(\zeta(k))=DT(\log^k(x))=k,\nn\\
&DT(x\cdot y)=DT(x)+DT(y)\;.
\end{align}
A function that is represented as a series in the dimensional
regularisation parameter $\eps$ is called a uniformly transcendental (UT)
function if  the coefficient of $\eps^k$ has weight $k$.
If one assigns weight $-1$ to $\eps$, a UT function is a function of
uniform weight zero.
The UT property allows for a straightforward solution of the system
of differential equations in terms of Chen iterated
integrals~\cite{ArkaniHamed:2010gh,Henn:2013pwa,Henn:2014qga}.
This is explained in more detail in Section~\ref{sec:DEanalytic}.

A convenient basis in this respect would be both UT and
quasi-finite~\cite{Badger:2016ozq,Schabinger:2018dyi}, but this is highly non-trivial to
achieve in an algorithmic way. Finding a UT basis is facilitated by
recent insights and developments. For example, in
Ref.~\cite{Chicherin:2018old}, a strategy to find a UT basis using an
analysis of leading singularities in the Baikov representation has
been used, in Ref.~\cite{Boehm:2020ijp} an algorithm is presented
which combines a powerful partial fractioning procedure with a UT basis
choice to achieve compact reduction coefficients.

\subsubsection{State of the art}
\label{sec:status_amp}

\subsubsection*{New directions in amplitude reduction}

As the reduction is often the bottleneck in the calculation of
amplitudes with several mass scales, there has been progress towards
avoiding it in several directions.

For example, it is possible to generate and apply the IBP relations numerically and carry
out the amplitude reconstruction directly, without the generation of
reduction tables.
The method is based on rational reconstruction, utilising
finite-field values for the kinematical invariants
\cite{vonManteuffel:2014ixa,Peraro:2016wsq,Klappert:2019emp,Smirnov:2019qkx} and then
performing the same reduction several times for the reconstruction of
the functional dependence on the kinematic invariants.
This method has been very successful  in tackling difficult problems~\cite{vonManteuffel:2017myy,Badger:2018enw, Abreu:2018zmy, Abreu:2019rpt, Badger:2019djh,vonManteuffel:2019wbj,vonManteuffel:2019gpr,Hartanto:2019uvl,Huber:2019fxe,vonManteuffel:2020vjv}.
Meanwhile public tools are available for  functional reconstruction
using finite fields, such as {\sc FiniteFlow}~\cite{Peraro:2019svx} or
{\sc FireFly}~\cite{Klappert:2019emp,Klappert:2020aqs}, 
implementations in {\sc Fire}~\cite{Smirnov:2019qkx} and  {\sc  Kira}~\cite{Klappert:2020nbg} are also
available.

Reduction at the integrand level has been instrumental at one loop
to reduce the algebraic complexity. Much progress to extend these
concepts beyond one loop has been achieved since,
using concepts of algebraic geometry~\cite{Zhang:2012ce,Mastrolia:2012an,Mastrolia:2012wf,Larsen:2015ped,Georgoudis:2017iza}.
A powerful novel IBP reduction method based on computational algebraic
geometry has been presented in Refs.~\cite{Bendle:2019csk,Boehm:2020ijp}.
It employs a Gr\"obner basis technique~\cite{Mastrolia:2012an,Larsen:2015ped} combined with a 
module intersection method~\cite{Boehm:2018fpv,Boehm:2017wjc}.
A Gr\"obner basis is a set of multivariate polynomials enjoying certain properties that allow for example
solving a system of polynomial equations as it occurs in IBP reduction, since
a Gr\"obner basis with respect to lexicographic ordering of multivariate  polynomials
leads to a triangular system which can be solved iteratively.

In Ref.~\cite{Boehm:2020ijp} it was demonstrated that with this combination of methods, an analytic reduction of the two-loop
five-point non-planar double pentagon diagram up to
numerator degree four can be achieved. 
Numerical unitarity has been extremely successful to calculate massless
two-loop  amplitudes~\cite{Ita:2015tya,Abreu:2017xsl,Abreu:2018jgq,Abreu:2019odu}, including the first computation of the
two-loop 4-gluon amplitude based on numerical
unitarity~\cite{Abreu:2017xsl} as well as progress towards two-loop
five-point processes with one massive external leg~\cite{Hartanto:2019uvl,Abreu:2020jxa}.

Another idea, which has been particularly  successful in $N=4$ Super Yang-Mills theory, is to bootstrap the amplitude from its
behaviour in multi-Regge and collinear limits~\cite{Dixon:2011pw,Dixon:2013eka,Dixon:2014iba,Dixon:2015iva,Caron-Huot:2016owq,Dixon:2016nkn,Almelid:2017qju,Chicherin:2017dob,Caron-Huot:2019vjl,Caron-Huot:2019bsq}.

An alternative reduction method, applied to massless two-loop five-parton scattering
amplitudes in Ref.~\cite{Guan:2019bcx} has been presented in Refs.~\cite{Liu:2018dmc,Liu:2017jxz, Wang:2019mnn},
using an expansion in the parameter $\eta$ of the Feynman $i\eta$-prescription.

From a more formal side, a very promising alternative to IBP reduction has been suggested based on the
intersection theory of differential
forms~\cite{Mizera:2017rqa,Mastrolia:2018uzb,Boehm:2018fpv,Frellesvig:2019kgj,Frellesvig:2019uqt,Chen:2020uyk,Frellesvig:2020qot}, 
for which the Baikov representation~\cite{Baikov:1996iu,Lee:2013hzt}  is particularly  suited.
In this approach,
a Feynman integral  written in terms of master integrals $M_i$ as
\begin{equation}
I=\sum_i c_i\, M_i
\end{equation}
is considered as a vector in a vector space spanned by the master
integrals.
Exploiting an inner product in this vector space, the coefficients $c_i$ can be extracted as follows~\cite{Mastrolia:2018uzb,Frellesvig:2019rso}
\begin{align}
  \bra{v}&=\sum_{i,j}\langle{v}\ket{v_j^*}(C^{-1})_{ji}\bra{v_i}\;\mbox{ with }\; C_{ij}=\langle{v_i}\ket{v_j^*}\nn\\
  &\Rightarrow c_i=\sum_{j}\langle{v}\ket{v_j^*}(C^{-1})_{ji}\;.
 \end{align} 
This opens the possibility to extract the coefficients of the master integrals without the need to solve huge linear systems as  they are generated by IBP relations.
The original formulation of this idea was based on the Baikov representation in
combination with maximal
cuts~\cite{Primo:2016ebd,Mastrolia:2018uzb,Frellesvig:2019kgj}. 
In Ref.~\cite{Frellesvig:2020qot}, multivariate intersection numbers
are discussed, and several strategies for integral reduction are presented, applicable to generic parametric representations of Feynman integrals.
 This method has been applied successfully to highly non-trivial topologies already, for example massless two-loop non-planar five-point diagrams or massive two-loop diagrams as they occur in $HH$ or $H+$jet production~\cite{Frellesvig:2019kgj,Frellesvig:2020qot}.


 \subsubsection*{Two-loop five-point amplitudes (and beyond)}

The first calculation of a cross section for a $2\to 3$ process in hadronic collisions, $pp\to 3\gamma$, has become available recently~\cite{Chawdhry:2019bji},
based on master integrals calculated
in~\cite{Gehrmann:2015bfy,Chawdhry:2018awn}, neglecting colour-suppressed non-planar contributions.
These NNLO results greatly improve the agreement with the data~\cite{Aaboud:2017lxm}
compared to the NLO description.
Fully differential NNLO predictions for  $pp\to 3\gamma$ based on $q_T$-subtraction and the
analytic results of Ref.~\cite{Abreu:2020cwb} also have been obtained very
recently~\cite{Kallweit:2020gcp}, finding full agreement with the
results of Ref.~\cite{Chawdhry:2019bji}.

Two-loop leading colour helicity amplitudes are also available
meanwhile, for triphoton~\cite{Chawdhry:2020for} as well as for diphoton+jet~\cite{Agarwal:2021grm,Chawdhry:2021mkw}.

Very recent highlights are given by the leading colour two-loop QCD
corrections to 3-jet production~\cite{Abreu:2021fuk} as well as the two-loop QCD corrections to $Wb\bar{b}$ production at hadron colliders~\cite{Badger:2021nhg}.

For the full-colour QCD case, the calculations of the  full two-loop
five-gluon all-plus helicity amplitude in analytic
form~\cite{Badger:2019djh,Dunbar:2019fcq}  are important milestones,
as well as  a public library for numerical evaluation of massless
pentagon functions~\cite{Chicherin:2020oor}.
An important development is also given by the availability of a public
C++ framework for the computation of multi-loop amplitudes with
numerical unitarity, {\sc Caravel}~\cite{Abreu:2020xvt}.
Preceding work to develop the techniques allowing such results showed an enormously rapid progress.
The master integrals were available first for the planar case~\cite{Gehrmann:2015bfy,Papadopoulos:2015jft,Gehrmann:2018yef}, followed quickly by
the non-planar master integrals~\cite{Abreu:2018rcw,Boehm:2018fpv,Abreu:2018aqd,Chicherin:2018mue,Chicherin:2018old}.

Helicity amplitudes for massless five-point amplitudes in QCD were first calculated semi-numerically using modular arithmetic~\cite{Badger:2013gxa,Dunbar:2016aux,Badger:2017jhb,
Abreu:2017hqn,Badger:2018gip,Abreu:2018jgq}. Based on these methods,  an analytic expression for the
single-minus helicity amplitudes~\cite{Badger:2018enw} was achieved, as well as complete leading colour
five-parton helicity amplitudes within the numerical unitarity framework~\cite{Abreu:2018zmy,Abreu:2019odu}.
Massless  two-loop five-point amplitudes including a 
non-planar sector have been pioneered in 
Super-Yang-Mills theory~\cite{Abreu:2018aqd,Chicherin:2018yne,Bourjaily:2019gqu} as well as
gravity~\cite{Abreu:2019rpt,Chicherin:2019xeg}, before they were applied to 
the all-plus sector of QCD~\cite{Badger:2019djh}.
The development of a reduction method based on convenient combinations of
propagators allowed the calculation of a family
of non-planar five-point two-loop master integrals with one external
off-shell particle~\cite{Papadopoulos:2019iam}.
Other remarkable steps forward are
the first calculation of a two-loop five-point amplitude with a massive leg, given by
the semi-numerical calculation of the planar two-loop helicity
amplitudes for $W$+4 partons~\cite{Hartanto:2019uvl}, 
the computation of a full set of planar five-point two-loop master
integrals with one external mass~\cite{Abreu:2020jxa,Canko:2020ylt}, and the analytic 
calculation of the full set of pentagon functions required for
all planar and non-planar massless five-point two-loop Feynman
integrals in the physical space, together with their implementation
into a public C++ library~\cite{Chicherin:2020oor}.
Recently, a compact analytic expression for the full colour two-loop six-gluon all-plus helicity amplitude has become available~\cite{Dalgleish:2020mof}, 
as well as an analytic form for an $n$-gluon all-plus helicity result for a certain colour partial amplitude~\cite{Dunbar:2020wdh},
which extends earlier work on leading colour all-plus  six- and
seven-point amplitudes~\cite{Dunbar:2016gjb,Dunbar:2017nfy}.
Non-planar six-particle amplitudes at two loops have been studied in Refs.~\cite{Bourjaily:2019iqr,Bourjaily:2019gqu}.

Two-loop QCD corrections to Higgs plus three parton amplitudes with
dimen\-sion-seven operators in Higgs effective field theory have been
calculated in Ref.~\cite{Jin:2019opr}, employing a new strategy of
combining unitarity cut-  and integration-by-parts  methods.

\subsection{Calculation of master integrals: analytic methods}
\label{sec:masters}


In the following we will give a brief overview on various methods to
calculate Feynman integrals, both analytically and numerically. 
In Sections~\ref{sec:numerical} and~\ref{sec:secdec} we will focus on the numerical evaluation of individual integrals,
which are usually the endpoints of some reduction procedure to master integrals,
as described in Section~\ref{subsec:reduction}.
Such integrals can be IR or UV divergent and therefore are defined as
integrals over $D$-dimensional loop momenta.
There are also approaches
 which try to avoid working in $D$ space-time dimensions, aiming to
 calculate the amplitudes fully numerically in momentum space.
 Such methods will be discussed in Section~\ref{subsec:4dim}.

As physicists, we would like to compare predictions with measurements,
so we are ultimately interested in numerical values for the integrals
which enter precision calculations of measurable quantities. So why do
we aim at an analytic representation of the integrals?
The answers are obvious: we would like to have maximal control over the
pole cancellation and the analytic regions, we need fast evaluation times, and
we would like to understand the mathematical structure.
However, the evaluation times with modern numerical methods (and
computing resources) are increasingly competitive with analytic
approaches,
and the numerical approaches have the advantage that the extension to
more loops or kinematic scales is more straightforward.
Therefore it seems fair to say that numerical methods are not any
longer the ``poor man's solution'' to problems where analytic results
are not yet available, they may even evolve to be the method of choice
for precision calculations in view of future lepton colliders where
electroweak corrections at two- and three-loop order, involving typically several mass scales, are
mandatory to match the experimental precision.
Some {\em Pro's and Con's} of analytic versus numerical methods are
listed in Table~\ref{tab:anaversusnum}.
Concerning the pole cancellation it should be mentioned that the pole
coefficients in general are much easier to evaluate than the finite
part, such that the numerical uncertainty on the pole coefficients is
usually not a problem to verify pole cancellation.
The automation of analytic calculations is particularly difficult if adding another
mass scale enlarges the function class the result is spanning, for
example exceeding the class of multiple polylogarithms.
Numerical methods have to struggle with more thresholds as the number
of mass scales increases, but they are less sensitive to new
mathematical structures and therefore might be more suitable
for the automated calculation of multi-loop amplitudes.
More details about various methods will be given below, see also
Refs.~\cite{Weinzierl:2006qs,Duhr:2014woa,Freitas:2016sty,Kotikov:2018wxe,Blumlein:2019svg} for reviews.

\begin{table}
  \hspace*{-1cm}
  \begin{tabular}{|l|c|c|}
    \hline
  &  analytic & numerical\\
    \hline
    pole cancellation& exact & with numerical uncertainty\\
    control of integrable singularities & analytic continuation
              & less straightforward\\
    fast and stable evaluation & yes (mostly)& depends\\
    extension to more scales/loops & difficult & promising\\
    automation & difficult & less difficult\\
    \hline    
  \end{tabular} 
  \caption{Strong and weak points of analytic versus numerical evaluations
    of loop integrals.}
  \label{tab:anaversusnum}
\end{table}

As many of the methods operate on the Schwinger- or Feynman parameter
representation of loop integrals, we will give a brief overview of these
representations in Section \ref{sec:FU}.


\subsubsection{Differential equations}

\label{sec:DEanalytic}

The method of differential equations (DE) to evaluate Feynman
integrals has first been suggested and developed in Refs.~\cite{Kotikov:1990kg,Remiddi:1997ny,Gehrmann:1999as}. A breakthrough of this method has been initiated by Ref.~\cite{Henn:2013pwa}, where an optimal, $\eps$-factorised form of these equations has been suggested, nowadays known as  {\em canonical form}.

The main idea of the DE method is to take derivatives of a given integral with respect to kinematic invariants and/or masses, which relates them to other integrals of a given family. This leads to a system of differential equations for the master integrals which can be solved given appropriate boundary conditions.
Below we briefly outline the algorithm, for more details we refer to
Refs.~\cite{Argeri:2007up,Henn:2014qga}.

\vspace*{3mm}

Consider a Feynman integral or an integral family as in \eqn{eq:dim_reg_int}, characterised by a set of propagators and  irreducible scalar products $P_i$ in $D$ space-time dimensions.
The dependence on the kinematics is given by invariants formed by
external momenta such as $p_i^2$, $(p_i+p_j)^2$ and masses, which we denote generically by $x_i$.
After successful IBP reduction, all integrals of the given family can be expressed in terms of a basis of $N$ independent  master integrals $I_k(D,x_i)$ with $k=1,...,N$.
Differential operators with respect to any of the external invariants $x_i$ can be constructed as
linear combinations of differentiations with respect to the external momenta $p_i^\mu$~\cite{Remiddi:1997ny}.
Therefore, acting with these differential operators on the  integrands of~\eqn{eq:dim_reg_int} leads to
linear combinations of integrals belonging to the same integral family and to
its sub-topologies. The latter can again be reduced to MIs, 
generating a system of
$N$ linear first order differential equations with rational coefficients in any of the invariants
$x_i$. Suppressing the dependence on the sub-topologies, which can be considered as a  
known inhomogeneous term in a bottom-up approach, the homogeneous part of the 
system can  be written as

\begin{align}
\frac{\partial}{\partial x_i} \left( \begin{array}{c} I_1(D,x_i) \\ \vdots \\ I_N(D,x_i) \end{array}\right)
&= \left( \begin{array}{ccc} a_{11}(D,x_i) & \ldots & a_{1N}(D,x_i)\\
\vdots & \ldots & \vdots \\
 a_{N1}(D,x_i) & \ldots & a_{NN}(D,x_i)
 \end{array}\right) 
\left( \begin{array}{c} I_1(D,x_i) \\ \vdots \\ I_N(D,x_i) \end{array} \right)\,,
\label{eq:sysdeq}
\end{align}
where the coefficients $a_{ij}$ are  rational functions of the dimension $D$ and of the 
kinematic invariants $x_i$. Introducing a vector of master integrals, $\vec{I}(D,x_i)$, and the matrix of
the coefficients, $A(D,x_i)$, we can rewrite the system in compact form,
\begin{align}
\frac{\partial}{\partial x_i} \, \vec{I}(D,x_i) = A(D,x_i)\, \vec{I}(D,x_i)\;. 
\label{eq:sysdeqmatrix}
\end{align}
This system of $N$ first order differential equations is in  general  coupled and 
therefore it can also be understood as an $N$-th order differential equation for any of the MIs.
Solving the system is simplified if the MIs are expanded as a Laurent series in $\eps=(4-D)/2$, 
\begin{align}
I_{k}(D,x_i) = \sum_{r=-p}^\infty \, I_{k}^{(r)}(4;x_i) \, (D-4)^{r}\,. \label{eq:ser4}
\end{align}
The leading pole  $1/\eps^p$ of an integral is usually easiest to evaluate, and 
this leads to a chained
system of $N$ differential equations where, at any order $r$, the previous orders can only appear
as inhomogeneous terms.

The system becomes particularly simple if we can achieve a basis such that the $\eps$-dependence factorises, i.e.
\begin{align}
\frac{\partial}{\partial x_i} \, \vec{I}(D,x_i) = \eps\,A(x_i)\, \vec{I}(D,x_i)\;. 
\label{eq:DEcanonical}
\end{align}
Such a basis is called a {\em canonical basis}~\cite{Henn:2013pwa}.
The most convenient form of Eq.~(\ref{eq:DEcanonical}) is the following, where we now drop the labels in $x_i$ and denote the $D$-dependence by $\eps$:
\begin{align}
  \label{eq:canonical2}
\rd\, \vec{I}(x,\eps) = \eps \, \left( \rd \, \tilde{A} \right) \,\vec{I}(x,\eps) \; ,\;
\tilde{A} = \sum_{k} A_{k} \ln \alpha_{k}(x) \,.
\end{align}
In this form the system can be integrated iteratively order by order in $\eps$ with suitably chosen boundary conditions.
In the case where this representation can be reached using only rational transformations, the $\alpha_k$ are of the form
$\alpha_{k} = x-x_{k} $\,,
where the $x_k$ are the locations of the singularities in the kinematic invariants.
In the general case they can be more complicated functions of $x$, for example involving square-roots.
The set of $\alpha_{k}$ is also called the {\em alphabet}. It characterises the function class the solution can belong to.

The solution to Eq.~(\ref{eq:canonical2})  with boundary value $\vec{I}_0(\eps)$ can be written as an iterated integral,
\be
\vec{I}(x,\eps) =\left(1  + \eps\int^x dt_1\, \tilde{A}(t_1)+\eps^2\int^x dt_1\int^{t_1}dt_2\, \tilde{A}(t_1)\tilde{A}(t_2)+\ldots\right)\,\vec{I}_0(\eps)\;.
\ee
More generally it can be written as
\begin{align}\label{eq:int_solution}
\vec{I}(\vec{x},\eps) = {\mathbb{P}} \exp \left[ \eps \int_{ \gamma}  \rd \, \tilde{A} \right] \vec{I}_{0}(\eps) \,,
\end{align}
where $ {\mathbb{P}}$ denotes path ordering along the integration contour $\gamma$,
and $\vec{I}_{0}(\eps)$ is a boundary value.
If the alphabet can be written in terms of rational functions,
one can write the answer in terms of multiple polylogarithms (MPLs),
also called  Goncharov
polylogarithms~\cite{Kummer,Nielsen,Goncharov:1998}, see also
Refs.~\cite{Goncharov:2010jf,Duhr:2011zq,Duhr:2012fh} for the relation
to the so-called {\em symbol} calculus.
The Goncharov polylogarithms can be defined iteratively as follows,
\begin{align}
&G(a_1 ;z) = \int_0^z \frac{dt}{t-a_{1}}  \;, \;G(;z) =1\;,\; (a_1\not=0)\;,\nn\\
&G(a_1,\ldots a_n ; z) = \int_0^z \frac{dt}{t-a_{1}} G(a_{2}, \ldots ,a_{n}; t) \,.
\end{align}
The number of elements $n$ in the vector $(a_1,\ldots a_n)$ is called the {\em weight} of the Goncharov polylogarithm.
Their properties are reviewed e.g. in Ref.~\cite{Duhr:2014woa}, the relations to logarithms and dilogarithms are given by
\begin{align}
  &G(\vec{0}_{n};z) = 1/n! \log^n(z)\;,\;G(\vec{a}_{n};z) = 1/n! \log^n(1-z/a)\;,\nn\\
  &G(\vec{0}_{n-1},1;z) = -Li_n(z)\;.
\end{align}
The subclass of functions where $a_i\in \{-1,0,1\}$ is called Harmonic Poly\-logarithms (HPLs), however the definition includes an extra sign: $H(\vec{a};z) =(-1)^p\,G(\vec{a};z) $, where $p$ is the number of elements equal to +1 in $\vec{a}$.
Several codes have been published which can evaluate MPLs
numerically~\cite{Gehrmann:2001pz,Gehrmann:2001jv,Vollinga:2004sn,Buehler:2011ev,Bogner:2015nda,Frellesvig:2016ske,Ablinger:2018sat,Duhr:2019tlz}.

A generalisation of MPLs which includes the class of functions
that appear in the massive sunrise diagram are the so-called elliptic multiple polylogarithms (eMPLs), which can be obtained
by considering iterated integrals of rational functions on a torus~\cite{Remiddi:2017har}.
While eMPLs are a rather large class of functions, 
for the particular case of the two-loop sunrise diagram a special subclass of eMPLs has been shown to be sufficient, the so-called
iterated Eisenstein integrals~\cite{Adams:2017ejb,Broedel:2018rwm,Duhr:2019rrs}, see also Section \ref{sec:elliptic}.

\subsubsection*{Example}

 A very simple example of an iterated integral is an integral where the differential equations have the following form (a form where $A(x)$ can be expressed in terms of derivatives of logarithms is also called {\em dlog} form):
 \be
 \rd\,I(x,\eps)=\eps\,A(x)\,I(x,\eps)\;,\; A(x)=\rd\ln{(x-1)}=\frac{\rd x}{x-1}\;.
 \ee
 Such a form is convenient because it can be solved  iteratively.
 With the boundary condition $I(0,\eps)=1$ we get, expanding $I(x,\eps)=\sum_{r=0}^\infty I^{(r)}(x)\,\eps^r$ and integrating iteratively each order in $\eps$:
 \be
  I(x,\eps)=1+\eps\,G(1;x)+\eps^2\,G(1,1;x)+\ldots\;.
 \ee

 Finding a transformation $T$ that brings the system to canonical form
 is highly non-trivial and several procedures and tools to achieve
 this can be found in the
 literature~\cite{Argeri:2014qva,Meyer:2016slj,Georgoudis:2016wff,Gituliar:2017vzm,Prausa:2017ltv,Henn:2020lye,Dlapa:2021qsl}. 
In Ref.~\cite{Dlapa:2020cwj} a program is presented which can construct a basis of integrals with the property of uniform
transcendentality (UT)  if one UT integral is known.
In Ref.~\cite{Hidding:2020ytt} the program  {\sc DiffExp} is presented,
which  iteratively solves  integrals from their differential equations
in terms of truncated one-dimensional series expansions along line
segments in phase-space. It can automatically detect the integration
sequence from the differential equations and contains an optimised
integration strategy for coupled integrals. 
 
It is commonly believed that for Feynman integrals which can
be expressed in terms of multiple polylogarithms  a canonical form of the DEs exists.
However the inverse is not true,   explicit examples have been found
where a canonical form exists, however the result cannot be expressed entirely 
 in terms of multiple polylogarithms~\cite{Brown:2020rda}.

 For Feynman integrals containing elliptic functions, a canonical form has been found for a few special cases~\cite{Adams:2018yfj,Bogner:2019lfa}.
Whether it would be possible to find a canonical form for all types of Feynman integrals is an open question.

\subsubsection{Elliptic integrals}
\label{sec:elliptic}

In addition to the ``two-loop multi-leg frontier'' (where ``multi-leg'' in this context means five or six external legs),
there is also the ``multi-scale loop integrals frontier'', where the
calculation of the integrals is highly non-trivial due to massive
propagators. 
The calculation of  two-loop amplitudes for
$2\to2$ scattering  with  massless
propagators and up to two off-shell legs can be considered as a solved
problem~\cite{Caola:2014iua,vonManteuffel:2015msa,Caola:2015ila,Gehrmann:2015ora}, as it can be tackled with the automated IBP reduction tools
mentioned above, in combination with the knowledge about master
integrals that lie in the function class of multiple polylogarithms
(MPLs)~\cite{Kummer,Nielsen,Goncharov:1998}, see Section \ref{sec:DEanalytic}.
The situation can change drastically when massive propagators are involved.
In such cases the functions required for an analytic representation of
the integrals may involve integrals of elliptic type,
and these functions are much less understood than generalised
polylogarithms, at least in the context of perturbative quantum field
theory.
Differential equations for elliptic integrals are harder to solve, as
a canonical form usually does not exist or cannot be found, and finding
appropriate boundary values is also highly non-trivial.
Furthermore, algorithms to evaluate elliptic functions numerically in
a fast and stable way for physical kinematics are just starting to be developed~\cite{Walden:2020odh}.

\begin{figure}[htb]
  \centering
\includegraphics[width=0.4\textwidth]{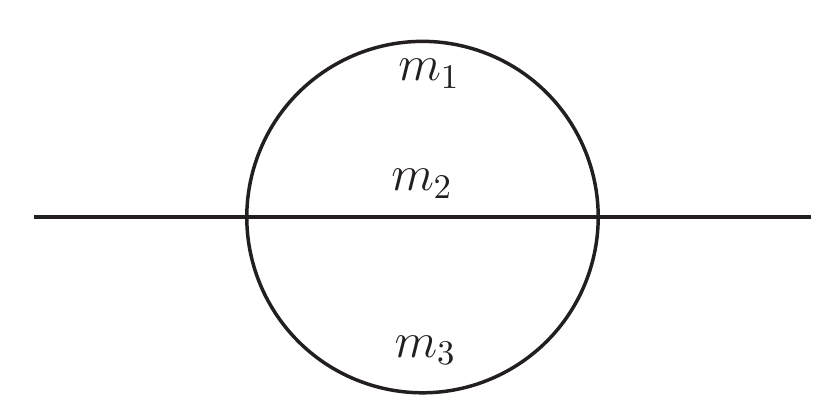}
\caption{Two-loop sunrise graph with three different masses.\label{fig:sunrise}}
\end{figure}

Elliptic integrals are characterised by the appearance of a 
(usually quartic or cubic) polynomial under a square-root in the integrand, i.e. they
are integrals of the form
\be
{\cal E}(x)=\int_0^x dt\,R(t, \sqrt{P(t)})\;,
\ee
where $R$ is a rational function of its arguments and $P(t)$ is a
polynomial of degree 3 or 4.
The square-root in the integrand is connected to an elliptic curve,
i.e. the points $(t,y)$ satisfying $y^2=P(t)$.
For example, elliptic integrals of first and second kind, $F(x)$ and
$E(x)$, in the so-called Legendre form are given by
\begin{align}
F(x;k)&=\int_0^x dt\,\frac{1}{\sqrt{(1-t^2)(1-k^2t^2)}}\;,\; E(x;k)=\int_0^x dt\,\frac{\sqrt{1-k^2t^2}}{\sqrt{1-t^2}}\;.
\end{align}  
In fact, already in 1962 new functions of elliptic type were
discovered in the calculation of the two-loop corrections to the
electron propagator in QED with massive electrons~\cite{Sabry1962}.
They can be described as iterated integrals on a complex surface of
genus one, i.e. a torus~\cite{Manin,LevinRacinet,Zagier,BrownLevin}.
The importance of these functions for higher order perturbative
calculations involving heavy particles such as top quarks or massive
vector bosons has been discovered more recently~\cite{Broadhurst:1987ei,Broadhurst:1993mw,Bauberger:1994by,Bauberger:1994hx,Caffo:1998du,Laporta:2004rb,Kniehl:2005bc,Aglietti:2007as,Czakon:2008ii,MullerStach:2011ru,CaronHuot:2012ab,Adams:2013nia,Bloch:2013tra,Remiddi:2013joa,Adams:2014vja,Adams:2015gva,
Adams:2015ydq,Bloch:2016izu,Remiddi:2016gno,vonManteuffel:2017hms,Remiddi:2017har,Bogner:2017vim,Ablinger:2017bjx,Groote:2018rpb,
Proceedings:2019hhn,Broedel:2019kmn,Bogner:2019lfa}.
The probably simplest Feynman integral where elliptic integrals occur
is the two-loop sunrise graph with three unequal masses, shown in
Fig.~\ref{fig:sunrise}.
While the discontinuity of this integral, where all three propagators are put on-shell, can be expressed in terms of complete elliptic integrals of the first kind~\cite{Laporta:2004rb}, the full integral can only be expressed in terms of iterated integrals over elliptic kernels~\cite{Bloch:2013tra} also called {\em elliptic multiple polylogarithms (eMPLs)}~\cite{Manin,LevinRacinet,Zagier,BrownLevin}.

Elliptic MPLs also appear in the calculation of one-loop  scattering amplitudes
in string theory~\cite{Broedel:2014vla,Broedel:2015hia,Broedel:2017jdo,Broedel:2018izr}.
Furthermore, an alternative formulation of elliptic MPLs, defined  by a polynomial equation
instead of on a torus, has been proposed~\cite{Broedel:2017kkb,Broedel:2017siw,Broedel:2018iwv}, which has made it possible to  compute
 multi-loop Feynman integrals that were considered out of reach
 previously~\cite{Broedel:2018qkq,Broedel:2019hyg}.

Among  prominent phenomenological applications of elliptic
integrals are the analytic results for the complete set of 2-loop master
integrals entering Higgs+jet production at NLO with full top quark
mass
dependence~\cite{Bonciani:2016qxi,Becchetti:2018xsk,Bonciani:2019jyb,Frellesvig:2019byn},
the  three-loop
corrections to the $\rho$ parameter with non-vanishing $b$-quark
mass~\cite{Grigo:2012ji,Blumlein:2018aeq,Abreu:2019fgk}, where Ref.~\cite{Abreu:2019fgk} provides a
closed analytic form, or the two-mass
contributions to  3-loop massive operator matrix elements
entering Wilson coefficients in deeply inelastic
scattering~\cite{Ablinger:2017err,Ablinger:2018yae,Blumlein:2019oas,Ablinger:2019gpu,Ablinger:2019etw,Ablinger:2020snj},
see also Ref.~\cite{Blumlein:2018cms} for an overview of methodology.


Reviews of recent work related to elliptic functions and modular
forms can be found in Refs.~\cite{Duhr:2019wtr,Proceedings:2019hhn}.


\subsubsection{Multi-loop results}
\label{sec:multi-loop}

Going beyond two loops, important progress has been made at three
loops including several mass scales, which often involve elliptic
integrals, see above, other recent multi-scale three-loop examples are 
the on-shell quark
 mass and wave function renormalisation constants  allowing  for  a  second  non-zero  quark  mass~\cite{Fael:2020bgs},
the N$^3$LO corrections to Higgs production via bottom quark fusion~\cite{Duhr:2020kzd},
the N$^3$LO corrections to charged-current Drell-Yan
production~\cite{Duhr:2020sdp} or the three-loop integrals for
massless four-particle scattering~\cite{Henn:2020lye}.

At four (and partly five) loops, impressive results have been achieved in the
calculation of the  four-loop cusp anomalous dimension, contributions to the
N$^3$LO splitting functions or the calculation of matching coefficients in heavy quark
effective field theory, see e.g. Refs.~\cite{vonManteuffel:2015gxa,
Davies:2016jie,
Luthe:2016xec,
vonManteuffel:2016xki,
Lee:2016ixa,
Ruijl:2017eht,
Lee:2017mip,
Moch:2017uml,
Moch:2018wjh,
Grozin:2018vdn,
Herzog:2018kwj,
Henn:2019rmi,
Bruser:2019auj,
Becher:2019avh,
vonManteuffel:2019wbj,
vonManteuffel:2019gpr,
Lee:2019zop,
Catani:2019rvy,
Henn:2019swt,Huber:2019fxe,Dlapa:2020cwj,Das:2019btv,
vonManteuffel:2020vjv,
Das:2020adl,Agarwal:2020nyc,
Grozin:2020jvt,Bruser:2020bsh,Fael:2020njb,Fael:2020tow,Grozin:2021tfi}.

Other multi-loop highlights are the 
five-loop QCD
beta-function~\cite{Baikov:2016tgj,Herzog:2017ohr,Luthe:2017ttg,Chetyrkin:2017bjc}
or scalar theories~\cite{Kompaniets:2021hwg},
the R-ratio for $e^+e^-$ to hadrons at
N$^{4}$LO~\cite{Baikov:2008jh,Baikov:2010je,Baikov:2012er,Herzog:2017dtz}
or the five-loop contributions to the anomalous magnetic moment of the
electron~\cite{Aoyama:2012wj,Aoyama:2014sxa,Aoyama:2017uqe,Volkov:2019phy}.

Results at six and more loops also have been achieved for single-scale
diagrams, see
e.g. Refs.~\cite{Broadhurst:1999ys,Panzer:2013cha,Batkovich:2016jus,Kompaniets:2017yct,Panzer:2019yxl,Borinsky:2020rqs},
where Ref.~\cite{Borinsky:2020rqs}, going up to 17 loops for a certain
class of dia\-grams, employs a new method using elements
from tropical geometry~\cite{tropical,Drummond:2019cxm} and geometric sector
decomposition~\cite{Ueda:2009xx,Kaneko:2009qx,Kaneko:2010kj,Schlenk:2016cwf}.
Other examples are the six-gluon MHV and NMHV amplitudes in $N=4$ Super-Yang-Mills
theory up to seven and six loops, respectively~\cite{Caron-Huot:2019vjl}.

Finally, the database Loopedia~\cite{Bogner:2017xhp,Loopedia_url} should be
mentioned, which is a search engine for  integrals calculated in the
literature, where the user can specify the topology to search for
existing results for the given type of Feynman diagram.



\subsubsection{Parametric representations}
\label{sec:FU}


The advantage of parametric representations of Feynman integrals is
that the integration over loop momenta in $D$ dimensions is carried
out analytically in a universal way, such that the remaining
integrations only involve scalar objects in Minkowski space.
The Baikov representation does not introduce additional parameters,
however it involves a transformation of the integration
over $D$-dimensional loop momenta  into an integration over kinematic invariants.

Our starting point is an $L$-loop integral as given in \eqn{eq:intfamily}, 
\begin{eqnarray}\label{eq0}
G(\nu_1\ldots\nu_N)  &=&
\int \prod\limits_{l=1}^{L} \frac{\rd^Dk_l}{i\pi^{\frac{D}{2}}}\;
\prod\limits_{j=1}^{N} 
\frac{1}{P_{j}^{\nu_j}(\{k\},\{p\},m_j^2)}\;, 
\end{eqnarray}
where the propagator powers $\nu_j$ need not all be positive. 
Multi-loop integrals can be transformed into integrals over Feynman or
Schwinger parameters for an arbitrary number of loops.
In order to keep the index structure simple, we focus on integrals
with no Lorentz tensors containing loop momenta in the numerator here.
Their inclusion is straightforward.

\subsubsection*{Schwinger parametrisation}

Consider the integral of \eqn{eq0}, where for simplicity of
notation we will assume that all propagator powers are positive.
The case of additional negative indices will be treated
later.

We can introduce a so-called Schwinger parameter $\alpha$ associated to each
propagator $D^\nu$ using the relation
\begin{align}
\frac{1}{D^\nu}=\frac{(-1)^\nu}{\Gamma(\nu)}\int_0^\infty\rd\alpha\,\alpha^{\nu-1}\,\exp{(\alpha
  D)}\;,\;\nu > 0\;.
\end{align}
Then the integral can be written as
\begin{align}
G(\nu_1\ldots\nu_N)  &=\prod\limits_{l=1}^{L} \int \frac{\rd^Dk_l}{i\pi^{\frac{D}{2}}}\; \prod\limits_{j=1}^{N}\, \frac{(-1)^{\nu_j}}{\Gamma(\nu_j)}\int_0^\infty  \rd\alpha_j\,\alpha_j^{\nu_j-1}\,\exp{\left(\sum_{k=1}^N\alpha_kD_k\right)}\;.
\end{align}
The argument of the exponential function is a quadratic form in the
$(L\times D)$-dimensional space of loop momenta, i.e.
$$\sum_{k=1}^N\alpha_kD_k=\sum\limits_{j,l=1}^{L} k_j\cdot k_l \, M_{jl} - 
2\sum\limits_{j=1}^{L} k_j\cdot Q_j +J +i\delta\;,$$
where $k_j\cdot k_l$ denotes the scalar product of two $D$-dimensional Lorentz-vectors.
After the shift $k_j=l_j+M_{jl}^{-1}Q_l$ we can use Gaussian integration in
$L\times D$ dimensions to obtain
\begin{align}
\prod\limits_{l=1}^{L} \int \frac{\rd^Dk_l}{i\pi^{\frac{D}{2}}}\,
  \exp{\left(\sum_{k=1}^N\alpha_kD_k\right)}&=\left(\det{M}\right)^{-\frac{D}{2}}\exp{\left(-\sum\limits_{j,l=1}^{L} Q_j\cdot Q_l \, M^{-1}_{jl} +J +i\delta\right)}\;.
\end{align}
Using
\begin{align}
{\cal U}&=\det(M)\quad , \quad N_{\nu}=\sum_{j=1}^N\nu_j\quad,\quad
{\cal F}=\det(M)\,\left[ \sum_{i,j=1}^L Q_i M_{ij}^{-1}Q_j-J -i\delta \right]\;,\label{eq:F}
\end{align}
we arrive at
\begin{align}
G(\nu_1\ldots\nu_N)  &= (-1)^{N_{\nu}}\prod\limits_{j=1}^{N}\,\left(\frac{1}{\Gamma(\nu_j)}
\int\limits_{0}^{\infty} \rd\alpha_j\,\alpha_j^{\nu_j-1} \right)\,{\cal U}^{-\frac{D}{2}}\exp{\left(-\frac{{\cal F}}{{\cal U}}\right)}\;.
\end{align}
The graph polynomials $\mathcal{U}(\vec{x})$ and $\mathcal{F}(\vec{x},s_{ij},m_i^2)$ are  also called first and second Symanzik polynomial.
${\cal U}$ and ${\cal F}$ are homogeneous of degree $L$ and $L+1$, respectively.

\subsubsection*{Feynman parametrisation}
Feynman- and Schwinger parametrisations are equivalent to each other. 
Starting again with an integral of the form  Eq.~(\ref{eq0}) and using
\begin{align}\label{eq:Feynmanpara}
  \prod\limits_{j=1}^{N} \frac{1}{P_{j}^{\nu_j}}&
  =\frac{\Gamma(N_\nu)}{\prod\limits_{j=1}^N\Gamma(\nu_j)}\int_0^\infty \left(\prod\limits_{j=1}^N\,dx_j\,x_j^{\nu_j-1}\right)
    \,\, \delta(1-\sum_{i=1}^N x_i)\,\left(\sum_{i=1}^N x_iP_i\right)^{-N_\nu}\;,
\end{align}
the integral has the form
\begin{align}
G(\vec{\nu}) &=  \frac{\Gamma(N_\nu)}{\prod\limits_{j=1}^{N}\Gamma(\nu_j)}\int_0^\infty \prod\limits_{j=1}^{N}\,dx_j\,x_j^{\nu_j-1}
 \,\, \delta(1-\sum_{i=1}^N x_i)\nn\\
&\times \int_{-\infty}^\infty d\kappa_1\dots d\kappa_L 
\left[ 
       \sum\limits_{j,l=1}^{L} k_j\cdot k_l \, M_{jl} - 
      2\sum\limits_{j=1}^{L} k_j\cdot Q_j +J +i\delta\right]^{-N_\nu}\;,\nn
\end{align}
where $\kappa_l=d^Dk_l/(i\pi^{\frac{D}{2}})$.
After momentum integration, this leads to
\begin{align}
  G(\vec{\nu})&= \frac{(-1)^{N_{\nu}}}{\prod_{j=1}^{N}\Gamma(\nu_j)}\Gamma(N_{\nu}-LD/2)
 \int\limits_{0}^{\infty} 
\,\prod\limits_{j=1}^{N}dx_j\,\,x_j^{\nu_j-1}\,\delta(1-\sum_{l=1}^N x_l)\,
\frac{\mathcal{U}^{N_{\nu}-(L+1) D/2}}{{\mathcal F}^{N_\nu-L D/2}}\;.\label{eq:feynint}
\end{align}

Negative indices, i.e. propagators in the numerator, can be treated using the identity
\begin{align}
\frac{P_n}{\left(\sum_{i=1}^N x_jP_j\right)^{\alpha+1}}=-\frac{1}{\alpha}\frac{\partial}{\partial x_n}\,\frac{1}{\left(\sum_{i=1}^N x_jP_j\right)^{\alpha}}\;,
\end{align}
where $n\leq N$. Denoting the positive propagator powers by $\tilde{\nu}_j$ and the negative ones by $\hat{\nu}_j$, and defining the set $\tilde{J}$ of labels belonging to positive powers and $\hat{J}$ as its complement with respect to $N$, 
we can write down a generalisation of the Feynman parametrisation formula, Eq.~(\ref{eq:Feynmanpara}), as
\begin{align}\label{eq:Feynmanpara_general}
  \frac{\prod\limits_{j\in \hat{J}}P_{j}^{\hat{\nu}_j}}{\prod\limits_{j\in\tilde{J}} P_{j}^{\tilde{\nu}_j}}
  &=\frac{\Gamma(N_\nu)}{\prod\limits_{j\in \tilde{J}}\Gamma(\tilde{\nu}_j)}\int_0^\infty \left(\prod\limits_{j\in\tilde{J}}\,dx_j\,x_j^{\tilde{\nu}_j-1}\right)
    \, \delta(1-\sum_{i=1}^N x_i)
  \, \left[\prod\limits_{n\in\hat{J}}  \left(-\frac{\partial}{\partial x_n}\right)^{\hat{\nu}_n}\right]\,\left(\sum_{i=1}^N x_iP_i\right)^{-N_\nu}\,.
\end{align}
%

\subsubsection*{Lee-Pomeransky representation}

Apart from the Schwinger representation, there is a closely related
representation derived by Lee and Pomeransky~\cite{Lee:2013hzt}:

\begin{equation}
 G(\nu_1\ldots\nu_N)  =\frac{(-1)^{N_\nu}\Gamma\left(D/2\right)}{\Gamma\left(\left(L+1\right)D/2-N_\nu\right)\prod_{j}\Gamma\left(\nu_{j}\right)}\int\limits _{0}^{\infty}\prod_{j=1}^Ndz_{j}\,z_{j}^{\nu_{j}-1}({\cal U}+{\cal F})^{-D/2}\,.\label{eq:LP}
\end{equation}
It can be proven to be equivalent to the form given in eq.~(\ref{eq:feynint}),
by inserting $1=\int_0^\infty\rd
\eta\,\delta(\eta-\sum_{j=1}^{N}z_j)$, substituting
$z_j=\eta\,x_j$  for $ j=1,\ldots, N$ and using
$\int_0^\infty d\eta\,\eta^{-1+a}\left(\eta^{-1}\,{\cal U}+{\cal
    F}\right)^b=\frac{\Gamma(-a)\Gamma(a-b)}{\Gamma(-b)}\;{\cal U}^a{\cal F}^{b-a}$.
This representation can be advantageous for example as a starting point for the method of expansion by regions~\cite{Smirnov:1991jn,Beneke:1997zp}, as it defines the integral in terms of just one polynomial raised to some power instead of two.

An interesting approach to the calculation of Feynman integrals based on the Lee-Pomeransky representation is presented in Ref.~\cite{delaCruz:2019skx}, showing that Feynman integrals, at least in Euclidean space, can be understood as a solution of a certain Gel’fand- Kapranov-Zelevinsky (GKZ) system. This leads to a canonical series representation which allows to evaluate integrals with arbitrary propagator powers as linear combinations of A-hypergeometric functions. Work based on similar ideas can also be found in Refs.~\cite{Klausen:2019hrg,Feng:2019bdx,Reichelt:2020vka,Bonisch:2020qmm,Kalmykov:2020cqz}.

\subsubsection*{Baikov representation}

A quite different representation for multi-loop
integrals is the one introduced by Baikov~\cite{Baikov:1996iu}.
In this representation the integration over the loop momenta is transformed into an integration over Lorentz invariants formed by $L$ loop momenta $k_i$  and external momenta $p_{1},\ldots,p_{E}$.
Following the notation of Ref.~\cite{Lee:2010wea} we define
\begin{equation}
s_{ij}=s_{ji}=k_{i}\cdot q_{j}\,;\quad i=1,\ldots,L;\quad j=1,\ldots,K,
\end{equation}
where $q_{1},\ldots,q_{L}=k_{1},\ldots,k_{L}$, $q_{L+1},\ldots,q_{L+E}=p_{1},\ldots,p_{E}$, and
$K=L+E$.
Then we replace
the integration over the loop momenta by the integration over 
\begin{equation}
s_{ij},\quad1\leqslant i\leqslant L,\quad i\leqslant j\leqslant K\,.\label{eq:svar}
\end{equation}
Assuming that the denominators $D_{1},\ldots,D_{M}$
are linearly independent, we can choose $N-M$
irreducible numerators $D_{M+1},\ldots D_{N}$. The resulting formula
reads
\begin{align}
G(\nu_1\ldots\nu_N) & =\frac{\pi^{\left(L-N\right)/2}S_{E}^{(E+1-D)/2}}{\left[\Gamma((D-E-L+1)/2)\right]_{L}}\nn\\
 & \times\int\left(\prod_{i=1}^{L}\prod_{j=i}^{L+E}ds_{ij}\right)S^{(D-E-L-1)/2}\prod_{j=1}^{N}D_{j}^{-\nu_{j}},
\end{align}
where $$\left[\Gamma(x)\right]_{L}\equiv \Gamma(x)\,\Gamma(x-1)\ldots \Gamma(x-L)$$ and  $\nu_{(j>M)}<0$.
The quantities $S$ and $S_{E}$ come from the Jacobian of the variable transformation and have the form 
\begin{align*}
S & =\det\left\{ \left.s_{ij}\right|_{i,j=1\ldots L+E}\right\} ,\quad S_{E}=\det\left\{ \left.s_{ij}\right|_{i,j=L+1\ldots L+E}\right\} \,.
\end{align*}
The functions $D_{j}$ are linear functions of the variables $s_{ij}$, so that\\ $\prod_{i=1}^{L}\prod_{j=i}^{L+E}ds_{ij}\propto dD_{1}\ldots dD_{N}$.
Thus, we have 
\begin{align*}
G\left(\vec{\nu}\right) & \propto\int\left(\prod_{j=1}^{N}D_{j}^{-\nu_{j}}dD_{j}\right)\,P^{(D-E-L-1)/2},
\end{align*}
where $P\left(D_{1},\ldots D_{N}\right)$ is obtained from $S$ by
expressing $s_{ij}$ via $D_{1},\ldots D_{N}$ and is often called {\it Baikov polynomial}.

An advantage of the Baikov representation is that it is conveniently
used for recurrence relations in the dimension $D$, because apart from
trivial functions only the Baikov polynomial $P$ depends on $D$.
Furthermore, it makes the cuts of an integral (putting propagators on-shell) more manifest.
%

\subsubsection*{Construction of the functions ${\cal F}$ and ${\cal U}$ from topological rules}\label{topologicalFU}

\begin{figure}[htb]
\begin{center}
\includegraphics[width=0.75\textwidth]{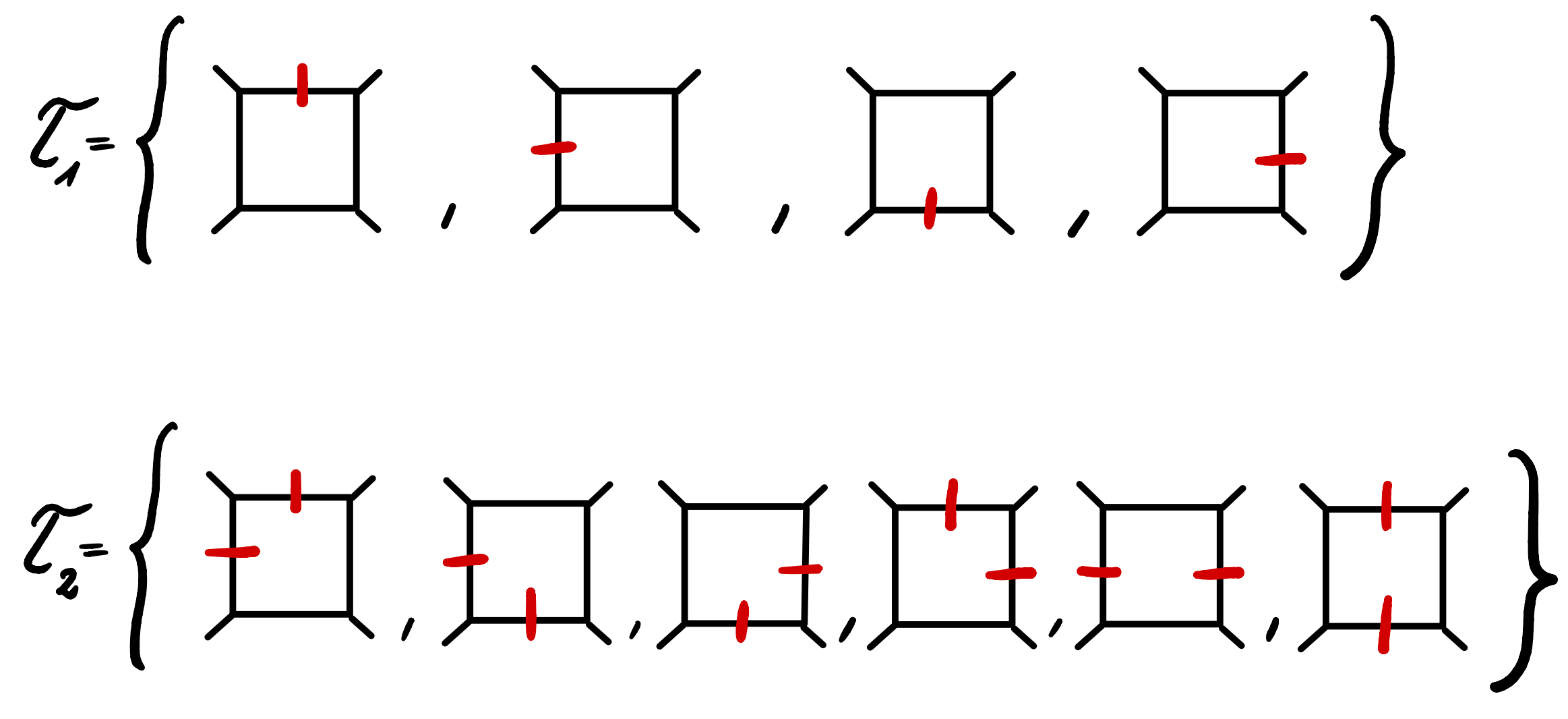}
\end{center}
\caption{1-trees and 2-trees for a one-loop box graph.\label{fig:FUgraphical}}
\end{figure}

The  functions ${\cal U}$ and ${\cal F}$ can also be constructed
from the topology of the corresponding Feynman graph.
An example for the graphical construction is shown in Fig.~\ref{fig:FUgraphical}.
The procedure is as  follows~\cite{nakanishi}:

Cutting $L$ lines of a given connected $L$-loop graph such that it becomes a connected
tree graph defines a {\em 1-tree} $T$. For each 1-tree $T$ we 
define a  {\em chord} ${\cal C}(T)$ as being the set of lines 
{\em not} belonging to this 1-tree.
The Feynman parameters associated with each chord 
define a monomial of degree $L$. 
The set of {\em 1-trees} is denoted by ${\cal T}_1$.  In the example shown in
To obtain ${\cal U}$,  we multiply the Feynman parameters belonging to a  chord ${\cal C}(T)$ for each 1-tree $T\in {\cal T}_1$:
 \be {\cal U}(\vec x) = \sum\limits_{T\in {\cal T}_1} \Bigl[\prod\limits_{j\in {\cal C}(T)}x_j\Bigr]\;.\ee
 
 Cutting one more line of a 1-tree leads to two disconnected trees, or a  {\em 2-tree} $\hat T\in {\cal T}_2$, where
${\cal T}_2$ is the set of all such  2-trees.
The corresponding chords (set of lines not belonging to this 2-tree) define  monomials of degree $L+1$.
Each 2-tree of a graph corresponds to a cut defined by cutting the lines which connected the two now disconnected trees
in the original graph. The momentum flow through the lines of such a cut defines a Lorentz invariant
$s_{\hat T} = ( \sum_{j\in \rm Cut(\hat T)} p_j )^2$.   
The function ${\cal F}_0$ is the second Symanzik polynomial of a graph with massless propagators.
${\cal F}_0$ is given by the sum over all such monomials in ${\cal
  C}(\hat T)$ times the corresponding invariant ``flowing through''
the cut (multiplied by $(-1)$).
If masses are present in the propagators, ${\cal F}(\vec x)$ gets an additional term  ${\cal U}\,\sum\limits_{j=1}^{N} x_j m_j^2$.
\begin{eqnarray}\label{eq0def}	
{\cal F}_0(\vec x) &=& \sum\limits_{\hat T\in {\cal T}_2}\;
\Bigl[ \prod\limits_{j\in {\cal C}(\hat T)} x_j \Bigr]\, (-s_{\hat T})\;,\nonumber\\
{\cal F}(\vec x) &=&  {\cal F}_0(\vec x) + {\cal U}(\vec x) \sum\limits_{j=1}^{N} x_j m_j^2\;.
\end{eqnarray}

\begin{figure}[htb]
\begin{center}
\includegraphics[width=0.26\textwidth]{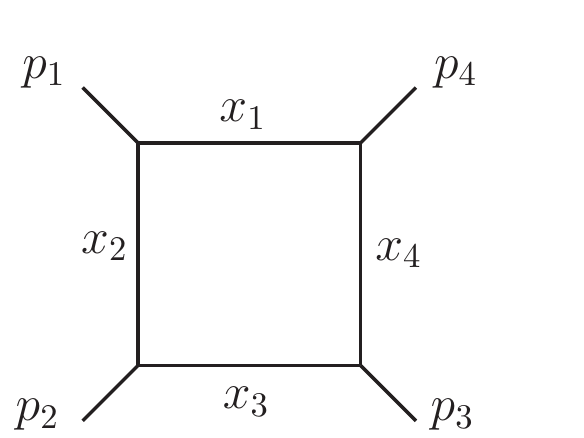}
\end{center}
\caption{Feynman parameters $x_i$ associated to the propagators of a
  one-loop box.\label{fig:box1L}}
\end{figure}

For the example of a one-loop box with labels as in Fig.~\ref{fig:box1L},
the 1-trees and 2-trees are shown in
Fig.~\ref{fig:FUgraphical}.
The chord associated with each 1-tree in ${\cal T}_1$ is
the cut line, associated with Feynman
parameter  $x_i$, corresponding to a monomial of
degree 1, such that ${\cal U}(\vec x)=x_1+x_2+x_3+x_4$.
The chords associated with the set of 2-trees shown in
Fig.~\ref{fig:FUgraphical} are the two lines which ``cut off'' the
kinematic invariant associated with the momenta flowing through the
cut lines, so the corresponding set of chords would be
$\left\{x_1x_2,x_2x_3,x_3x_4,x_4x_1,x_2x_4,x_1x_3\right\}$, such that
\be
{\cal F}_0(\vec x) =-p_1^2x_1x_2-p_2^2x_2x_3-p_3^2x_3x_4-p_4^2x_4x_1-s_{23}x_2x_4-s_{12}x_1x_3\;.
\ee

\subsubsection{Mellin-Barnes representation}
\label{sec:MBanalytic}

%

Mellin-Barnes representations have been very successful for obtaining the
first analytic results for  two-loop box integrals,
for the planar case~\cite{Smirnov:1999gc,Smirnov:1999wz} as well as
the non-planar case~\cite{Tausk:1999vh}.
Two-loop box integrals with massive propagators also have been
calculated based on  Mellin-Barnes representations, for example some
planar master
integrals entering Bhabha scattering~\cite{Heinrich:2004iq,Czakon:2005gi}, however
more recently the method of differential equations has been used to
calculate such integrals~\cite{Bonciani:2004gi,Henn:2013woa,Argeri:2014qva,Mastrolia:2017pfy,DiVita:2018nnh,Becchetti:2019tjy,DiVita:2019lpl}.
Packages to generate Mellin-Barnes representations automatically 
have been constructed in Refs.~\cite{Czakon:2005rk,Gluza:2007rt,Smirnov:2009up}
and applied successfully, in particular in combination with numerical
evaluations, see Section~\ref{sec:MBnumerical}.

The basic formula underlying the 
Mellin-Barnes representation of a (multi-) loop integral with
denominators $D_1,\ldots, D_n$ reads
\begin{align}
&\left(D_1 + D_2 + ... + D_n \right)^{-\lambda} 
 = 
 \frac{1}{\Gamma(\lambda)} \frac{1}{\left(2\pi i\right)^{n-1}} 
 \int\limits_{c-i\infty}^{c+i\infty} dz_1 ... \int\limits_{c-i\infty}^{c+i\infty} dz_{n-1}
 \label{MB}\\
 & 
 \times 
 \Gamma(-z_1) ... \Gamma(-z_{n-1}) \Gamma(z_1+...+z_{n-1}+\lambda)
 \; 
 D_1^{z_1} ...  D_{n-1}^{z_{n-1}} D_n^{-z_1-...-z_{n-1}-\lambda}  \;.
 \nonumber 
\end{align}
Each contour is chosen such that the poles of $\Gamma(-z_i)$ are to the right and the poles
of $\Gamma(\ldots +z)$ are to the left.


The representation in Eq.~(\ref{MB}) can be used to convert the sum of monomials contained in 
the functions ${\cal U}$ and ${\cal F}$ into
products, such that all Feynman parameter integrals become integrations over $\Gamma$-functions. However, we are still left with the complex contour integrals.
The latter can be performed by closing the contour at infinity and summing up all 
residues which lie inside the contour.
In general this procedure leads to multiple sums over  hyper\-geo\-metric expressions.
Several techniques and packages have been developed to manipulate these sums, see e.g. Refs.~\cite{Moch:2001zr,Davydychev:2003mv,Weinzierl:2004bn,Kalmykov:2007dk,Blumlein:2018cms,McLeod:2020dxg,Kotikov:2020ccc}
and the programs \texttt{Summer}~\cite{Vermaseren:1998uu},
\texttt{nestedsums}~\cite{Weinzierl:2002hv},
\texttt{XSummer}~\cite{Moch:2005uc},
\texttt{HarmonicSums}~\cite{Ablinger:2013cf}, \texttt{SumProduction}
or
\texttt{EvaluateMultiSums}~\cite{Schneider:2013zna,Ablinger:2013eba},
the latter two being  built on
\texttt{Sigma}~\cite{Moch:2007jk,Schneider:2013zna}.

In Ref.~\cite{Kalmykov:2012rr} it is shown that Mellin-Barnes
representations of Feynman diagrams can be used to derive
linear systems of homogeneous differential equations for the original
Feynman diagrams with arbitrary powers of propagators,
without the need for IBP relations.
This method in addition can be used to deduce extra relation between
master integrals which are not found using
IBP reduction~\cite{Kalmykov:2011yy,Kalmykov:2016lxx}, see also Refs.~\cite{Lee:2013hzt,Bitoun:2017nre,Bitoun:2018afx,Lee:2019lsr}.

In simple cases the contour integrals in the Mellin-Barnes
representation can be performed in closed form with
the help of two lemmata by Barnes~\cite{BarnesLemma}.
Barnes' first lemma states that
\begin{align}
\frac{1}{2\pi i} \int\limits_{-i\infty}^{i\infty} dz\,
\Gamma(a+z) \Gamma(b+z) \Gamma(c-z) \Gamma(d-z) 
 =  
\frac{\Gamma(a+c) \Gamma(a+d) \Gamma(b+c) \Gamma(b+d)}{\Gamma(a+b+c+d)}
\nonumber 
\end{align}
if none of the poles of $\Gamma(a+z) \Gamma(b+z)$ coincide with the
ones from $\Gamma(c-z) \Gamma(d-z)$,
Barnes' second lemma reads
\begin{align}
&\frac{1}{2\pi i} \int\limits_{-i\infty}^{i\infty} dz
\frac{\Gamma(a+z) \Gamma(b+z) \Gamma(c+z) \Gamma(d-z) \Gamma(e-z)}{\Gamma(a+b+c+d+e+z)}  \nonumber \\
& = 
\frac{\Gamma(a+d) \Gamma(b+d) \Gamma(c+d) 
 \Gamma(a+e) \Gamma(b+e) \Gamma(c+e)}{\Gamma(a+b+d+e) \Gamma(a+c+d+e) \Gamma(b+c+d+e)}\;.
\end{align}
For further details we refer to Refs.~\cite{Smirnov:2006ry,Smirnov:2009up}.

\subsubsection{Direct integration}

The direct evaluation of Feynman integrals in parameter space can
become very cumbersome with increasing numbers of parameters. However,
this approach has seen major progress due to the work of
Brown~\cite{Brown:2009ta} and Panzer~\cite{Panzer:2015ida}, who introduced the concept of {\em linear reducibility}.
This property for Feynman integrals basically means that, for some ordering of integrations over the parameters, 
the integrand can be successively linearised in one of the integration parameters.
Through the linearisation procedure the integral can be mapped to multiple polylogarithmic functions of the form
\be
G(a_1, \ldots, a_n;z)=\int_0^z\frac{dz_1}{z_1-a_1}\int_0^{z_1}\frac{dz_2}{z_2-a_2}\ldots \int_0^{z_{n-1}}\frac{dz_n}{z_n-a_n}\;.
\ee
The program {\sc HyperInt}~\cite{Panzer:2014caa} automates the iterative integration over the parameters occurring linearly.
However, if  a linear factorisation can only be achieved at the expense of introducing algebraic roots for remaining integration variables, this method is not viable, and it is usually a sign that the integral can not be expressed in terms of MPLs.

In Ref.~\cite{Hidding:2017jkk} it has been observed that in many cases the irreducible square roots appear only in the integration with respect to the last parameter. When the inner integration kernel depends on one elliptic curve and no other algebraic functions, this class of Feynman integrals can be algorithmically solved in terms of elliptic MPLs, and is called linearly reducible elliptic Feynman integral in Ref.~\cite{Hidding:2017jkk}.

In Ref.~\cite{Heller:2019gkq}, the criterion of linear reducibility was used to show  that a complete solution in terms of standard MPLs can be obtained even in the presence of unrationalisable symbol letters. This was shown in the context of calculating the two-loop master integrals for the mixed EW-QCD corrections to Drell-Yan lepton pair production in the physical region, see also Refs.~\cite{Bonciani:2016ypc,vonManteuffel:2017myy}.

Rationalising all occurring square roots by a suitable variable change is crucial, and the program {\tt RationalizeRoots}~\cite{Besier:2019kco} offers an automated procedure. In Ref.~\cite{Besier:2020hjf}, a rigorous definition of rationalisability for square roots of ratios of polynomials is given and it is 
shown that the problem of deciding whether a single square root is rationalisable can be reformulated in geometrical terms.

Furthermore,  an algebraic structure known as {\em coaction}~\cite{Brown:2015fyf} has recently played an interesting role in understanding the analytic structure of Feynman diagrams that are expressible in terms of MPLs.
Formally, the coaction is a map $m$ acting on elements of a commutative algebra ${\cal H}$,
associating to each element of ${\cal H}$ an element in the tensor product of ${\cal H}$ with itself, $m: {\cal H}\to {\cal H}\otimes {\cal H}$.
In the context of Feynman diagrams, the coaction can be seen as a mathematical operation that exposes properties of MPLs through a decomposition into simpler functions. In particular, since the coaction is naturally compatible with the actions of differential operators and taking discontinuities across branch cuts, it captures information about discontinuities.
Therefore it can be useful as a computational tool, for example to obtain canonical forms of differential equations~\cite{Abreu:2017enx,Abreu:2017mtm,Abreu:2019wzk,Duhr:2019wtr}.

\subsubsection{Asymptotic expansions}

A systematic approach to the expansion of Feynman integrals in various
limits, characterised by a scaling parameter which is small in each limit, is
the so-called {\em expansion by regions}.
The method has been pioneered in
Refs.~\cite{Smirnov:1991jn,Beneke:1997zp}, where it was formulated
in terms of the momenta involved in a loop integral.
Later it was also formulated in Feynman parameter space, where it
allows a geometric interpretation~\cite{Smirnov:1999bza,Pak:2010pt,Ananthanarayan:2018tog,Ananthanarayan:2020ptw}.
A remarkable feature of this method is that in most cases, it allows to reproduce the
exact result for the original integral if  all regions are
identified correctly and summed up, even though the integrations
 in the various regions are performed over the full integration domain.
In Ref.~\cite{Jantzen:2011nz}, the relations between the expanded
integrals and the original integral have been investigated in detail,
with the aim to clearly identify the conditions under which the original integral
can be correctly reproduced by summing over the individual regions. It was shown that
overlap contributions usually yield scaleless integrals which can be
set to zero if they are regulated appropriately.

Still,  the strategy of regions has not been proven strictly
mathematically, in particular in the presence of Glauber or potential
regions~\cite{Jantzen:2012mw}, or more generally in cases
where  the graph polynomials can change sign.
Nonetheless, the method has been extremely useful in a multitude of
calculations so far, see e.g.~\cite{Smirnov:2012gma,Mishima:2018olh}
for further references. Characterising the method by a statement from
Ref.~\cite{Semenova:2018cwy}:
``Although this strategy certainly looks suspicious for mathematicians it was successfully applied in numerous calculations.
It has the status of experimental mathematics and should be applied with care''.
The method also has been implemented in the code {\tt
asy2.m}~\cite{Jantzen:2012mw} which is part of the program {\sc
Fiesta}~\cite{Smirnov:2013eza,Smirnov:2015mct}.

\begin{figure}[htb]
  \centering
  \includegraphics[width=0.3\textwidth]{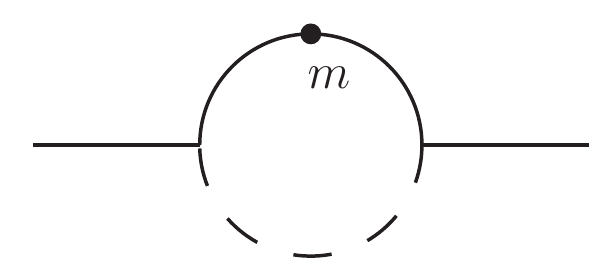}
  \caption{One-loop two-point function with the massive propagator squared.}
  \label{fig:dotted_bubble}
\end{figure}

Let us consider a simple example, discussed in more detail in Ref.~\cite{Jantzen:2011nz}.
We consider the large-momentum expansion of the integral
\be
I_2=\mu^{2\eps}\int d\kappa\frac{1}{(k+p)^2(k^2-m^2)^2}\;,
\ee
depicted in Fig.~\ref{fig:dotted_bubble}.

In the limit $|p^2|\gg m^2$ we can expand the integrand in $m^2/p^2$.
For the loop momentum $k$ we can identify two regions, a {\em hard region (h)} with $k\sim p$ and a {\em soft region (s)} with $k\sim m$.
Expansion leads to
\begin{align}
  (h): \quad \frac{1}{(k+p)^2(k^2-m^2)^2}&\to \frac{1}{(k+p)^2\,(k^2)^2}\left( 1+2\frac{m^2}{k^2}+\ldots\right)\;,\nn\\
  (s): \quad \frac{1}{(k+p)^2(k^2-m^2)^2}&\to \frac{1}{(k^2-m^2)^2\,p^2}\left( 1-2\,\frac{p\cdot k}{p^2}-\frac{k^2}{p^2}+\ldots\right)\;.
\end{align}
Integrating the leading contributions in the expansions above results in
\begin{align}
I_2^{(h)}&=\frac{1}{p^2}\left[ -\frac{1}{\eps}+\ln\left( \frac{-p^2}{\mu^2}\right) \right]+{\cal O}(\eps)\;,\label{eq:epsIR}\\ 
I_2^{(s)}&=\frac{1}{p^2}\left[ \frac{1}{\eps}-\ln\left( \frac{m^2}{\mu^2}\right) \right]+{\cal O}(\eps)\;,\label{eq:epsUV}
\end{align}
such that
\be
I_2=I_2^{(h)}+I_2^{(s)}=\frac{1}{p^2}\,\ln\left( \frac{-p^2}{m^2}\right)+{\cal O}(\eps)\;.\label{eq:combi}
\ee
The above result is actually valid exactly, i.e. not only in the limit $|p^2|\gg m^2$, for a proof we refer to Ref.~\cite{Jantzen:2011nz}.
Even though we integrated over the whole integration domain, no double counting has occurred.
This is because the integrals in the ``overlap region'' can be shown to correspond to  scaleless integrals which vanish in dimensional regularisation. The leading term in the overlap region reads
\be
I_2^{(h,s)}=\frac{1}{p^2}\int d^Dk \frac{1}{(k^2)^2}\;,\label{eq:overlap}
\ee
higher terms in the expansion add powers of $m^2$ and $p\cdot k$ in the numerator and higher powers of $k^2$ and $p^2$
in the denominator, so do not change the fact that the integral over the momentum $k$ is scaleless.
In fact, \eqn{eq:overlap} can be written as
\be
I_2^{(h,s)}=\frac{1}{p^2}\left( \frac{1}{\eps_{UV}}- \frac{1}{\eps_{IR}}\right)\;.
\ee
Note that the (spurious) pole in Eq.~(\ref{eq:epsIR}) is of infrared nature, while the pole in  Eq.~(\ref{eq:epsUV}) is of UV nature, so formally the combinations $I_2^{(h)}-I_2^{(h,s)}$ and $I_2^{(s)}-I_2^{(h,s)}$ are separately IR and UV finite.
The cancellation of the poles between Eqs.~(\ref{eq:epsIR}) and (\ref{eq:epsUV}) is therefore also related to the fact that scaleless integrals vanish in dimensional regularisation.

For more general cases it is convenient to classify the scaling properties in the so-called Sudakov parametrisation of momenta.
For example, if we consider cases where two directions are defined by two momenta $p$ and $l$,
we can define two light-like momenta $n^\mu$ and $\bar{n}^\mu$ with $n\cdot \bar{n}\not=0$ and
choose the coordinate system such that
\begin{align}
n^\mu & =\frac{1}{\sqrt{2}} ( 1,0,0,1) \sim p^\mu/p^0 \;,\;
\bar{n}^\mu  = \frac{1}{\sqrt{2}}( 1,0,0,-1) \sim l^\mu/l^0\;.
\end{align}
Any Lorentz vector $r$ can then be written in the form
\begin{align} 
r^\mu &= (n\cdot r) \,\bar{n}^\mu+(\bar{n}\cdot r) \,n^\mu+r^\mu_\perp 
\equiv r_+^\mu + r_-^\mu+ r^\mu_\perp\;.
\label{eq:light-cone-param}
\end{align}
To discuss the scaling properties, we 
consider the massless triangle integral
\begin{equation}
I_3  =  \int d^Dk \frac{1}{k^2(k+p)^2 (k+l)^2}\;,
\label{eq:I3}
\end{equation}
where we define a small parameter $\lambda$ by
\be
\lambda^2\sim \frac{|p^2|}{Q^2}\sim  \frac{|l^2|}{Q^2}\ll 1\;, \mbox{where } Q^2=(p-l)^2\;.
\ee
The components of the external momenta then scale like
\begin{align}
p^\mu \sim  (\lambda^2 , 1 , \lambda ) \,Q \; ,\;
l^\mu \sim   (1 ,  \lambda^2 ,  \lambda) \,Q\;.
\end{align}
For the loop integral (\ref{eq:I3}), we can identify regions according to the scaling properties of the loop momentum
$k$ in each region~\cite{Becher:2014oda,Becher:2018gno}:
\begin{equation}
\begin{aligned}
  &	&	 & (n\cdot k, \bar{n}\cdot k, k_\perp^\mu\,)  \\
&\text{hard} &(h)\quad &   (\,\;\; 1\; \;,\; \;1\; \;\;,\;\; 1 \;) \,Q \,, \\
&\text{collinear to $p^\mu$} &(c)\quad &  (\,\;\; \lambda^2 \,,\; \;1\; \;\;,\;\; \lambda \,) \, Q\,,  \\
&\text{collinear to $l^\mu$} &(\bar{c})\quad &  (\,\;\; 1 \;\;, \;\; \lambda^2\;\;  , \; \lambda\, \;) \, Q \,, \\
&\text{soft} &(s)\quad &  (\,\; \lambda^2 \;\,, \;\; \lambda^2\;\;  , \; \lambda^2 ) \, Q\,.
\end{aligned}
\label{eq:Sudakov_regions}
\end{equation}
Hence, for the considered example, in the soft region 
$k_{\rm{soft}}^2 \sim \lambda^4 Q^2 $, therefore this mode is sometimes also called {\em ultra-soft}.
For other cases,  the soft mode scales as $(\lambda , \lambda, \lambda)\,Q$.

To cover  all possible regions for  integrals of Sudakov form factor type, two more regions should in principle be considered, which are the {\em Glauber} region and the {\em collinear plane} region~\cite{Jantzen:2011nz}, given by
\begin{equation}
\begin{aligned}
	 & (n\cdot k, \bar{n}\cdot k, k_\perp^\mu\,)  \\
\text{Glauber: } &   (\;\lambda^2\;\;,\; \lambda^2\; ,\; \lambda\,) \,Q \,, \\
\text{collinear plane: } &  (\;1 \;\;,\; \;1\; \;,\;\; \lambda \,) \, Q\;. \\
\end{aligned}
\label{eq:Glauber_collp}
\end{equation}
However, it can be shown~\cite{Jantzen:2011nz} that they only lead to scaleless integrals and therefore can be dropped.

In non-relativistic QCD, for heavy quarkonia with velocity $v\ll1$, one can identify regions defined by the smallness parameter $v$, such as~\cite{Beneke:1997zp}
\begin{equation}
\begin{aligned}
\text{hard: } &k^0\sim m, \vec{k}\sim m\;,\\
\text{soft: } &k^0\sim m v, \vec{k}\sim m v\;,\\
\text{potential: } &   k^0\sim m v^2, \vec{k}\sim m v\;,\\
\text{ultrasoft: } &k^0\sim m v^2, \vec{k}\sim m v^2\;.\\
\end{aligned}
\label{eq:NRQCD}
\end{equation}

\subsection{Numerical methods: overview}
\label{sec:numerical}

\subsubsection{Numerical Mellin-Barnes integration}
\label{sec:MBnumerical}

Mellin-Barnes representations as discussed in Section
\ref{sec:MBanalytic} also have been used as a starting point for a
subsequent numerical evaluation of the integral.

In the Euclidean region, this has been developed first in
Ref.~\cite{Czakon:2005rk} and implemented in the package {\tt MB.m},
which can resolve singularities, expand in dimensional and analytic
regulators, perform an asymptotic expansion, add
up residues in terms of multi-fold sums (see also
Ref.~\cite{Ochman:2015fho}) and numerically evaluate the integrals in
the Euclidean region. 
The program can be found on the {\sc hepforge} database {\tt MB
Tools}~\cite{mbtools}, together with further packages, for example
{\tt MBresolve.m}~\cite{Smirnov:2009up}, a tool realising another
strategy of resolving singularities of Mellin-Barnes integrals, as
well as {\tt AMBRE}~\cite{Gluza:2007rt,Gluza:2010rn,Blumlein:2014maa,Dubovyk:2016ocz}
and {\tt barnesroutines.m}~\cite{Gluza:2016fwh}.
The numerical evaluation of Mellin-Barnes integral representations
with physical kinematics has been brought forward in
Ref.~\cite{Anastasiou:2005cb} for massless integrals.
However, as has been shown in Ref.~\cite{Czakon:2005rk}, massive
propagators lead to an oscillatory behaviour of the integrand which
complicates the numerical integration. Nonetheless, numerical
Mellin-Barnes integration in the physical region has seen significant progress
recently~\cite{Dubovyk:2016ocz,Gluza:2016fwh,Dubovyk:2017cqw,Usovitsch:2018shx,Usovitsch:2018qdk}, and has
been applied to calculate electroweak pseudo-observables and $Z$-boson
form factors at two-loop
accuracy~\cite{Dubovyk:2016aqv,Dubovyk:2018rlg,Dubovyk:2019szj}.
An interesting development is presented in Ref.~\cite{Prausa:2017frh},
where the ``method of brackets''~\cite{Gonzalez:2010zzb} has been
extended to produce low-dimensional MB representations, applicable to
both planar and non-planar Feynman diagrams.
The method of brackets defines a set of simple rules which, when
applied to a Schwinger parametrised Feynman integral,
yields a set of multi-dimensional MB integrals.

\subsubsection{Numerical solutions of differential equations}

Differential equations are a very powerful tool to achieve analytic representations of multi-loop Feynman integrals, as has been explained in Section \ref{sec:DEanalytic}.
However, numerical solutions of differential equations also have been used successfully to evaluate Feynman integrals, and they offer  interesting possibilities for future developments.

Early work on numerical solutions of differential equations for massive two-loop self-energy diagrams can be found in Refs.~\cite{Caffo:2002ch,Caffo:2003ma,Martin:2003qz,Pozzorini:2005ff}.
In the case of massive two-point functions, there is only one kinematic invariant $p^2$ in addition to the internal masses.
Choosing the boundary value $p^2/m^2=0$, the boundary integrals reduce to vacuum integrals which usually can be evaluated
analytically.

The numerical integration will fail if the integrand vanishes on the real axis, for example due to thresholds.
As these singularities are in general integrable,  a deformation of the integration contour into the complex plane can be used to achieve numerical convergence.
The numerical evaluation of differential equations for massive two-loop self-energy 
integrals has been implemented in the public programs {\tt TSIL}~\cite{Martin:2005qm} and {\tt BoKaSun}~\cite{Caffo:2008aw}.

Recent programs to tackle massive one- and two-point functions up to three loops are {\tt TIVD}~\cite{Bauberger:2019heh,Bauberger:2017nct, Freitas:2016zmy} and {\tt 3VIL}~\cite{Martin:2016bgz}.
In addition to numerical solutions of differential equations, these programs partly use the numerical evaluation of dispersion relations~\cite{Bauberger:1994by,Bauberger:1994hx}
as well as a combination of (partial) analytic and numerical integrations based on ideas outlined in Ref.~\cite{Ghinculov:1996vd}.

Numerical solutions of differential equations for multi-scale integrals have been pioneered in Refs.~\cite{Boughezal:2007ny,Czakon:2007qi,Czakon:2008zk}, in the context of two-loop integrals entering top quark pair production at NNLO.
For the boundary conditions,  the high-energy limit has been employed, i.e. an expansion in the parameter $m_t^2/k$,
where $k\in \{s,|t|,|u|\}$ is large compared to $m_t^2$. Choosing the high-energy limit for the 
boundary conditions has the advantage that no physical threshold is crossed when integrating the system from the boundary to a physical kinematic point.
However, spurious singularities still need to be avoided by contour deformation.
Further, in the subleading colour contributions, there is a Coulomb singularity due to exchange of a Glauber gluon between the top quarks, which needs to be treated by resummation in order to achieve a reliable result.
The method developed along these lines led to the successful calculation of $t\bar{t}$ production in hadronic collisions at NNLO accuracy~\cite{Czakon:2013goa} as well as the 3-loop Higgs-gluon form factor with full quark mass dependence~\cite{Czakon:2020vql}.

Calculating integrals via numerical solutions of differential equations has the advantage that, in constrast to multi-loop Feynman integrals, the integrals are of 
low dimensionality and therefore can be evaluated efficiently and with high precision.
However, the approach relies on a successful reduction of the amplitude to a basis of master integrals, which may not be easily achieved for multi-loop multi-scale problems.
Further, it relies on the fact that the boundary terms can be obtained to high precision with some (analytic or numerical) method, which is not self-understood for problems of high complexity, even though they involve less scales.

Another powerful strategy to solve differential equations for Feynman integrals by series expansions near singular points has been developed in Refs.~\cite{Lee:2017qql,Lee:2018ojn}, see also~\cite{Liu:2017jxz}.
This work also inspired further progress in the semi-numerical evaluation of two-loop integrals with massive propagators which are known to contain elliptic functions~\cite{Hidding:2017jkk,Bonciani:2018uvv,Mandal:2018cdj,Frellesvig:2019byn}.

In Ref.~\cite{Mandal:2018cdj}, an approach has been developed which solves differential equations numerically and uses boundary conditions at Euclidean points, obtained by sector decomposition. As the initial conditions are in the unphysical region, a suitable integration contour has to be found to continue result to the physical region. This procedure requires the knowledge of the singular points and therefore is not easily automated.

\subsubsection{Four-dimensional methods}
\label{subsec:4dim}


Infrared singularities as we encounter them in perturbative
calculations involving massless particles are a consequence of the fact that we calculate with final states of a fixed number of particles, distinguishing quantum states involving $n$ particles from quantum states with $n$ plus a number of soft or collinear particles, even though they are physically equivalent.
The use of dimensional regularisation for both virtual and real corrections allows to deal with these ``spurious'' singularities in a gauge- and Lorentz invariant as well as algorithmic way. 
However there are many obstacles related to calculations in $D$ dimensions, in particular beyond the next-to-leading order in perturbation theory, ranging from the question of a consistent treatment of $\gamma_5$ to the necessity of complicated subtraction schemes for IR divergent real radiation at NNLO and beyond.
Therefore it seems natural to investigate whether perturbative calculations can be formulated differently, performing the summation over degenerate soft and collinear states at integrand level, such that the singularities cancel before they would show up explicitly as $1/\eps$--poles, which allows to calculate entirely in the four physical space-time dimensions. Of course this also requires the subtraction of UV divergences at integrand level.
A summary of recent developments about various approaches in $D$ versus four dimensions can be found in Refs.~\cite{Gnendiger:2017pys,Bruque:2018bmy,Gnendiger:2019vnp}.

The idea to cancel the singularities at integrand level and to perform all the integrations as a phase space integral numerically goes back to Dave Soper~\cite{Soper:1998ye,Soper:1999xk}, and a similar approach has been pursued to calculate NLO amplitudes numerically by several groups~\cite{Catani:2008xa,Gong:2008ww,Kilian:2009wy,Becker:2010ng,Becker:2011vg,Becker:2012aqa,Becker:2012nk,Duplancic:2016lzh}.
Methods to calculate one-loop integrals in a parameter representation numerically also have been worked out~\cite{Passarino:2001jd,Binoth:2002xh,Kurihara:2005ja,Binoth:2005ff,Kurihara:2006kt,Nagy:2006xy,Lazopoulos:2007ix,Yuasa:2009ym,Yuasa:2011kt}, partially with extensions beyond one loop~\cite{Binoth:2000ps,Binoth:2003ak,Kurihara:2005ja,Anastasiou:2005cb,Anastasiou:2006hc,Anastasiou:2007qb,Bierenbaum:2010cy,Yuasa:2011ff,Becker:2012bi}, see also Sections~\ref{sec:other_numerical} and~\ref{sec:secdec} for more details.
A four-dimensional subtraction scheme operating on two-loop diagrams has been worked out in Ref.~\cite{Anastasiou:2018rib} and extended to operate on amplitude level in QED in Ref.~\cite{Anastasiou:2020sdt}.

The direct numerical integration over loop momenta has been further developed by various groups, in particular using {\em loop-tree duality}~\cite{Catani:2008xa,Kilian:2009wy,Bierenbaum:2010cy,Bierenbaum:2012th,Buchta:2014dfa,Hernandez-Pinto:2015ysa,Buchta:2015wna,Sborlini:2016gbr,Sborlini:2016hat,Driencourt-Mangin:2017gop}.
The method relies on a dual representation of a loop integral or amplitude, represented by a product of propagators, which is
obtained by cutting the internal lines one by one and
applying the residue theorem accordingly.
Considering a generic $N$-point integral at $L$ loops, it can be written as~\cite{Aguilera-Verdugo:2020kzc}
\begin{equation}
{\cal A}_{N}^{\left(L\right)}(1,\ldots,n)=\int d^Dl_1\ldots d^Dl_L\,{\cal N}\,\times\,G_{F}(1,\ldots,n)\;,
\label{eq:AN}
\end{equation}  
with
${\cal N}$ being a polynomial function of the loop and external momenta
and $G_F$ products of propagators raised to some powers $\alpha_i$,
\begin{equation}
G_{F}\left(1,\ldots,n\right)=\prod_{i\in 1\cup 2\cup \ldots\cup n}\,\left(G_{F}(q_{i})\right)^{\alpha_{i}}\;,
\label{eq:ProductoGFs}
\end{equation}
where  the product of the Feynman propagators can run over one set or the union of several sets of propagators and
\begin{align}
G_{F}\left(q_{i}\right) & =\frac{1}{q_{i}^{2}-m_{i}^{2}+i\eps}=\frac{1}{\left(q_{i,0}+q_{i,0}^{\left(+\right)}\right)\left(q_{i,0}-q_{i,0}^{\left(+\right)}\right)}\;,\nn\\
q_{i,0}^{\left(+\right)}&=\sqrt{\mathbf{q}_{i}^{2}+m_{i}^{2}-i\eps}
\;.
\label{eq:gf}
\end{align}
The energy component $q_{i,0}$ of the loop momentum can be integrated out explicitly, using 
the Cauchy residue theorem iteratively.
Setting the propagators that belong to the set $s$ consecutively on-shell leads to the dual function
\begin{equation}
G_D(s;t) = -2\pi i\sum_{i_s \in s} {\rm Res}\left(G_F(s;t),{\rm Im}(q_{i_s,0})<0 \right) \; ,
\label{eq:cauchyG}
\end{equation} 
where $G_F(s, t)$ represents the product of the Feynman propagators that belong to the two sets $s$ and $t$. 
Iterating this procedure,
the final loop-tree-duality representation is given by the sum of all the nested residues, which corresponds to setting $L$ lines on-shell simultaneously, such that the loop amplitude is represented by non-disjoint tree amplitudes.
\begin{align}
{\cal A}_D(1,\ldots,r;r+1,\ldots,n) &=
\sum_{i_r \in r} {\rm Res}\left({\cal A}_D(1,\ldots,r-1;r,\ldots,n),{\rm Im}(q_{i_{r},0})<0 \right) \,.
\label{eq:IteratedResidue}
\end{align}


Methods based on loop-tree duality have seen very rapid progress recently~\cite{Driencourt-Mangin:2019aix,Driencourt-Mangin:2019sfl,Runkel:2019zbm,Capatti:2019ypt,Capatti:2019edf,Aguilera-Verdugo:2019kbz,Baumeister:2019rmh,Runkel:2019yrs,Verdugo:2020kzh,Aguilera-Verdugo:2020kzc,Ramirez-Uribe:2020hes,Plenter:2020lop}.
As these methods exploit the Cauchy residue theorem to reduce the dimension of the loop integrations,
and do not require the introduction of Feynman parameters, their complexity exhibits a very favourable scaling with increasing number of loops.
The so-called {\em Four-Dimensional Unsubtraction (FDU)} method has been developed in Refs.~\cite{Hernandez-Pinto:2015ysa,Sborlini:2016gbr,Sborlini:2016hat,Driencourt-Mangin:2019sfl}, whose aim is to combine real and virtual corrections into a single, numerically stable integral, thus avoiding the need for IR subtractions.

Alternative representations based on loop-tree duality have been proposed in Refs.~\cite{Runkel:2019zbm,Baumeister:2019rmh,Capatti:2019ypt,Capatti:2019edf,Runkel:2019yrs,Capatti:2020ytd,Capatti:2020xjc}, where Ref.~\cite{Capatti:2020xjc} presents an explicit framework to calculate higher order cross sections numerically such that the IR cancellations are realised fully locally.

In Refs.~\cite{Aguilera-Verdugo:2020kzc,Verdugo:2020kzh,Ramirez-Uribe:2020hes,Capatti:2020ytd}, a new strategy is proposed to generalise loop-tree-duality, showing that it leads to integrand representations which are manifestly free of non-causal singularities to all orders.

Explicit applications to four-loop diagrams, including non-planar topologies, are shown in Ref.~\cite{Ramirez-Uribe:2020hes}, a mathematical proof has been presented in Ref.~\cite{Aguilera-Verdugo:2020nrp}.
A geometric interpretation of the causal structure of multi-loop Feynman integrals and amplitudes has been worked out in Ref.~\cite{Sborlini:2021owe}, exhibiting a deep connection between geometry and causality.
In Ref.~\cite{TorresBobadilla:2020ekr} it was shown algebraically that the causal representation of multi-loop Feynman integrals can also be formulated in terms of cusps and edges.
The relation between this algebraic approach and the geometric approach of  Ref.~\cite{Sborlini:2021owe} is currently under investigation.


Among four-dimensional methods where two-loop applications already have been achieved is also the Four-Dimensional Regularisation/Renormalisation (FDR) approach~\cite{Pittau:2012zd,Donati:2013voa,Zirke:2015spg,Page:2015zca,Page:2018ljf,Pittau:2019oqe}, which subtracts UV divergences at integrand level and re\-gu\-lates IR singularities by a dimensionful scale in the propagators.


Another four-dimensional method is {\em FDF}, a  Four-Dimensional Formulation of dimensionally regulated amplitudes where the $(D-4)$-dimensional components are mapped to terms related to mass terms~\cite{Fazio:2014xea}. This scheme is particularly suited for the unitarity-based construction of $D$-dimensional one-loop amplitudes. 

An important ingredient for two-loop amplitudes constructed from on-shell cuts in four dimensions are the so-called rational parts $R_2$,
which, in dimensional regularisation, are attributed to ${\cal O}(\eps)$ terms hitting UV poles in individual diagrams\footnote{However, we should point out that amplitudes which are UV finite, such as the 4-photon amplitude, also can have rational parts, as the UV poles occurring at intermediate stages of the calculation cancel, while the  ${\cal O}(\eps)$ terms multiplying the poles are dependent on different kinematic invariants and therefore do not cancel.}.
The first calculation of the full set of two-loop rational counterterms in Yang-Mills theories has been presented in Refs.~\cite{Pozzorini:2020hkx,Lang:2020nnl}.

\subsubsection{Other methods}
\label{sec:other_numerical}

In this section, a selection of other numerical methods is presented,
which were very successful in particular applications but are less
widely used so far, mostly because they are more tailored to a particular
application, see also Ref.~\cite{Freitas:2016sty} for an overview.

\subsubsection*{Bernshtein-Sato-Tkachov method and related approaches}

Techniques for the numerical calculation of two-loop three-point
functions with several mass scales based on the Bernshtein-Sato-Tkachov (BST)
Theorem~\cite{Bernstein,Tkachov:1996wh,Sato} have been developed in
Refs.~\cite{Passarino:2001jd,Ferroglia:2002mz,Ferroglia:2003wc,Ferroglia:2003yj,Actis:2004bp,Passarino:2006gv,Actis:2008ts,Passarino:2010qk}
and applied to calculate the complete electroweak corrections to Higgs
boson production at hadron colliders~\cite{Actis:2008ug} and the decay of a Higgs
boson into two photons~\cite{Passarino:2007fp}.

The basic idea of the BST theorem can be explained easiest by looking
at Eq.~(\ref{eq:feynint}), the representation of a loop integral after
the introduction of Feynman parameters. The polynomial ${\mathcal F}$
carries information about the kinematic singularities such as
thresholds.
The numerical integration of such singularities (in the case where they
are integrable), will be more difficult the more negative the power of
${\mathcal F}$ in the integrand is. Therefore trying to increase the
power of ${\mathcal F}$ is a good strategy.
At one-loop, ${\mathcal F}$ is a quadratic polynomial in the Feynman
parameters. In this case, a differential operator can be defined~\cite{Tkachov:1996wh} which
increases the power of ${\mathcal F}$ (denoted generically by
$-b-\eps$):
\begin{align}
{\cal F}^{-b-\epsilon} = \frac{1}{B} \biggl [ 1- \frac{(\vec{x}+\vec{A})^\top
\vec{\partial}_x}{2(1-b-\epsilon)}\biggr ] {\cal F}^{-b-\epsilon+1}\;,
\label{eq:BST}
\end{align}
where (see also eq.~(\ref{eq:F}))
\begin{align}
B &= J - \vec{Q}^\top {\cal M}^{-1} \vec{Q}\;,\;
\vec{A} = {\cal M}^{-1} \vec{Q}\;.
\end{align}
This operation can be repeated until the form ${\mathcal F}^{-\eps}$
in the integrand is achieved. The numerical integration over
logarithmic singularities is usually well-behaved.
Potential IR singularities should be extracted beforehand. In
Refs.~\cite{Ferroglia:2002mz,Passarino:2006gv} this has
been achieved using sector decomposition or Mellin-Barnes representations.

The application of the BST theorem beyond one-loop faces difficulties
because the polynomial ${\mathcal F}$ is not a quadratic form in the
Feynman parameters anymore. Nonetheless, the theorem can be applied to
one-loop subdiagrams based on a loop-by-loop Feynman parametrisation.
Applying this operation repeatedly, sometimes assisted by analytic
integrations of simpler subexpressions, again an at most logarithmic
singularity behaviour can be
achieved~\cite{Ferroglia:2003yj,Passarino:2006gv}, see also 
Ref.~\cite{Radionov:2020avv} for recent developments regarding the
extension to two loops.

The main problems of the BST-approach are related to the factors $B$
in the denominator generated by Eq.~(\ref{eq:BST}). They can become
very small in the vicinity of certain kinematic configurations
corresponding to physical or spurious singularities.
These singularities are usually tamed by expressions in the numerator
which go to zero when such a limit is approached. However, depending
on the organisation of the various terms, this can lead to very large
numerical cancellations. Methods to overcome this problem, for example
by an expansion around $B=0$ or a modified BST-relation, have been
worked out in Refs.~\cite{Ferroglia:2002mz,Ferroglia:2003yj,Uccirati:2004vy}.

While these methods work very well for the two-loop vertex
calculations they have been applied to, they are difficult to
generalise to problems with more legs or loops, where the number of
Feynman parameters and thus the dimension of the numerical
integrations is increased.

A general approach to calculate two-loop integrals by doing the integration
over Feynman parameters partially analytically and partially
numerically has been proposed in Refs.~\cite{Guillet:2019hfo,Guillet:2020rhg}.
It makes use of a relation similar to the BST-relation (\ref{eq:BST})
as part of the two-loop integrations. The remaining numerical
integrations then have a relatively low dimensionality and therefore
are computationally less costly.

\subsubsection*{Expansions}

The program {\sc TayInt}~\cite{Borowka:2018dsa} also tries to find an ideal compromise between  analytic and  numerical integration parts.
It operates on quasi-finite integrals in the Feynman parameter
representation and first performs sector decomposition to obtain parameter integrals
which are in a convenient form.
Conformal mappings and partitionings of the integrand are used before the integrand is Taylor expanded in the Feynman parameters.
To deal with thresholds, a variable transformation which implements the correct ana\-lytic continuation of the integrand to the complex plane is performed, which requires integration over one parameter before the Taylor expansion. The method has been applied to non-trivial two-loop examples with several mass scales.

Asymptotic expansions of Feynman amplitudes at integrand level within the loop-tree duality formalism have been presented recently in Refs.~\cite{Plenter:2019jyj,Plenter:2020lop}.

In the context of expansions, the method developed in Refs.~\cite{Francesco:2019yqt,Frellesvig:2019byn,Hidding:2020ytt}, already described in Section~\ref{sec:DEanalytic} and also applied in Ref.~\cite{Abreu:2020jxa}, should be mentioned again.
It operates on systems of differential equations and parametrises the
integrals along line segments, which allow to find solutions in 
terms of truncated one-dimensional series expansions.
It by far outperforms purely numerical methods in precision and speed.

\subsubsection*{Direct Computation Method}

Numerical methods to evaluate
multi-loop integrals without the need to introduce contour deformation
have been developed in
Refs.~\cite{deDoncker:2004fb,Yuasa:2009ym,Yuasa:2011ff,deDoncker:2014tya,deDoncker:2017gnb,Kato:2018ium}.
The so-called ``Direct Computation Method'' (DCM) relies on the fact
that threshold singularities can be regulated by introducing a
``finite width'' using the replacement
\be
m^2\to m^2(1-i\rho)\;.\label{eq:extrapol}
\ee
For non-zero
and not too small values of $\rho$, a relatively fast convergence of
the integral can be achieved. The art is to devise an extrapolation
method such that the series of integrals for $\rho\to 0$ converges to
the correct value while keeping the computation time low.
For the latter purpose, Quasi-Monte-Carlo integration based on
rank-one lattice rules has been employed~\cite{deDoncker:2018nqe} (see
also Section \ref{sec:qmc}).
In the presence of UV or IR divergences, the dimensional regularisation
parameter $\eps$ is also kept finite, and a double extrapolation
method is used~\cite{deDoncker:2012qk}.

A method based on the replacement (\ref{eq:extrapol}) and subsequent
extrapolation $\rho\to 0$  also has been adopted in
Refs.~\cite{Baglio:2018lrj,Baglio:2020ini} to calculate gluon fusion into Higgs pairs
at NLO QCD with full top quark mass dependence.
For the extrapolation, the method of Richardson~\cite{Richardson} has
been employed.

\subsection{Sector decomposition}
\label{sec:secdec}

Sector decomposition as it is used today is a constructive method to
isolate endpoint singularities in dimensional regularisation. The coefficients of the  resulting Laurent series in $\eps$ are finite parameter integrals which lend themselves to numerical integration.

\subsubsection*{Some history}
The idea of sector decomposition goes back to Hepp~\cite{Hepp:1966eg} 
(see also Speer~\cite{Speer:1974hd,Speer:1975dc}) to analyse the
ultraviolet (and infrared) convergence of Feynman integrals in Euclidean space in order to prove renormalisation theorems established by Bogoliubov, Parasiuk, Hepp and 
Zimmermann~\cite{Bogoliubov:1957gp,Hepp:1966eg,Zimmermann:1969jj}.
A recursive so-called $R$-operation has been defined which subtracts
appropriate counter-terms from a Feynman graph in order to render it UV
finite. It also has been partly used in  modern formulations of
renormalisation theorems~\cite{EbrahimiFard:2005gx, Kennedy:2007sj,Bergbauer:2009yu}.

However, in quantum field theories involving massless particles, infrared (IR) divergences are also present.\footnote{We will use ``infrared" divergences
to denote both soft and collinear divergences. In Euclidean space, only soft IR divergences are present.}
Early work to define IR finite Green's functions can be found e.g. in Refs.~\cite{Breitenlohner:1975hg,Breitenlohner:1976te,Speer:1975dc,Speer:1977uf}. 
For reviews of these methods we refer to Refs.~\cite{Smirnov:1991jn,Smirnov:2008aw,Smirnov:2012gma}.

A highly algorithmic way to subtract IR singularities in Euclidean space has been developed as the so-called $R^{\star}$-operation~\cite{Chetyrkin:1982nn,Chetyrkin:1984xa,Smirnov:1986me,Herzog:2017bjx,Beekveldt:2020kzk}.
In Refs.~\cite{Herzog:2017bjx,Herzog:2017jgk} the
$R^{\star}$-operation has been extended to Feynman graphs of arbitrary
tensorial rank, serving to calculate the 5-loop
beta-function~\cite{Herzog:2017ohr,Chetyrkin:2017bjc}, see also
Refs.~\cite{Luthe:2016ima,Luthe:2017ttc,Luthe:2017ttg}. In
Ref.~\cite{Beekveldt:2020kzk} the Hopf-algebraic structure of the
$R^{\star}$-operation is elucidated.

The first phenomenological application of sector decomposition in Minkowski space can be found in Ref.~\cite{Roth:1996pd}, where it has been employed to 
extract logarithmic mass singularities from multi-scale two-loop
integrals in the high energy limit. In Refs.~\cite{Pozzorini:2004rm,Denner:2004iz} it also has been applied successfully in the context of electroweak corrections.

In Ref.~\cite{Binoth:2000ps}, the concept of sector decomposition 
has been elaborated into an automated algorithm, 
allowing to isolate end-point singularities of UV as well as IR nature in the context of dimensional regularisation.
The original iterative algorithm is described in detail in Refs.~\cite{Binoth:2000ps,Heinrich:2008si}, here we only sketch it and focus on the more recent developments.

\subsubsection{Basic concept of sector decomposition}\label{sec:basics}

\begin{figure}[htb]
  \centering
  \includegraphics[width=0.8\textwidth]{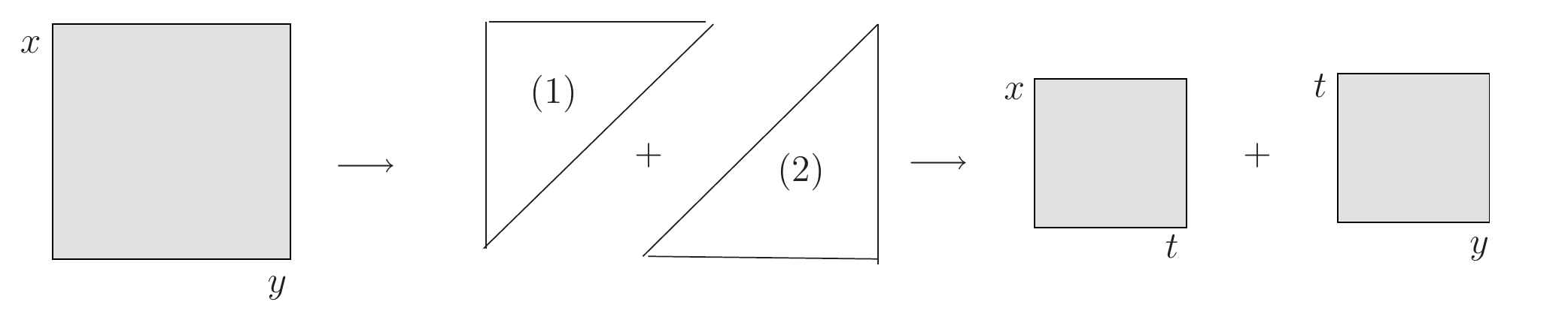}
\caption{Sector decomposition schematically.\label{sectors}}
\end{figure}

To introduce the basic concept of sector decomposition, let us look at the simple example of a two-dimensional 
parameter integral of the following form:
\begin{eqnarray}
I&=&
\int_0^1 dx\,\int_0^1dy \,(x+y)^{-2+\epsilon}\;.
\end{eqnarray}
The integral contains a singular region where $x$ and $y$ vanish simultaneously, 
i.e. the singularities for $x\to 0$ and $y\to 0$ are overlapping.
Our aim is to factorise these singularities. 
Therefore we divide the integration range into two 
sectors where $x$ and $y$ are ordered (see Fig.~\ref{sectors})
\begin{eqnarray*}
I&=&
\int_0^1 dx\,\int_0^1dy \,(x+y)^{-2+\epsilon}\,
[\underbrace{\Theta(x-y)}_{(1)}+\underbrace{\Theta(y-x)}_{(2)}]\;.
\end{eqnarray*}
Now we substitute  $y=x\,t$ in sector (1) and $x=y\,t$ in sector (2) 
to remap the integration range to the unit square and obtain
\begin{align}
I&=\int_0^1 dx\,x^{-1+\epsilon}\int_0^1 dt\,(1+t)^{-2+\eps}
+\int_0^1 dy
\,y^{-1+\epsilon}\int_0^1 dt\, (t+1)^{-2+\eps}\nn\\
&=2 \int_0^1 dx\,x^{-1+\epsilon}\int_0^1 dt\,(1+t)^{-2+\eps}\;.
\end{align}
We observe that the singularities are now factorised such that 
they can be read off from the powers of simple monomials in the integration 
variables, while the remaining polynomial  goes to a constant if the 
integration variable $t$ approaches zero.
The same concept can be applied to 
$N$-dimensional parameter integrals over polynomials raised to 
some power, as they occur for example in Feynman integrals, where the procedure in general 
has to be iterated to achieve  complete factorisation. This iteration
can lead to a large number of subsector integrals, which is the main
drawback of sector decomposition. 

\subsubsection{Decomposition algorithms}
\label{sec:geomethod}

The decomposition algorithms for loop integrals operate on the representation of an integral with $L$ loops in $D$ dimensions in terms of Feynman parameters, as introduced in Section~\ref{sec:FU},
\begin{align}
G(\vec{\nu}) &=  (-1)^{N_{\nu}}\frac{\Gamma(N_{\nu}-LD/2)}{\prod_{j=1}^{N}\Gamma(\nu_j)}
\prod\limits_{j=1}^{N}\, \int\limits_{0}^{\infty} dx_j\,x_j^{\nu_j-1} 
\delta(1-\sum_{i=1}^N x_i)\,\frac{{\cal U}^{N_{\nu}-(L+1) D/2}}
               {{\cal F}^{N_{\nu}-L D/2}}\;.\label{eq:FUsecdec}
\end{align}

The Euclidean region is defined as the kinematic region where all kinematic invariants $s_{ij}$ are negative  and all masses $m^2_i$ are positive or zero.
In this region the  polynomial $\mathcal{F}$ is a positive semi-definite function of the Feynman para\-meters.
The polynomial $\mathcal{U}$ does not depend on the kinematics and therefore is always positive semi-definite.
The graph polynomials can vanish when subsets of the Feynman
parameters go to zero. In the Euclidean region, such end-point
singularities are the only source of singularities (apart from overall
singularities residing in the prefactor composed of $\Gamma$-functions).

Zeroes of the $\mathcal{U}$-polynomial can lead to UV divergences,
which can be understood from the fact that increasing the dimension $D$ leads to a more negative power of $\mathcal{U}$.
A necessary (but not sufficient) condition for IR divergences is $\mathcal{F}=0$.
Both types of end-point singularities are regulated by the dimensional regularisation parameter $\eps$.

For non-Euclidean kinematics, Feynman integrals can develop
additional, kinematic dependent singularities, such as threshold
singularities, which are usually integrable. This will be discussed in
more detail in the context of contour deformation in Section
\ref{subsec:contour}.

\subsubsection*{The iterative decomposition algorithm}

The algorithm described below to achieve factorisation of the pole
structure in terms of monomials of Feynman parameters with negative
powers is the original one of Ref.~\cite{Binoth:2000ps}. 

\begin{enumerate}
\item{\bf Primary decomposition}


Before the decomposition is started, we eliminate one of the Feynman
parameters using the constraint $\delta(1-\sum_{l=1}^N x_l)$.
In the more general case of decomposing polynomials where the
integrand does not comprise
such a constraint, this step is absent.

The $\delta$-constraint  can be eliminated in two ways, either by the
so-called ``primary sector decomposition''  or by using the Cheng-Wu-theorem~\cite{Cheng:1987ga,Smirnov:2006ry}.
The latter will be discussed later in the context of the geometric
decomposition method.

For primary sector decomposition
 we decompose the integration over $N$ Feynman parameters into $N$ 
sectors, where in each sector $l$, $x_l$ is largest (note that 
the remaining $x_{j\not=l}$ are not  further ordered):
\begin{eqnarray}
\int_0^{\infty}d^N \vec{x} =
\sum\limits_{l=1}^{N} \int_0^{\infty}d^N \vec{x}
\prod\limits_{\stackrel{j=1}{j\ne l}}^{N}\theta(x_l- x_j)\;.
\end{eqnarray} 
The integral is now split into $N$ domains corresponding   
to $N$ integrals $G_l$ (from which we extract a common factor
containing $\Gamma$-functions for convenience),
\be
G=\frac{(-1)^{N_\nu}}{\prod_j\Gamma(\nu_j)} \Gamma(N_\nu-LD/2) \sum_{l=1}^{N} G_l\;.
\ee
 In the  integrals $G_l$ we substitute 
\begin{eqnarray}
x_j = \left\{ \begin{array}{lll} x_l t_j     & \mbox{for} & j<l \\
                                   x_l         & \mbox{for} & j=l \\
                                   x_l t_{j-1} & \mbox{for} & j>l \end{array}\right.
\end{eqnarray} 
and then  integrate out $x_l$ using the $\delta$-distribution.
As ${\cal U},{\cal F}$ are homogeneous of degree $L$,\,$L+1$, respectively,  
and  $x_l$ factorises completely, we have
${\cal U}(\vec x) \rightarrow {\cal U}_l(\vec t\,)\, x_l^L$ and
${\cal F}(\vec x) \rightarrow {\cal F}_l(\vec t\,)\, x_l^{L+1}$
and thus, using $\int dx_l/x_l\,\delta(1-x_l(1+\sum_{k=1}^{N-1}t_k ))=1$, we obtain 
\begin{align}\label{EQ:primary_sectors}
 G_l &= \int\limits_{0}^{1} \left(\prod_{j=1}^{N-1}{\rd}t_j\,t_j^{\nu_j-1}\right)\,\,
\frac{ {\cal U}_l^{N_\nu-(L+1)D/2}(\vec{t}\,)}{ {\cal F}_l^{N_\nu-L D/2}(\vec{t}\,)} 
\quad , \quad l=1,\dots, N \;.
\end{align} 
Note that the singular behaviour leading to $1/\epsilon$\,-poles
still comes  from regions where a set of parameters $\{t_i\}$ 
goes to zero. This feature would be lost 
 if  the $\delta$-distribution was integrated out 
in a different way, since this would produce poles at  upper 
limits of the parameter integral as well. 
The $N$ sectors generated this way
are called {\em primary} sectors.

\item{\bf Iterations}

Starting from Eq.~(\ref{EQ:primary_sectors}) we repeat the following 
steps until a complete separation of overlapping regions is achieved.
\begin{description}
\item[(i)] Determine a minimal set of parameters, say 
${\cal S}=\{t_{\alpha_1},\dots ,t_{\alpha_r}\}$, such that  
${\cal U}_l$, respectively  ${\cal F}_l$, vanish 
if the parameters of ${\cal S}$ are set to zero. Note that ${\cal S}$ is in general not unique.
\item[(ii)] Decompose the corresponding $r$-cube into $r$  subsectors
by defining a largest parameter in each subsector, not requiring complete ordering of all parameters:
\begin{eqnarray}
\prod\limits_{j=1}^r \theta(1-t_{\alpha_j}) \theta(t_{\alpha_j})=
\sum\limits_{k=1}^r \prod\limits_{\stackrel{j=1}{j\ne k}}^r 
\theta(t_{\alpha_k}-t_{\alpha_j})\theta(t_{\alpha_j})\theta(t_{\alpha_k})\;.
\end{eqnarray}
 \item[(iii)] Remap  the variables to the unit hypercube in each new 
 subsector by the substitution
 \begin{eqnarray}
t_{\alpha_j} \rightarrow 
\left\{ \begin{array}{lll} t_{\alpha_k} t_{\alpha_j} &\mbox{for}&j\not =k \\
                           t_{\alpha_k}              &\mbox{for}& j=k\,.  \end{array}\right.
\end{eqnarray}
This gives a Jacobian factor of $t_{\alpha_k}^{r-1}$. By construction
$t_{\alpha_k}$ factorises from at least one of the functions 
${\cal U}_l$, ${\cal F}_l$. The resulting subsector integrals have the 
general form
\begin{eqnarray}\label{eq:subsec_form}
G_{l} &=& \int\limits_{0}^{1} \left( \prod_{j=1}^{N-1}{\rd}t_j
\; t_j^{a_j-b_j\epsilon}  \right)
\frac{{\cal U}_{l}^{N_\nu-(L+1)D/2}}{{\cal F}_{l}^{N_\nu-LD/2}}\;.
\end{eqnarray}
\end{description}
For each subsector the above steps have to be repeated 
as long as a set ${\cal S}$ 
can be found such that one of the functions 
${\cal U}_{l}$ or ${\cal F}_{l}$ vanishes 
if the elements of ${\cal S}$ are set to zero. 
This way  new subsectors are created in each subsector 
of the previous iteration, resulting in a tree-like structure. 
The iteration stops if the functions 
${\cal U}_{l}$ or ${\cal F}_{l}$
contain a constant term, i.e. if they are of the  form
\begin{eqnarray}\label{EQ:subsec_UF}
{\cal U}_{l} &=& 1 +  u(\vec t\,) \\
{\cal F}_{l} &=& -s_{0} + 
\sum\limits_{\beta} (-s_{\beta}) f_\beta(\vec t\,)\;, \nonumber
\end{eqnarray}
where $u(\vec t\,)$  are polynomials in the
variables $t_j$, $f_\beta(\vec t\,)$ are monomials in $t_j$ and
$s_0, s_{\beta}$ denote kinematic invariants or masses (appearing
with positive sign).
Thus, after a certain number of iterations, the integral 
is represented as a sum of subsector integrals where the endpoint singularities are factored out.
The singular behaviour of the integrand  can be 
read off from the exponents $a_j$, $b_j$ of each subsector integral,
see Eq.~(\ref{eq:subsec_form}).
The above form, where the graph polynomials are of the form ``constant
plus some function of the Feynman parameters'', will be called
``standard form'' in the following.
In Euclidean space, this form of the graph polynomials will always be positive definite.

\item{\bf Extraction of the poles}

To make the pole structure manifest, we apply a subtraction procedure
which for logarithmic poles, i.e. $a_j=-1$ in
Eq.~(\ref{eq:subsec_form}), corresponds to the ``plus prescription''~\cite{Gelfand}. For each
integration parameter $t_j$ we use
\begin{align}
I_j=-\frac{1}{b_j\epsilon}\,{\cal I}_j(0,\{t_{i\not=j}\},\epsilon)+
\int\limits_0^1 dt_j\,t_j^{-1-b_j\epsilon}\,
\Big( {\cal I}(t_j,\{t_{i\not=j}\},\epsilon) -{\cal
  I}_j(0,\{t_{i\not=j}\},\epsilon)\Big)\;.
\end{align}
For poles which are higher than logarithmic ($a_j<-1$), the procedure
is analogous, however the $p$-th order of a Taylor series around
$t_j=0$ has to be subtracted:
\begin{align}
 &I_j = \sum\limits_{p=0}^{|a_j|-1} \frac{1}{a_j+p+1-b_j\epsilon}
 \frac{{\cal I}_j^{(p)}(0,\{t_{i\not=j}\},\epsilon)}{p!} 
+ \int\limits_{0}^{1} dt_j \, t_j^{a_j-b_j \epsilon} R(\vec{t},\epsilon) \;,\nn\\
 &R(\vec{t},\epsilon) =
{\cal I}(t_j,\{t_{i\not=j}\},\epsilon) -\sum\limits_{p=0}^{|a_j|-1}
{\cal I}_j^{(p)}(0)\frac{t_j^p}{p!} 
\;,\label{tjsubtr}\\
&{\cal I}_j^{(p)}(0)=
\frac{\partial^p}{\partial t_j^p} {\cal I}(t_j,\{t_{i\not=j}\},\epsilon)\Big|_{t_j=0}\;.\nn
\end{align}
By construction, the integral containing the remainder term 
$R(\vec{t},\epsilon)$ is finite in the limit $t_j\to 0$. 
If $a_j<-1$, for practical purposes it is advisable to use integration
by parts until the power is raised to $a_j=-1$ before performing the
subtractions as outlined above. This increases the stability of the
numerical integration and avoids large expressions due to higher orders in the Taylor expansion.

The procedure is repeated for all integration variables, such that after $N-1$ steps
all poles are made manifest and the remaining finite 
expressions can be expanded in $\epsilon$ to obtain
a Laurent series in $\epsilon$. 
Since each loop can contribute at most one soft and collinear 
$1/\epsilon^2$ term, the highest possible infrared pole of an $L-$loop graph  is $1/\epsilon^{2L}$. 
Expanding to order $\eps^r$, one obtains
\begin{align}\label{EQ:eps_series_Glk}
 G_{l} &= \sum\limits_{m=-r}^{2L} \frac{C_{lm}}{\epsilon^m} + {\cal O}(\epsilon^{r+1})\;.
\end{align}
The coefficients $C_{lm}$ are  $(N-1-m)$--dimensional 
integrals over parameters $t_j$ which are finite for $t_j\to 0.$
\end{enumerate}

\subsubsection*{The geometric decomposition algorithm}
The iterative decomposition algorithm described above suffers from the problem that in some cases an infinite recursion can occur.
On the other hand, as Feynman integrals (in integer dimensions) mathematically are objects known as {\em periods}~\cite{Bogner:2007mn}, it follows that decomposition strategies must exist which do terminate.
%
%
The first proof of this fact was presented in Ref.~\cite{Bogner:2007cr}, where the decomposition procedure was mapped to
the mathematical problem of Hironaka's polyhedra game~\cite{hironaka}. It
allowed to define strategies that are mathematically proven to
terminate, but usually produce a large number of sectors.
A related strategy was implemented in Ref.~\cite{Smirnov:2008py}.
A different approach based on convex geometry was introduced by Kaneko
and Ueda in Refs.~\cite{Ueda:2009xx,Kaneko:2009qx,Kaneko:2010kj}.
A variant of this algorithm which produces a low number of sectors has
been worked out by J.~Schlenk~\cite{Schlenk:2016cwf,Schlenk:2016a} and
is implemented in the programs \secdec-3~\cite{Borowka:2015mxa} and \pysecdec{}~\cite{Borowka:2017idc}.

In contrast to the iterative sector decomposition algorithm, where a primary
decomposition is performed, the Cheng-Wu
theorem~\cite{Cheng:1987ga,Smirnov:2006ry} is used to integrate out
the Dirac--$\delta$ in \eqn{eq:feynint}.
The Cheng-Wu theorem states that in Feynman integrals over the $x_i$ from
zero to infinity, the $\delta$-distribution
$\delta(1-\sum_{i=1}^N x_i)$  can be replaced by $\delta(1-H(x_i))$ where $H(x_i)$
is a homogeneous function with $H(\alpha x_i)=\alpha\,H(x_i)$, without
changing the value of the integral.
In the geometric decomposition method implemented in \pysecdec{}, the
Cheng-Wu theorem is used  to replace the $\delta$-distribution by $\delta(1-x_N)$.
The method starts from  Newton polytope $\Delta$ of the polynomial
$\mathcal{U}\cdot\mathcal{F}\cdot\mathcal{N}=\sum_{j=1}^m,
c_j\mathbf{x}^{\mathbf{v}_j}$, where $\mathcal{N}$ is a numerator
polynomial, and the multi-index notation
$\mathbf{x}^{\mathbf{v}_j}=\prod_ix_i^{(\mathbf{v}_j)_i}$ is used.
The Newton polytope is a convex polytope where each vertex corresponds to a term of the polynomial, so it is
the convex hull of the $(N-1)$-dimensional exponent vectors $\mathbf{v}_j$,
$\Delta=\text{ConvHull}(\mathbf{v}_1, \dots ,\mathbf{v}_m)$.
Transforming to facet variables and aiming at  simplicial cones provides a decomposition of the graph polynomials to standard form.


\subsubsection{Public programs}

\begin{figure}[htb]
  \centering
 \includegraphics[width=11.5cm]{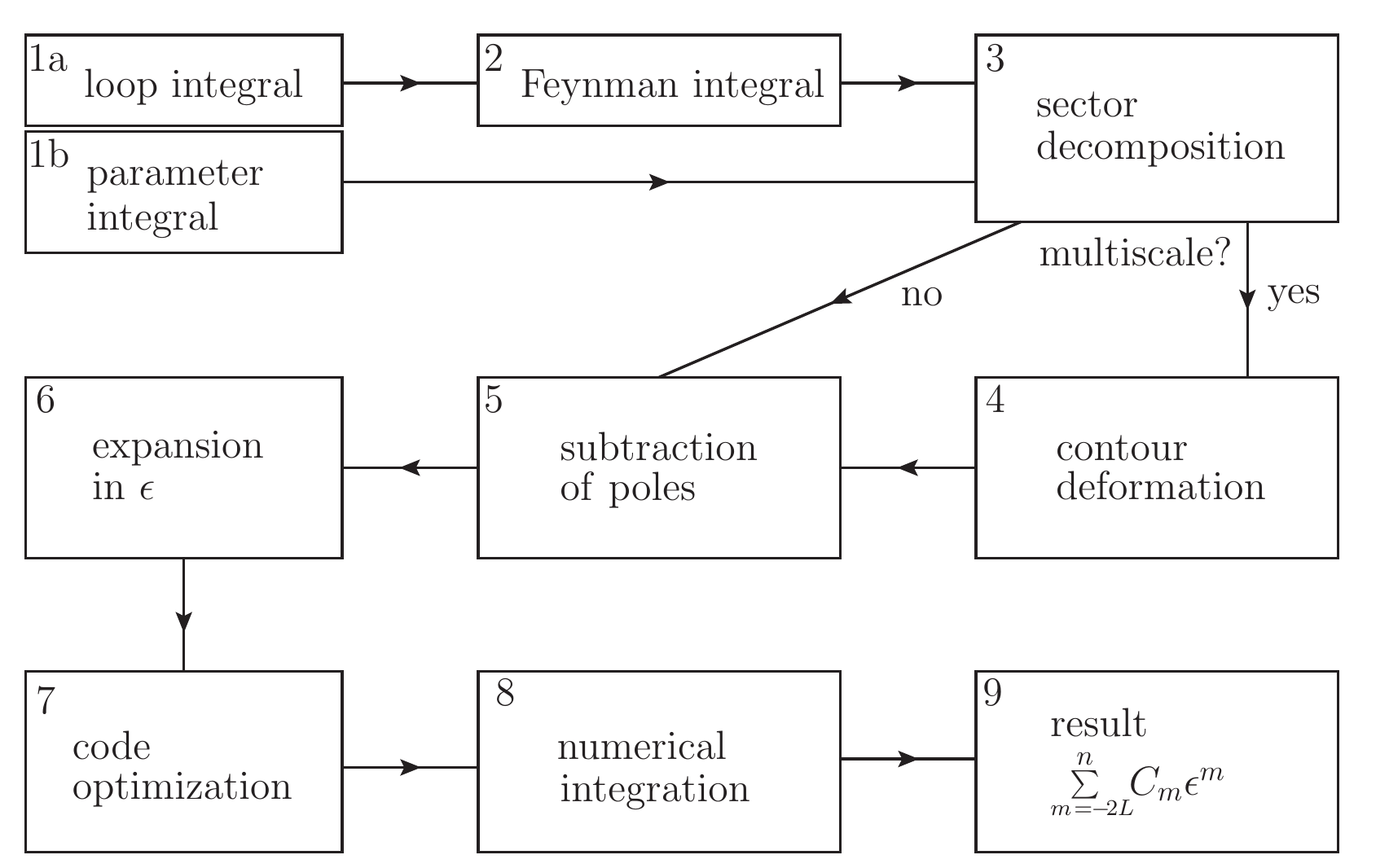}
\caption{Workflow of the program \pysecdec.\label{fig:secdec_workflow}}
\end{figure}

The first implementations of the sector decomposition algorithm into a
public code can be found in {\tt
  sector\_decomposition}~\cite{Bogner:2007cr} and early versions of {\sc
  Fiesta}~\cite{Smirnov:2008py, Smirnov:2009pb}  and \secdec~\cite{Carter:2010hi}.
However, these implementations were limited to the Euclidean region, where the integrand can only vanish at the endpoints of the Feynman parameter integrals.
The restriction to Euclidean kinematics was lifted by combining sector decomposition with a method to deform 
the multi-dimensional integration contour into the complex plane~\cite{Soper:1999xk,Binoth:2005ff,Beerli:2008zz, Borowka:2012yc, Borowka:2014aaa}.
The programs \secdec-2~\cite{Borowka:2012yc,Borowka:2013cma}, {\sc Fiesta3/4}~\cite{Smirnov:2013eza,Smirnov:2015mct}, \secdec-3~\cite{Borowka:2015mxa} and \pysecdec~\cite{Borowka:2017idc,Borowka:2018goh} are all capable of contour deformation and therefore can be used for non-Euclidean kinematics.
The workflow of the program \pysecdec{} is shown in Fig.~\ref{fig:secdec_workflow}.

\subsubsection{Contour deformation}\label{subsec:contour}

As mentioned above, the function ${\cal F}(\vec{x}, s_{ij},m_i^2)$ is
in general not of definite sign for physical values of kinematic
invariants.
The function  ${\cal F}$ can  vanish within the integration 
region on a hyper-surface given by  solutions of the Landau
equations~\cite{Landau:1959fi,ELOP,Collins:2020euz}, corresponding for example to physical thresholds.
The Landau equations however are more general, including also endpoint singularities. In momentum space, they can be formulated as follows. If the $N$ propagators are denoted by $P_i=q_i^2(\{k,p\})-m_i^2+i\delta$ and  $x_i$ are the Feynman parameters associated with propagator $P_i$, they read
\begin{align}
\begin{split}
x_i\,(q_i ^2(\{k,p\})-m_i^2)&=0\quad \forall\, i\in \{1,\ldots,N\}\\
\frac{\partial}{\partial k_l^\mu}\sum_{i\in {\rm loop}\, l}\, x_i\,(q_i ^2(\{k,p\})-m_i^2)&=0\quad \forall\, l \in \{1,\ldots,L\}\;.
\end{split}
\end{align}
The Landau equations are necessary, but in general not sufficient conditions for a singularity to be produced.
The first condition contains endpoint singularities ($x_i=0$) as well
as kinematic singularities, related to a propagator  going on-shell, ($q_i^2=m_i^2$).
The second condition entails that the singularities in the complex plane trap the integration contour, such that the contour cannot be deformed away from the singularity. Therefore this situation is also called ``pinch singularity''.
In Feynman parameter space the Landau equations translate to 
\begin{align}
  \begin{split}
&  {\cal F} = 0 \mbox{ and }\\
&\left(\mbox{ either } x_i = 0 \;
\mbox{ or } \frac{\partial}{\partial x_i} {\cal F} = 0\; \right)\,\forall i\;\;.
\end{split}
\end{align}
A singularity with all $x_i\not=0$ is called {\em leading Landau singularity}. 
%
\begin{figure}[h]
  \begin{center}
    \includegraphics[width=0.4\textwidth]{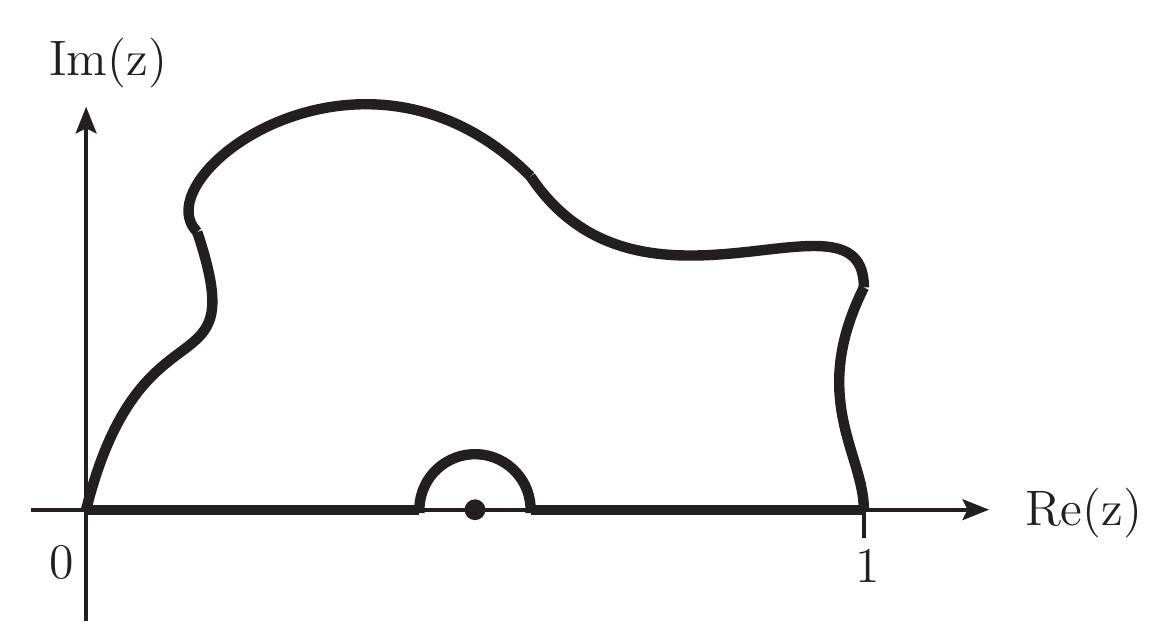}
  \end{center}
	\caption{Schematic picture of the closed contour avoiding poles on the real axis.}
	\label{fig:scenario1}  
\end{figure}
The solutions of the Landau equations corresponding to thresholds  are integrable
singularities and we can make use of Cauchy's theorem to
avoid the non-physical poles on the real axis by a deformation of the integration contour 
into the complex plane, see Fig.~\ref{fig:scenario1}.
A valid deformation must be in accordance with the causal $i\delta$-prescription 
for Feynman propagators, and no poles should be enclosed in the integration
path.
Using the fact that the integral over the closed contour $c$ 
 is zero, we have 
\begin{align}
  0&=\oint_c  \prod\limits_{j=1}^{N}\rd z_j {\cal I}(\vec{z}) 
  =  \int_0^1 \prod\limits_{j=1}^{N}\rd x_j \,{\cal I}(\vec{x})+\int_\gamma \prod\limits_{j=1}^{N}\rd z_j \,{\cal I}(\vec{z}(\vec{x})) \;,
    \label{eq:contour}
\end{align}
where in the second line we have split the contour $c$ into an
integral along the real axis from zero to one and a path $\gamma$
closing the contour. The path $\gamma$ is parametrised by $\vec{z}(\vec{x})$
and fulfils $\vec{z}(0)=\vec{z}(1)=0$.
The integral from zero to one in Eq.~(\ref{eq:contour}) is the
original Feynman integral, which now is written as
\begin{align}
\int_0^1 \prod\limits_{j=1}^{N}\rd x_j &=-\int_\gamma \prod\limits_{j=1}^{N}\rd z_j {\cal I}(\vec{z}) 
=\int_\gamma \prod\limits_{j=1}^{N}\rd x_j\left\vert \left(\frac{\partial z_k(\vec{x})}{\partial x_j}\right)\right\vert {\cal
I}(\vec{z}(\vec{x}))\;,
\end{align}
where $\left\vert \left(\frac{\partial z_k(\vec{x})}{\partial x_j}\right)\right\vert$ denotes the Jacobian of the transformation $\vec{x}\to \vec{z}(\vec{x})$.
The $i\delta$-prescription for the Feynman propagators, resulting in a term $-i\delta$ in eq.~(\ref{eq:F}),
tells us that the contour deformation into the complex plane should be such that 
the imaginary part of ${\cal F}$ is negative. 
For real masses and Mandelstam invariants $s_{ij}$, the following Ansatz\,\cite{Soper:1999xk,Nagy:2006xy,Binoth:2005ff,Beerli:2008zz,Borowka:2012yc,Borowka:2014aaa}
is therefore convenient:
\begin{align}
\label{eq:condef}
z_k( \vec x) &= x_k- i\;  \tau_k(\vec{x})\;,\;
\tau_k = \lambda\, x_k (1-x_k)
 \, \frac{\partial {\cal F}(\vec{x})}{\partial x_k}  \;.
\end{align}
The real parameter $\lambda$ should be small and positive and can be
used to tune the integration contour for the numerical integration.
In \pysecdec{} there is an individual parameter $\lambda_k$ for each
$x_k$.
Unless we are faced with a leading Landau singularity, 
the deformation leads to a well behaved integral at the points on the
real axis where the function ${\cal F}$ vanishes.
The validity of \eqn{eq:contour}, starting from a closed integration
contour, is guaranteed by the factors $x_k$ and $(1-x_k)$, which keep the endpoints fixed. 
In terms of the new variables, we thus obtain, expanding around
$\lambda=0$, 
\begin{equation}
\label{eq:newF}
{\cal F}(\vec{z}(\vec{x}))={\cal F}(\vec{x})
-i\,\lambda\,\sum\limits_{j} \, x_j (1-x_j)\, \left(\frac{\partial {\cal F}}{\partial x_j}  \right)^2 + {\cal O}(\lambda^2)\;,
\end{equation}
such that ${\cal F}$ acquires a negative imaginary part of order
$\lambda$. For large values of $\lambda$, higher order terms in the
above expansion can change the sign of the imaginary part. The program
\pysecdec{} checks the sign of the imaginary part and stops with an
error message if such a situation occurs, which usually  can be cured by
decreasing the value of $\lambda$. Note that the transformation
(\ref{eq:condef}) also introduces a kinematic dependence into the
${\cal U}$-polynomial through the derivative of ${\cal F}$.
Therefore the program \pysecdec{} also checks whether $Re({\cal U})\geq 0$ is fulfilled after the contour deformation.

\subsubsection{Complex masses}

Calculating radiative corrections involving unstable particles, the widths of the particles need to be included.
A consistent treatment of the width in combination with NLO
calculations is provided by the complex-mass scheme~\cite{Denner:1999gp,Denner:2005fg}, 
where the width $\Gamma$ is included as a negative imaginary part of the mass via the replacement
\begin{align}
m^2 \to m_c^2 \equiv m^2\left(1-i\frac{\Gamma}{m}\right).
\end{align}
The graph polynomial $\mathcal{F}$ then has the form
\begin{align}
\mathcal{F} = \mathcal{F}_0 + \mathcal{U} \sum_j x_j m_j^2 \left(1- i
  \,\frac{\Gamma_j }{m_j} \right),
  \label{eq:ImF}
\end{align}
i.e. the widths induce a negative imaginary part,
\begin{align}
\imag\mathcal{F} = -\,\mathcal{U} \sum_j x_j  m_j \Gamma_j\;.
\end{align}
For zero widths, the zeroes of $\mathcal{F}$ on the real
axis are avoided by a suitable deformation of the integration contour
as described above.
A non-zero width can help to avoid these singular regions as well, but
this does not lead to a stable numerical integration in all cases, in
particular if the width is small.
Therefore it is advisable to combine the two in a consistent way, which is possible since both the contour deformation and the complex masses are required to produce only negative imaginary parts in order to fulfil the Feynman $i\delta$-prescription.
In \pysecdec{} we use the following generalisation of Eq.~(\ref{eq:condef}):
\begin{subequations}
\begin{align}
\vec{z}(\vec{x}) &= \vec{x} - i \vec\tau(\vec{x})\;,\;
\tau_k = \lambda x_k (1-x_k)  \frac{\partial\mathcal{\real F}}{\partial x_k},
\end{align}
\end{subequations}
which means the widths do not enter the definition of the deformed contour.
For small deformations we then have
\begin{align}
 & \mathcal{F}(\vec{z}(\vec{x})) = \real\mathcal{F}(\vec{x}) + i \imag\mathcal{F}(\vec{x})\nn\\
&-i\lambda \sum_k x_k(1-x_k) \left[\left(\frac{\partial\mathcal{\real F}}{\partial x_k}\right)^2
 + i \frac{\partial\real\mathcal{F}}{\partial x_k}\frac{\partial\imag\mathcal{F}}{\partial x_k}\right]
\nonumber\\
&- \frac{\lambda^2}{2} \sum_{k,l} x_k(1-x_k) x_l (1-x_l)
\frac{\partial\mathcal{\real F}}{\partial x_k} \frac{\partial\mathcal{\real F}}{\partial x_l} 
 \left[ 
\frac{\partial^2\real\mathcal{F}}{\partial x_k \partial x_l} 
+i\frac{\partial^2\imag\mathcal{F}}{\partial x_k \partial x_l} 
\right]
+ \mathcal{O}(\lambda^3).
\end{align}
Up to order $\lambda$, the imaginary parts induced by the widths and the contour deformation are both negative
as they should. The term involving
$\frac{\partial\imag\mathcal{F}}{\partial x_k}$ does not change the
sign of the imaginary part because it is purely real.
At order $\lambda^2$, however, $\imag\mathcal{F}$ leads to an
imaginary part of indefinite sign, which for real masses
would  have been the case at one order higher in $\lambda$.
On the other hand, this term is proportional to $\Gamma_j/m_j$
compared to the terms $\sim m_j^2$, see Eq.~(\ref{eq:ImF}), and thus suppressed
since the widths should be small compared to the corresponding masses.
Therefore we conclude that for a sufficiently small value of $\lambda$, one can consistently combine complex masses and contour deformation.


\subsubsection{Numerical integration}
\label{sec:qmc}

The  finite integrals which are produced by the sector decomposition algorithm are usually too complicated to be integrated analytically.
However, they are usually well suited for a numerical integration because by construction, they are polynomials that start with a constant term.
Nonetheless, for integrals with a complicated infrared singularity structure,
the subtractions can make the integrand less well behaved with regards
to numerical integration.
Therefore, if \pysecdec{} is used to calculate an amplitude where the basis of master integrals can be changed,
it is advisable to use a so-called quasi-finite basis~\cite{vonManteuffel:2014qoa,vonManteuffel:2015gxa}.
Such a basis contains all poles in the regulator $\eps$ only in the prefactors of the integrals, while the integrand itself is such that no further $1/\eps\,$-poles arise, 
and therefore no subtractions in the integrand are necessary.

As the integrals where sector decomposition is useful are mostly integrals over a relatively large number of Feynman parameters,
Monte Carlo integration has the advantage of the error estimate being independent of the number of dimensions (i.e. integration variables),
which is not the case for deterministic integration rules.
One of the most widely used tools for numerical integration is the 
{\sc Cuba} package~\cite{Hahn:2004fe,Hahn:2014fua}, which implements several
numerical integration routines relying on pseudo-random sampling, quasi-random sampling or
cubature rules.

Monte Carlo errors scale like $1/\sqrt{n}$ (for square-integrable functions and a large number of samples $n$).
In numerical mathematics, it has been known for some time already that  Quasi-Monte-Carlo (QMC) methods have a more favourable error scaling behaviour if the integrand functions fulfill certain requirements, see e.g. Refs.~\cite{QMCActaNumerica,KuoNuyensPractical}.
Recently it was realised that this feature can also be exploited for Feynman integrals~\cite{Li:2015foa,deDoncker:2018nqe,De_Doncker_2018}.

Let us consider an integral over the $d$-dimensional hypercube of an integrand function $f(\vec{x})\equiv  f(\mathbf{x})$, 
\begin{equation}
I_d[ f ] \equiv \int_{[0,1]^d} \mathrm{d} \mathbf{x} \ f(\mathbf{x}) \equiv \int_0^1 \mathrm{d} x_1 \cdots \mathrm{d} x_d \ f(x_1,\ldots,x_d)\;.
\end{equation}
Monte Carlo integration approximates the integral by the estimator
\begin{align}
  Q_{n,d}[f]=\frac{1}{n} \sum_{i=0}^{n-1} f(\vec{t}_i)\;,
  \label{eq:MC}
\end{align}
where the in the standard Monte Carlo case the $\vec{t}_i$ are points in the $d$-dimensional hypercube which are randomly chosen.
In this case the error estimate for large $n$ is given by $\sigma(f)/\sqrt{n}$ where $\sigma(f)^2$ is the variance of $f$.
Quasi-Monte Carlo methods also use an estimator of the form (\ref{eq:MC}), however the $\vec{t}_i$ are chosen deterministically.
There are various possibilities to define the $\vec{t}_i$.
For applications to Feynman integrals, Rank-1 Shifted Lattice (R1SL) rules seem to have particularly good convergence properties.
A rank-1 lattice is defined by a generating vector $\mathbf{z} \in \mathbb{Z}^d$ which is a fixed $d$-dimensional vector of integers coprime to $n$.
The lattice is shifted $m$ times by vectors $\boldsymbol{\Delta}_k \in [0,1)^d$, which are $d$-dimensional vectors consisting of independent, uniformly distributed random real numbers in the interval $[0,1)$. Therefore the estimator is given by
\begin{align}
&I_d[f] \approx \bar{Q}_{n,m,d}[f] \equiv  \frac{1}{m} \sum_{k=0}^{m-1} Q_{n,d}^{(k)}[f], &
&Q_{n}^{(k)}[f] \equiv \frac{1}{n} \sum_{i=0}^{n-1} f \left( \left\{ \frac{i \mathbf{z}}{n} + \boldsymbol{\Delta}_k \right\} \right) \;,\label{eq:lattice}
\end{align}
where $\{\}$ denotes ``mod 1'', such that all arguments of $f$ remain in the $d$-dimensional unit hypercube.
The estimate of the integral depends on the number of lattice points $n$ and the number of random shifts $m$.
For certain classes of 1-periodic functions and generating vectors
a linear error scaling $\sim 1/n$  can be achieved~\cite{QMCActaNumerica}.
The scaling in the number of lattice shifts, as they are random, is ${\cal O}(1/\sqrt{m})$. Therefore it is advisable to
choose a relatively small number of shifts $m$ and a large number of lattice points $n$.
An efficient algorithm to construct generating vectors is the component-by-component construction~\cite{nuyens2006fast}, where a
generating vector in $d$ dimensions is iteratively obtained from a $(d-1)$-dimensional
one by selecting the additional component such that the worst-case error is minimal.
Generating vectors obtained in this way are optimised for a fixed lattice size $n$, therefore in Ref.~\cite{Borowka:2018goh}
each lattice is associated with its individual optimal generating vector.

The integrands as provided by the sector decomposition algorithm are in general not periodic. They
can be periodised by a suitable change of variables
$\mathbf{x}=\phi(\mathbf{u})$,
\begin{equation}
I[ f ] \equiv \int_{[0,1]^d} \mathrm{d} \mathbf{x} \ f(\mathbf{x}) = \int_{[0,1]^d} \mathrm{d} \mathbf{u} \ \omega_d(\mathbf{u}) f(\phi({\mathbf{u})})\;,
\end{equation}
where 
\begin{align} 
&\phi(\mathbf{u})= (\phi(u_1),\ldots,\phi(u_d)), & &  \omega_d(\mathbf{u}) =  \prod_{j=1}^d \omega(u_j) & & \mathrm{and} & & \omega(u) = \phi^\prime(u)\;.
\end{align}
In practice, the periodising transform may be specified in terms of the weight function $\omega$, such that the change of variables is given by
\begin{equation}
\phi(u) \equiv \int_0^u \mathrm{d} t \ \omega(t)\;,
\end{equation}
 as for example given by the Korobov transform~\cite{Korobov1963,LAURIE1996337,Kuo2007},
\begin{align}
&\omega_{r_0,r_1}(u)= \frac{u^{r_0}(1-u)^{r_1}}{\int_0^1 \mathrm{d} t \  t^{r_0} (1-t)^{r_1}} = (r_0+r_1+1) \binom{r_0+r_1}{r_0} u^{r_0} (1-u)^{r_1}, \label{eq:korobov}
\end{align}
The weight parameters $r_0,r_1$ are usually chosen to be equal. 
%
Another useful transformation is the baker's transform~\cite{baker_trafo} (also called ``tent transform''), given by
\begin{align}
\phi(u) = 1- \big| 2u-1\big|=&\left\{
\begin{tabular}{ll}
$2u$ & $\mbox{ if } u\leq \frac{1}{2},$\\
$2-2u$ & $\mbox{ if } u> \frac{1}{2}\;.$
\end{tabular}
\right.                   \label{eq:baker}           
\end{align}
This transform periodises the integrand by mirroring it rather than forcing it to a particular value on the integration boundary.

The latest version of \pysecdec{}~\cite{Borowka:2018goh} provides a public implementation of a R1SL QMC-integrator~\cite{qmc}.
In addition, the implementation is capable of performing numerical
integration on multiple {\sc CUDA} compatible Graphics Processing Units (GPUs),
which can accelerate the evaluation of the integrand significantly.

\section{Real radiation: Schemes to isolate IR divergences}
\label{sec:IRsubtraction}

At NLO, several schemes exist to subtract IR divergent real
radiation. The most widely used ones are the Catani-Seymour (CS) dipole
subtraction scheme~\cite{Catani:1996vz,Catani:2002hc} and the
Frixione-Kunszt-Signer (FKS) subtraction
scheme~\cite{Frixione:1995ms,Frixione:1997np}. Another scheme, particularly developed in view of matching to parton showers, is the Nagy-Soper subtraction scheme~\cite{Nagy:2003qn,Nagy:2007ty,Chung:2010fx,Chung:2012rq,Bevilacqua:2013iha}.
The dipole subtraction scheme has been automated in Refs.~\cite{Gleisberg:2007md,Frederix:2008hu,Hasegawa:2009tx,Bellm:2015jjp}, the FKS scheme in Refs.~\cite{Frederix:2009yq,Alioli:2010xd}.

In contrast, at NNLO and beyond, the development of efficient and general schemes to isolate IR singularities in real emissions
is a field which is far from being settled, even though it has seen enormous progress in
the last years.
We can roughly classify these methods into two categories: (i) methods
based on subtraction, and (ii) methods based on partitions of the phase
space into IR-sensitive regions and hard regions, sometimes also called ``slicing methods''.
Subtraction methods aim at a local subtraction of the IR singular
structures,  i.e. a cancellation of singularities pointwise in phase
space, while in slicing methods the cancellation of the
contributions from the IR
sensitive and the hard regions is intrinsically non-local.
Slicing methods which have been applied successfully at NNLO are the $q_T$-method~\cite{Catani:2007vq}
and $N$-jettiness~\cite{Stewart:2010tn,Boughezal:2015dva,Gaunt:2015pea}.

Controlling large cancellations is an issue within the slicing methods, however
power corrections in the resolution variable can be calculated in an
expansion around the singular limits, such that  the numerical
performance of the subtractions can be systematically improved.
Next-to-leading  power  corrections have  been computed at NLO and
NNLO~\cite{Moult:2016fqy,Boughezal:2016zws,Moult:2017jsg,Boughezal:2018mvf,Ebert:2018lzn,Cieri:2019tfv,Boughezal:2019ggi}.
The $q_T$-method extended to N$^3$LO already has been applied to the calculation of differential Higgs boson production at N$^3$LO~\cite{Cieri:2018oms} and Higgs boson pair production at  N$^3$LO~\cite{Chen:2019lzz,Chen:2019fhs}.
Extensions of the $N$-jettiness and $q_T$-methods to N$^3$LO are under
development~\cite{Melnikov:2018jxb,Melnikov:2019pdm,Billis:2019vxg,Behring:2019quf,Luo:2019szz,Baranowski:2020xlp,Ebert:2020dfc,Chen:2020uvt,Ebert:2020yqt,Ebert:2020unb,Ebert:2020lxs}
and led to a recent highlight in the field of precision
calculations, represented by predictions for the gluon-fusion Higgs $q_T$ spectrum at N$^3$LL$^\prime$+N$^3$LO  including fiducial cuts~\cite{Billis:2021ecs}.
Extensions of subtraction methods based on momentum mappings beyond NNLO are also under investigation~\cite{DelDuca:2019ctm}.


The $q_T$-method~\cite{Catani:2007vq}, originally limited
to final states which do not carry colour, recently has been extended
to allow the calculation of top quark pair production at
NNLO~\cite{Bonciani:2015sha,Catani:2019iny,Catani:2019hip}, as well as
bottom-quark pair production~\cite{Catani:2020kkl} and
the flavour off-diagonal channels of $t\bar{t}H$ production at
NNLO~\cite{Catani:2021cbl}.
The $q_T$-method also has been used recently for a $2\to 3$ process
calculated at NNLO, triphoton production~\cite{Kallweit:2020gcp}.

The first combined QCD and QED  $q_T$-resummation up to NLL+NLO in QED and NNLL+NNLO in QCD
has been presented in Ref.~\cite{Cieri:2018sfk}. 
In Ref. ~\cite{Buonocore:2019puv} the $q_T$-formalism is applied to
the computation of the electroweak corrections to massive lepton pairs
through the Drell-Yan mechanism and used
as a case study to investigate the power suppressed contributions in
the parameter $R_{\mathrm cut}$.
The complete ${\cal O}(\alpha_s\alpha)$ mixed QED\,$\otimes$\,QCD to
off-shell $Z$-boson production have been calculated based on the
$q_T$-formalism in Ref.~\cite{Cieri:2020ikq}, for the case of an
on-shell $Z$-boson these corrections also have been calculated in
Ref.~\cite{Delto:2019ewv} using the nested soft-collinear subtraction method~\cite{Caola:2017dug}.

Antenna subtraction~\cite{GehrmannDeRidder:2005cm,Currie:2013vh} led
to the first NNLO calculation of $e^+e^-\to
3$\,jets~\cite{Gehrmann-DeRidder:2007nzq},  of single-jet Inclusive production~\cite{Currie:2016bfm},
of di-jet production~\cite{Currie:2017eqf} in
the leading colour approximation, as well as the production of dijet
final states in deep inelastic scattering~\cite{Currie:2016ytq} as well as a number of processes involving vector bosons and jets, see Table~\ref{tab:processes}.

Sector-improved residue subtraction~\cite{Czakon:2010td,Czakon:2011ve,Boughezal:2011jf,Czakon:2014oma} has been inspired by sector
decomposition applied to phase space
integrals~\cite{Heinrich:2002rc,GehrmannDeRidder:2003bm,Anastasiou:2003gr,Anastasiou:2004qd,Binoth:2004jv,Anastasiou:2004xq,Anastasiou:2005qj,Melnikov:2006di,Melnikov:2006kv}
and FKS subtraction~\cite{Frixione:1995ms, Frederix:2009yq}. It has led
to the first calculation of $t\bar{t}$ production at
NNLO~\cite{Czakon:2013goa}, as well as the NNLO calculation of single-jet inclusive rates with
exact color~\cite{Czakon:2019tmo}.
The latter calculation confirmed that the difference due to approximation applied in Ref.~\cite{Currie:2016bfm}, i.e. using the
leading-colour approximation in the case of channels involving quarks and exact colour
in the all-gluon channel, is negligible for phenomenological applications.
The calculation of the NNLO QCD corrections to 3-photon production at the LHC, neglecting colour-suppressed non-planar contributions, has also been achieved based on sector-improved residue subtraction for the double real radiation contributions~\cite{Chawdhry:2019bji}.

Nested soft-collinear subtraction~\cite{Caola:2017dug} has elements of the sector-improved residue subtraction, for example the feature to partition the phase space, however it provides analytic results for the integrated subtraction terms~\cite{Caola:2018pxp,Caola:2019nzf,Caola:2019pfz,Delto:2019asp,Asteriadis:2019dte,Bizon:2020tzr} and aims at a generic local subtraction scheme providing the basic building blocks for the subtraction terms similar to the CS-dipole subtraction scheme at NLO.

Another subtraction scheme aiming at fully local and analytic subtractions is the one developed in Refs.~\cite{Magnea:2018hab,Magnea:2018ebr}.
It provides definitions for local soft and collinear counterterms written in terms of gauge invariant matrix elements of fields and Wilson lines,
and aims at a scheme valid to all orders in perturbation theory.

The ``projection-to-Born''
method~\cite{Han:1992hr,Brucherseifer:2014ama,Cacciari:2015jma,Dreyer:2016oyx},
also called ``structure function approximation'', is inspired by
VBF-type kinematics.
The VBF process can be considered as being formed by two quark currents, 
connected through a vertex involving weak bosons and/or Higgs bosons.
The QCD corrections to each of the two quark currents can be considered separately, mapping the momenta such that
the momentum of the produced boson is unaffected, i.e. has Born-like kinematics.
This factorisation into corrections to two ``structure functions''
relies on the fact that exchanges of coloured particles between the
quark lines are both kinematically and colour suppressed.
In Ref.~\cite{Chen:2021isd}, the projection-to-Born method has been extended to
the production of an arbitrary colour-neutral final state in
hadronic collisions, to achieve the calculation of N$^3$LO predictions
for fiducial distributions for the two-photon final state from
gluon-fusion Higgs boson decay.

In Ref.~\cite{Herzog:2018ily} a geometric method for final state radiation, employing a FKS-like residue subtraction procedure based on a Feynman diagram dependent slicing observable.
In Ref.~\cite{Ma:2019hjq}, a forest formula  in momentum space to subtract IR singularities for wide-angle scattering has been developed.

 A summary of various schemes to treat unresolved real radiation at NNLO is given in Table~\ref{tab:realsub}, processes calculated with the
 different methods are collected in Table~\ref{tab:processes}. For $H$, $H+$jet and $HH$ production in gluon fusion the heavy top limit is understood for the highest loop order entering the calculation.
Partial inclusions of heavy quark mass effects in some of the results listed  in Table~\ref{tab:processes} are discussed in Section~\ref{sec:Higgspheno}.
\begin{table}[htb]
  \centering
\begin{tabular}{|l|l|}
\hline\hline
method &  analytic integration \\
             &of subtraction terms \\
\hline\hline
subtraction&\\
\hline
antenna subtraction~\cite{GehrmannDeRidder:2005cm,GehrmannDeRidder:2005aw,GehrmannDeRidder:2005hi,Daleo:2006xa,Daleo:2009yj,Gehrmann:2011wi,Boughezal:2010mc,GehrmannDeRidder:2012ja,Currie:2013vh}& yes\\
sector-improved residue subtraction~\cite{Czakon:2010td,Czakon:2011ve,Boughezal:2011jf,Czakon:2014oma}&no\\
nested soft-collinear subtraction~\cite{Caola:2017dug,Caola:2018pxp,Caola:2019nzf,Caola:2019pfz,Delto:2019asp,Asteriadis:2019dte,Bizon:2020tzr}&yes\\
ColorFulNNLO~\cite{Somogyi:2005xz,Somogyi:2006da,Somogyi:2006db,Somogyi:2008fc,Aglietti:2008fe,Somogyi:2009ri,Bolzoni:2009ye,Bolzoni:2010bt,DelDuca:2013kw,Somogyi:2013yk}&partial
\\
projection to Born~\cite{Han:1992hr,Brucherseifer:2014ama,Cacciari:2015jma,Dreyer:2016oyx}
&yes\\
local analytic subtraction~\cite{Magnea:2018hab,Magnea:2018ebr}&
yes\\
\hline\hline
slicing&\\
\hline
$q_T$~\cite{Catani:2007vq,Catani:2009sm}&yes \\
N-jettiness~\cite{Stewart:2010tn,Boughezal:2015dva,Gaunt:2015pea}& yes\\
geometric subtraction~\cite{Herzog:2018ily} &yes \\
\hline
\end{tabular}
\caption{\label{tab:realsub}Methods for the isolation of IR divergent real
  radiation at NNLO.}
\end{table}

\begin{table}[htb]
\vspace*{-2cm}
\hspace*{-1.7cm}
\begin{tabular}{|l|l|}
\hline\hline
method &  processes (NNLO unless stated otherwise)\\
  \hline\hline
  antenna subtraction&$e^+e^-\to 3$\,jets~\cite{Gehrmann-DeRidder:2007nzq,GehrmannDeRidder:2007hr,Weinzierl:2008iv,Weinzierl:2009ms,Gehrmann:2017xfb}, $q\bar{q}\to t\bar{t}$~\cite{Abelof:2015lna}, $pp\to H+$jet~\cite{Chen:2014gva,Chen:2016zka,Chen:2019wxf}, \\
       &$pp\to $\,jet+X~\cite{Currie:2016bfm,Currie:2017ctp,Currie:2018xkj}, dijets~\cite{Currie:2017eqf,Gehrmann-DeRidder:2019ibf},\\
   &$pp\to Z+$jet~\cite{Ridder:2015dxa,Ridder:2016nkl}, $pp\to \gamma+$jet~\cite{Currie:2016ytq,Currie:2017tpe,Chen:2019zmr}, $pp\to Z+$b-jet~\cite{Gauld:2020deh},\\
       &DIS dijet~\cite{Currie:2016ytq}, DIS jet+X~\cite{Abelof:2016pby,Currie:2017tpe}, DIS 2j diffractive~\cite{Britzger:2018zvv}, \\
  & DIS charged current~\cite{Niehues:2018was}, DIS event shapes~\cite{Gehrmann:2019hwf},\\
  &VBF Higgs production~\cite{Cruz-Martinez:2018rod},  $pp\to VH (b\bar{b})$~\cite{Gauld:2019yng}\\
 \hline
  sector-improved residue subtraction&$pp\to t\bar{t}$~\cite{Czakon:2013goa,Czakon:2014xsa,Czakon:2016ckf,Czakon:2017wor,Czakon:2020qbd}, $pp\to H+$jet~\cite{Boughezal:2013uia,Boughezal:2015dra}, top decay~\cite{Brucherseifer:2013iv},\\
       & $pp\to$\,jet+X~\cite{Czakon:2019tmo}, $pp\to 3\,\gamma$~\cite{Chawdhry:2019bji}\\
  \hline
  nested soft-collinear subtraction&Drell-Yan~\cite{Caola:2017dug}, $pp\to WH (H\to b\bar{b})$~\cite{Caola:2019nzf}, DIS~\cite{Asteriadis:2019dte},\\
       &$pp\to Z,pp\to H$~\cite{Caola:2019nzf}, $H\to gg, H\to b\bar{b}$~\cite{Caola:2019pfz}, $H\to b\bar{b}, m_b\not=0$~\cite{Behring:2019oci}\\
  \hline
  ColorFulNNLO&$e^+e^-\to 3$\,jets~\cite{DelDuca:2016ily}, $H\to b\bar{b}$~\cite{DelDuca:2015zqa,Somogyi:2020mmk}\\
  \hline
  projection to Born&single top \cite{Brucherseifer:2014ama}, VBF Higgs production~\cite{Cacciari:2015jma,Dreyer:2020urf}, \\
  &VBF Higgs pair production~\cite{Dreyer:2018rfu,Dreyer:2020xaj},\\
       &N$^3$LO VBF Higgs production~\cite{Dreyer:2016oyx},  \\
       &N$^3$LO VBF $HH$~\cite{Dreyer:2018qbw}, N$^{3}$LO  jet production in DIS~\cite{Currie:2018fgr},\\
  &charged current DIS N$^3$LO~\cite{Gehrmann:2018odt}, $pp\to H\to \gamma\gamma$ N$^3$LO ~\cite{Chen:2021isd}\\
  \hline
  local analytic subtraction&Drell-Yan~\cite{Magnea:2018hab,Magnea:2018ebr}\\
  \hline
  $q_T$&$pp\to H$~\cite{Catani:2007vq}, Drell-Yan~\cite{Catani:2009sm}, $pp\to\gamma\gamma$~\cite{Catani:2011qz},  $pp\to\gamma\gamma\gamma$~\cite{Kallweit:2020gcp},\\
       &$pp\to HH$~\cite{deFlorian:2016uhr,Grazzini:2018bsd}, $pp\to ZH$~\cite{Ferrera:2014lca}, $pp\to WH$~\cite{Ferrera:2013yga}, \\
  &processes in {\sc Matrix}~\cite{Grazzini:2017mhc}: $pp\to V, pp\to H$, $pp\to Z\gamma$~\cite{Grazzini:2013bna}, \\
  &$pp\to W\gamma$~\cite{Grazzini:2015nwa}, $pp\to ZZ$~\cite{Gehrmann:2015ora,vonManteuffel:2015msa,Cascioli:2014yka,Grazzini:2015hta,Kallweit:2018nyv}, \\
       &$pp\to WW$~\cite{Gehrmann:2014fva,Grazzini:2016ctr,Kallweit:2019zez}, $pp\to WZ$~\cite{Grazzini:2016swo,Grazzini:2017ckn}; \\
  &$pp\to WHH$~\cite{Li:2016nrr}, $pp\to ZHH$~\cite{Li:2017lbf}, \\
  &  $pp\to t\bar{t}$~\cite{Catani:2019iny,Catani:2019hip,Catani:2020tko}, $pp\to b\bar{b}$~\cite{Catani:2020kkl},\\
  &$pp\to H$ N$^3$LO~\cite{Cieri:2018oms}, $pp\to HH$ N$^3$LO~\cite{Chen:2019lzz,Chen:2019fhs}\\
  \hline
  N-jettiness&top decay~\cite{Gao:2012ja}, $pp\to W+$jet~\cite{Boughezal:2015dva}, $pp\to H$~\cite{Gaunt:2015pea}, single top~\cite{Berger:2016oht}\\
       &$pp\to H+$jet~\cite{Boughezal:2015aha,Campbell:2019gmd}, $pp\to W$+jet~\cite{Boughezal:2015dva,Boughezal:2016dtm}, $pp\to ZZ$~\cite{Heinrich:2017bvg}\\
   &processes in {\sc Mcfm}: $pp\to Z,W,H,HZ,HW,\gamma\gamma$~\cite{Boughezal:2016wmq,Campbell:2016jau,Campbell:2016yrh}, \\
  &$pp\rightarrow Z$+jet~\cite{Boughezal:2015ded,Boughezal:2016isb,Boughezal:2019ggi}, $pp\to \gamma+$jet~\cite{Campbell:2016lzl}\\
  \hline
  geometric method& $H\to gg$~\cite{Herzog:2018ily} \\
\hline
\end{tabular}
\caption{\label{tab:processes}Processes calculated with the different
  methods to isolate  IR divergent real
  radiation at NNLO and beyond.}
\end{table}

\clearpage


\section{Summary and outlook}
\label{sec:summary}

The precision frontier is one of the main pillars of particle physics and will play a major role in shaping its future.
This review  aims to give an overview on the recent developments in precision calculations,
focusing on Higgs phenomenology in the first part, in particular on recent advances in fixed order calculations.
Going through various channels of Higgs boson production, the purpose is on one hand to highlight important results,
on the other hand also to point to directions where the precision should be increased in view of current or anticipated experimental data.
Obviously this cannot be done in an exhaustive way here. For example, among the most pressing issues are the matching of NNLO corrections
to parton shower Monte Carlo programs as well as the analytic resummation of large logarithms in certain kinematic regions.
These important subjects are only treated in passing in this review.
Furthermore, precision calculations within Effective Field Theory parametrisations of physics beyond the Standard Model (BSM) are not discussed,
nor higher order corrections within concrete BSM models or recent progress using machine learning tools, as these subjects deserve a review on their own.

The main directions for fixed order calculations to improve the precision can be very roughly divided into two categories:
one is to increase the perturbative order for processes with a low number of scales, 
as for example in Higgs boson production in the heavy top limit or the Drell-Yan process.
Here the main challenges are to achieve fully differential results for the highest available perturbative order.
The other category concerns an increasing number of physical scales like Mandelstam-type kinematic invariants and masses.
The number of kinematic invariants naturally increases with increasing particle multiplicity.
Therefore $2\to 3$ processes at NNLO certainly form one of the current frontiers, posing challenges from both the loop amplitude- as well as the real radiation side,
with remarkable recent progress concerning such processes with massless particles or one massive external leg.
The other challenge in the second category is given by processes with a large number of massive propagators, as they occur for example in electroweak corrections or QCD processes with massive quark loops, or a combination of both.
The more scales are involved, the more complicated the corresponding analytic expressions for the integrals can be, and the study of differential equations and integrals  whose solutions leave the class of multiple poylogarithms is one of the most vibrant current topics.
Numerical methods scale better with increasing numbers of mass scales, however they have other challenges, and tackling those has also seen much progress in the last few years.

To account for all these recent advances, the second part of this review is dedicated to techniques which are at the core of pushing the precision frontier further in both, SM and BSM calculations.
This part aims to give a broad overview on established as well as newly emerging methods, rather than an in-depth discussion of a particular method.
For some of these methods, what is called the frontier now may soon be outdated, given their potential to tackle problems that were not feasible before.
For example, in the area of amplitude reduction, methods based on finite fields or intersection numbers are very promising.
Or, concerning the calculation of multi-scale two-loop integrals, new expansion techniques as well as (semi-)numerical techniques have emerged.
Fully numerical methods, avoiding the conundrums of dimensional regularisation, also have seen a lot of progress recently.

Last but not least, a brief overview on recent developments regarding the subtraction of infrared divergent real radiation, at NNLO and beyond, is given.
These developments point to the fact that general schemes with some potential for automation, much like we have them at NLO today, are on the horizon.

Finally, with all this very encouraging progress from the mathematical as well as computational side, it should be emphasised that all this is part of a bigger picture,
which is the preparation for the future of high energy physics, in the hope that the LHC as well as future colliders will allow us to further unravel the mysteries of Nature.

\section*{Acknowledgements}

First of all I would like to thank all my collaborators for many fruitful discussions and collaborations over the last years.
I am also grateful to Ansgar Denner, Stefan Dittmaier, Aude Gehrmann-De Ridder, Stephen Jones, Matthias Kerner, German Rodrigo, Ludovic Scyboz and Lorenzo Tancredi for valuable comments on parts of the manuscript. Furthermore, I would like to thank my former colleagues at the Max Planck Institute for Physics, as well as my current colleagues at KIT, for the pleasant working atmosphere while parts of this review were written.
This research was supported by the Deutsche Forschungsgemeinschaft (DFG, German Research Foundation) under grant  396021762 - TRR 257
and by the COST Action CA16201 (`Particleface') of the European Union.

\bibliography{Precision.bib}

\end{document}